\documentclass[journal]{IEEEtran}

\ifCLASSINFOpdf
\else
\fi
\usepackage{array}
\usepackage{multirow}
\usepackage{pdflscape}
\usepackage{adjustbox}
\usepackage{amsmath}
\usepackage{amssymb}
\usepackage{mathtools}
\usepackage{mathptmx}
\usepackage{cite}
\usepackage{times}
\usepackage{graphicx}
\graphicspath{{figures/}}
\usepackage{algorithm}
\usepackage[noend]{algorithmic}
\usepackage{subfig}
\usepackage{stfloats}
\usepackage{caption}
\usepackage{color}
\usepackage{fancyvrb}
\usepackage{makeidx}
\usepackage{mdwmath}
\usepackage{mdwtab}
\usepackage{moreverb}
\usepackage{etoolbox}
\usepackage[nolist]{acronym}
\usepackage{url}

\begin{acronym}[BEREC]
\acro{10G-EPON}{10 Gigabit EPON}
\acro{3GPP}{3rd Generation Partnership Project}
\acro{5G}{Fifth Generation}
\acro{6G}{Sixth Generation}
\acro{A-CPI}{A Controller Plane Interface}
\acro{A-RoF}{Analog Radio-over-Fiber}
\acro{ABNO}{Application-Based Network Operations}
\acro{ADC}{Analog-to-Digital Converter}
\acro{AE}{Allocative Efficiency}
\acro{amc}{adaptive modulation and coding}
\acro{AON}{Active Optical Network}
\acro{AI}{Artificial Intelligence}
\acro{ML}{Machine Learning}
\acro{API}{Application Programming Interface}
\acro{AWG}{Arrayed Waveguide Grating}
\acro{B2B}{Business to Business}
\acro{B2C}{Business to Consumer}
\acro{B-RAS}{Broadband Remote Access Servers}
\acro{BaaS}{Blockchain as a Service}
\acro{BB}{Budget Balance}
\acro{BBF}{BroadBand Forum}
\acro{BBU}{Baseband Unit}
\acro{BER}{Bit Error Rate}
\acro{BEREC}{Body of European Regulators for Electronic Communications}
\acro{BF}{beamforming}
\acro{BFT}{Byzantine-Fault-Tolerant}
\acro{BGP}{Border Gateway Protocol}
\acro{BMap}{Bandwidth Map}
\acro{BPM}{Business Process Management}
\acro{BS}{Base Station}
\acro{BSC}{Base Station Controller}
\acro{BtB}{Back to Back}
\acro{BW}{Bandwidth}
\acro{C2C}{Consumer to Consumer}
\acro{C-RAN}{Cloud Radio Access Network}
\acro{C-RoFN}{Cloud-based Radio over optical Fiber Network}
\acro{CA}{Certificate Authority}
\acro{CapEx}{Capital Expenditure}
\acro{CDMA}{code division multiple access}
\acro{CDR}{Clock Data Recovery}
\acro{CFO}{carrier frequency offset}
\acro{CIR}{Committed Information Rate}
\acro{CO}{Central Office}
\acro{CoMP}{Coordinated Multipoint}
\acro{COP}{Control Orchestration Protocol}
\acro{CORD}{Central Office Rearchitected as a Data Centre}
\acro{CPE}{Customer Premises Equipment}
\acro{CPRI}{Common Public Radio Interface}
\acro{CPU}{Central Processing Unit}
\acro{CMC}{Conflict Mitigation Controller}
\acro{CMS}{Conflict Mitigation System}
\acro{CQI}{channel quality indicator}
\acro{CR}{Conflict Resolution}
\acro{CS}{Central Station}
\acro{CSI}{channel state information}
\acro{D2D}{Device to Device}
\acro{D-CPI}{D Controller Plane Interface}
\acro{D-RoF}{Digital Radio-over-Fiber}
\acro{DAC}{Digital-to-Analog Converter}
\acro{DAS}{Distributed Antenna Systems}
\acro{DBA}{Dynamic Bandwidth Allocation}
\acro{DBRu}{Dynamic Bandwidth Report upstream}
\acro{DC}{Direct Current}
\acro{DL}{Downlink}
\acro{DLT}{Distributed Ledger Technology}
\acro{DMT}{Discrete Multitone}
\acro{DNS}{Domain Name Server}
\acro{DPDK}{Data Plane Development Kit}
\acro{DSB}{Double Side Band}
\acro{DSL}{Digital Subscriber Line}
\acro{DSLAM}{Digital Subscriber Line Access Multiplexer}
\acro{DSP}{Digital Signal Processing}
\acro{DSS}{Distributed Synchronization Service}
\acro{DU}{Distributed Unit}
\acro{DUE}{D2D user equipment}
\acro{DWDM}{Dense Wave Division Multiplexing}
\acro{E-CORD}{Enterprise CORD}
\acro{EAL}{Environment Abstraction Layer}
\acro{EDF}{Erbium-Doped Fiber}
\acro{EFM}{Ethernet in the First Mile}
\acro{EMBS}{Elastic Mobile Broadband Service}
\acro{EON}{Elastic Optical Network}
\acro{EPC}{Evolved Packet Core}
\acro{EPON}{Ethernet Passive Optical Network}
\acro{eMBB}{Enhanced Mobile Broadband}
\acro{ETSI}{European Telecommunications Standards Institute}
\acro{EVM}{Error Vector Magnitude}
\acro{FANS}{Fixed Access Network Sharing}
\acro{FASA}{Flexible Access System Architecture}
\acro{FBMC}{Filterbank Multicarrier}
\acro{FCC}{Federal Communications Commission}
\acro{FDD}{Frequency Division Duplex}
\acro{FDMA}{frequency-division multiple access}
\acro{FEC}{Forward Error Correction}
\acro{FFR}{Fractional Frequency Reuse}
\acro{FFT}{Fast Fourier Transform}
\acro{FPGA}{Field-Programmable Gate Array}
\acro{FSO}{Free-Space-Optics}
\acro{FSR}{Free Spectral Range}
\acro{FTN}{faster-than-nyquist}
\acro{FTTC}{Fiber-to-the-Curb}
\acro{FTTH}{Fiber-to-the-Home}
\acro{FTTx}{Fiber-to-the-x}
\acro{G.Fast}{Fast Access to Subscriber Terminals}
\acro{GEM}{G-PON encapsulation method}
\acro{GFDM}{Generalized Frequency Division Multiplexing}
\acro{GMPLS}{Generalized Multi-Protocol Label Switching}
\acro{GPON}{Gigabit Passive Optical Network}
\acro{GPP}{General Purpose Processor}
\acro{GPRS}{General Packet Radio Service}
\acro{GSM}{Global System for Mobile Communications}
\acro{GTP}{GPRS Tunneling Protocol}
\acro{GUI}{Graphical User Interface}
\acro{HA}{Hardware Accelerator}
\acro{HARQ}{Hybrid-Automatic Repeat Request}
\acro{HD}{half duplex}
\acro{HetNet}{heterogeneous networks}
\acro{IaaS}{Infrastructure as a Service}
\acro{I-CPI}{I Controller Plane Interface}
\acro{I2RS}{Interface 2 Routing System}
\acro{IA}{Interference Alignment}
\acro{IAM}{Identity and Access Management}
\acro{IBFD}{in-band full duplex}
\acro{IC}{Incentive Compatibility}
\acro{ICIC}{inter-cell interference coordination}
\acro{ICT}{information and communications technology}
\acro{IEEE}{Institute of Electrical and Electronics Engineers}
\acro{IETF}{Internet Engineering Task Force}
\acro{IF}{Intermediate Frequency}
\acro{IFFT}{inverse FFT}
\acro{ITS}{Intelligent Transport Systems}
\acro{InP}{Infrastructure Provider}
\acro{IoT}{Internet of Things}
\acro{IP}{Internet Protocol}
\acro{IQ}{In-phase/Quadrature}
\acro{IR}{Individual Rationality}
\acro{IRC}{Interference Rejection Combining}
\acro{ISI}{inter-symbol interference}
\acro{ISP}{Intent Specification Platform}
\acro{IV}{Intelligent Vehicle}
\acro{ITU}{International Telecommunication Union}
\acro{KPI}{Key Performance Indicator}
\acro{KVM}{Kernel-based Virtual Machine}
\acro{L2}{Layer-2}
\acro{L3}{Layer-3}
\acro{LAN}{Local Area Network}
\acro{LO}{Local Oscillator}
\acro{LOS}{Line Of Sight}
\acro{LR-PON}{Long Reach PON}
\acro{LSP-DB}{Label Switched Path Database}
\acro{LTE-A}{Long Term Evolution Advanced}
\acro{LTE}{Long Term Evolution}
\acro{M-CORD}{Mobile CORD}
\acro{M-MIMO}{massive MIMO}
\acro{M2M}{machine-to-machine}
\acro{MAC}{Medium Access Control}
\acro{ME}{Merging Engine}
\acro{MIMO}{multiple input multiple output}
\acro{MME}{Mobility Management Entity}
\acro{mmWave}{millimeter wave}
\acro{MNO}{Mobile Network Operator}
\acro{MPLS}{Multiprotocol Label Switching}
\acro{MRC}{Maximum Ratio Combining}
\acro{MSC}{Mobile Switching Centre}
\acro{MSP}{Membership Service Providers}
\acro{MSR}{Multi-Stratum Resources}
\acro{MTC}{machine type communication}
\acro{mMTC}{massive Machine Type Communications}
\acro{MVNO}{Mobile Virtual Network Operator}
\acro{MVNP}{mobile virtual network provider}
\acro{MZI}{Mach-Zehnder Interferometer}
\acro{MZM}{Mach-Zehnder Modulator}
\acro{NE}{Network Element}
\acro{NFV}{Network Function Virtualization}
\acro{NFVaaS}{Network Function Virtualization as a Service}
\acro{NG-PON2}{Next-Generation Passive Optical Network 2}
\acro{NGA}{Next Generation Access}
\acro{NLOS}{None Line Of Sight}
\acro{NMS}{Network Management System}
\acro{NRZ}{Non Return-to-Zero}
\acro{NTT}{Nippon Telegraph and Telephone}
\acro{OBPF}{Optical BandPass Filter}
\acro{OBSAI}{Open Base Station Architecture Initiative}
\acro{ODN}{Optical Distribution Network}
\acro{OFC}{Optical Frequency Comb}
\acro{OFDM}{orthogonal frequency-division multiplexing}
\acro{OFDMA}{orthogonal frequency-division multiple access}
\acro{OLO}{Other Licensed Operator}
\acro{OLT}{Optical Line Terminal}
\acro{ONF}{Open Networking Foundation}
\acro{ONU}{Optical Network Unit}
\acro{OOB}{out-of-band}
\acro{OOK}{On-off Keying}
\acro{OpenCord}{Central Office Re-Architected as a Data Center}
\acro{OpEx}{Operating Expenditure}
\acro{OSI}{Open Systems Interconnection}
\acro{OSS}{Operations Support System}
\acro{OTT}{Over-the-Top}
\acro{OXM}{OpenFlow Extensible Match}
\acro{P2MP}{Ethernet over point-to-multipoint }
\acro{P2P}{Point-to-Point }
\acro{PAM}{Pulse Amplitude Modulation}
\acro{PAPR}{Peak-to-Average Power Ratio}
\acro{PBMA}{Priority Based Merging Algorithm}
\acro{PCE}{Path Computation Elements}
\acro{PCEP}{Path Computation Element Protocol}
\acro{PCF}{Photonic Crystal Fiber}
\acro{PD}{Photodiode}
\acro{PDCP}{RD Control Protocol}
\acro{PDF}{probability distribution function}
\acro{PGW}{Packet Gateway}
\acro{PHY}{physical Layer}
\acro{PIR}{Peak Information Rate}
\acro{PMD}{Polarization Division Multiplexing}
\acro{PON}{Passive Optical Network}
\acro{POTS}{Plain Old Telephone Service}
\acro{PoET}{Proof of Elapsed Time}
\acro{PoW}{Proof of Work}
\acro{PoS}{Proof of Stake}
\acro{PTP}{Precision Time Protocol}
\acro{PWM}{Pulse Width Modulation}
\acro{QAM}{Quadrature Amplitude Modulation}
\acro{QoE}{Quality of Experience}
\acro{QoS}{Quality of Service}
\acro{QPSK}{Quadrature Phase Shift Keying}
\acro{R-CORD}{Residential CORD}
\acro{RA}{Resource Allocation}
\acro{RAN}{Radio Access Network}
\acro{RAT}{Radio Access Technology}
\acro{RB}{resource block}
\acro{RFIC}{radio frequency integrated circuit}
\acro{RIC}{RAN Intelligent Controller}
\acro{RLC}{Radio Link Control}
\acro{RN}{Remote Node}
\acro{ROADM}{Reconfigurable Optical Add Drop Multiplexer}
\acro{RoF}{Radio-over-Fiber}
\acro{RRC}{Radio Resource Control}
\acro{RRH}{Remote Radio Head}
\acro{RRPH}{Remote Radio and PHY Head}
\acro{RRU}{Remote Radio Unit}
\acro{RSOA}{Reflective Semiconductor Optical Amplifier}
\acro{RU}{Radio Unit}
\acro{Rx}{receiver}
\acro{SC-FDMA}{single carrier frequency division multiple access}
\acro{SCM}{single carrier modulation}
\acro{SD-RAN}{Software Defined Radio Access Network}
\acro{SDAN}{Software Defined Access Network}
\acro{SDMA}{Semi-Distributed Mobility Anchoring}
\acro{SDN}{Software Defined Networking}
\acro{SDR}{Software Defined Radio}
\acro{SFBD}{Single Fiber Bi-Direction}
\acro{SGW}{Serving Gateway}
\acro{SHAP}{SHapley Additive exPlanations}
\acro{SI}{self-interference}
\acro{SIC}{self-interference cancellation}
\acro{SIMO}{Single Input Multiple Output}
\acro{SLA}{Service Level Agreement}
\acro{SMF}{Single Mode Fiber}
\acro{SNMP}{Simple Network Management Protocol}
\acro{SNR}{Signal-to-Noise Ratio}
\acro{SP}{Service Providers}
\acro{Split-PHY}{Split Physical Layer}
\acro{SRI-OV}{Single Root Input/Output Virtualization}
\acro{SU}{secondary user}
\acro{SUT}{System Under Test}
\acro{T-CONT}{Transmission Container}
\acro{TC}{Transmission Convergence}
\acro{TCO}{Total Cost of Ownership}
\acro{TD-LTE}{Time Division LTE}
\acro{TDD}{Time Division Duplex}
\acro{TDM}{Time Division Multiplexing}
\acro{TDMA}{Time Division Multiple Access}
\acro{TED}{Traffic Engineering Database}
\acro{TEID}{Tunnel endpoint identifier}
\acro{TLS}{Transport Layer Security}
\acro{TPS}{Transactions per Second}
\acro{TTI}{transmission time interval}
\acro{TWDM}{Time and Wavelength Division Multiplexing}
\acro{Tx}{transmitter}
\acro{UD-CRAN}{Ultra-Dense Cloud Radio Access Network}
\acro{UDP}{User Datagram Protocol}
\acro{UE}{User Equipment}
\acro{UF-OFDM}{Universally Filtered OFDM}
\acro{UFMC}{Universally Filtered Multicarrier}
\acro{UL}{Uplink}
\acro{UMTS}{Universal Mobile Telecommunications System}
\acro{URLLC}{Ultra-Reliable Low-Latency Communication}
\acro{USRP}{Universal Software Radio Peripheral}
\acro{V2I}{Vehicle to Infrastructure}
\acro{V2V}{Vehicle to Vehicle}
\acro{vBBU}{virtualized BBU}
\acro{vBMap}{Virtual Bandwidth Map}
\acro{vBS}{virtual Base station}
\acro{VCG}{Vickrey--Clarke--Groves}
\acro{vCPE}{virtual CPE}
\acro{vCPU}{Virtual Central Processing Unit}
\acro{vDBA}{virtual DBA}
\acro{VDSL2}{Very-high-bit-rate digital subscriber line 2}
\acro{VLAN}{Virtual Local Area Network}
\acro{VM}{Virtual Machine}
\acro{VNF}{Virtual Network Function}
\acro{VNI}{visual networking index}
\acro{VNO}{Virtual Network Operator}
\acro{VNTM}{Virtual Network Topology Manager}
\acro{vOLT}{virtual OLT}
\acro{VPE}{virtual Provider Edge}
\acro{VPN}{Virtual Private Network}
\acro{V2X}{Vehicle to Everything}
\acro{VTN}{Virtual Tenant Network}
\acro{VULA}{Virtual Unblunded Local Access}
\acro{WAN}{Wide Area Network}
\acro{WAP}{Wireless Access Point}
\acro{WCDMA}{wide-band code division multiple access}
\acro{WDM-PON}{Wavelength Division Multiplexing - Passive Optical Network}
\acro{WDM}{Wavelength Division Multiplexing}
\acro{WiMAX}{Worldwide Interoperability for Microwave Access}
\acro{WRPR}{Wired-to-RF Power Ratio}
\acro{XG-PON}{10 Gigabit PON}
\acro{XGEM}{XG-PON encapsulation method}
\acro{XOS}{XaaS Operating System}
\acro{Near-RT-RIC}{Near Real-Time RAN Intelligent Controller}
\acro{Non-RT-RIC}{Non-Real-Time RAN Intelligent Controller}
\acro{xApp}{Near-RT RIC Application}
\acro{rApp}{Non-RT RIC Application}
\acro{CMF}{Conflict Mitigation Framework}
\acro{EGS}{Eisenberg-Gale Solution}
\acro{NSWF}{Nash Social Welfare Function}
\acro{CDC}{Conflict Detection Controller}
\acro{PMon}{Performance Monitoring}
\acro{RCP}{Recently Changed Parameter}
\acro{PGD}{Parameter Group Definition}
\acro{RCPG}{Recently Changed Parameter Group}
\acro{DCKD}{Decision Correlated with KPI Degradation}
\acro{KDO}{KPI Degradation Occurrences}
\acro{SDL}{Shared Data Layer}
\acro{SON}{Self-Organising Network}
\acro{SF}{SON Function}
\acro{SONCO}{SON Coordinator}
\acro{AM}{Arithmetic Mean}
\acro{PKR}{Parameter and KPI Ranges}
\acro{KPM}{Key Performance Metric}
\acro{ICP}{Input Control Parameter}
\acro{RF}{Radio Frequency}
\acro{NCP}{Network Configuration Parameter}
\acro{CIO}{Cell Individual Offset}
\acro{MRO}{Mobility Robustness Optimisation}
\acro{MLB}{Mobility Load Balancing}
\acro{CCO}{Coverage and Capacity Optimisation}
\acro{ES}{Energy Saving}
\acro{CAN}{Cognitive Autonomous Network}
\acro{CF}{Cognitive Function}
\acro{OCRS}{Optimal Configuration Range Set}
\acro{CW}{Config-Weight}
\acro{SCL}{Synchronized Cooperative Learning}
\acro{HOAP}{Handover Aggregate Performance}
\acro{DDQN}{Double Deep Q-Network}
\acro{PER}{Prioritised Experience Replay}
\acro{CADC}{Conflict-Aware Distributed Coordination}
\acro{SLC2}{Scalable Low-Complexity Coordination}
\acro{DRL}{Deep Reinforcement Learning}
\acro{MARL}{Multi-Agent Reinforcement Learning}
\acro{GNN}{Graph Neural Network}
\acro{GCN}{Graph Convolutional Network}
\acro{COTS}{Commercial Off-The-Shelf}
\acro{PG}{Parameter Group}
\acro{QACM}{QoS-Aware Conflict Mitigation}
\acro{MATL}{Multi-Agent Team Learning}
\acro{GenC}{Synthetic Conflict Generation Framework}
\acro{XAI}{Explainable Artificial Intelligence}
\acro{FL}{Federated Learning}
\acro{DT}{Digital Twin}
\acro{PPO}{Proximal Policy Optimisation}
\acro{MAB}{Multi-Armed Bandit}
\acro{UCB}{Upper Confidence Bound}
\acro{RLF}{Radio Link Failure}
\acro{LLM}{Large Language Model}
\acro{RAG}{Retrieval-Augmented Generation}
\acro{IBN}{Intent-Based Networking}
\acro{SMO}{Service Management and Orchestration}
\acro{O-RAN}{Open Radio Access Network}
\acro{vRAN}{Virtualised Radio Access Network}
\acro{D-RAN}{Distributed Radio Access Network}
\acro{T-RAN}{Traditional Radio Access Network}
\acro{H-CRAN}{Heterogeneous Cloud Radio Access Network}
\acro{O-CU}{O-RAN Central Unit}
\acro{O-DU}{O-RAN Distributed Unit}
\acro{O-RU}{O-RAN Radio Unit}
\acro{Near-RT}{Near-Real-Time}
\acro{Non-RT}{Non-Real-Time}
\acro{dApp}{Distributed RAN Application}
\acro{RL}{Reinforcement Learning}
\acro{GT}{Game Theory}
\acro{MDP}{Markov Decision Process}
\acro{SVM}{Support Vector Machine}
\acro{NSGA-II}{Non-dominated Sorting Genetic Algorithm II}
\acro{PCA}{Principal Component Analysis}
\acro{ANN}{Artificial Neural Network}
\acro{LSTM}{Long Short-Term Memory}
\acro{Bi-LSTM}{Bidirectional Long Short-Term Memory}
\acro{GRU}{Gated Recurrent Unit}
\acro{SMOTE}{Synthetic Minority Oversampling Technique}
\acro{CTDE}{Centralised Training with Decentralised Execution}
\acro{DQN}{Deep Q-Network}
\acro{A2C}{Advantage Actor-Critic}
\acro{SPSA}{Simultaneous Perturbation Stochastic Approximation}
\acro{PDDL}{Planning Domain Definition Language}
\acro{NBS}{Nash Bargaining Solution}
\acro{WNBS}{Weighted Nash Bargaining Solution}
\acro{KSBS}{Kalai--Smorodinsky Bargaining Solution}
\acro{SEBS}{Shannon Entropy Bargaining Solution}
\acro{PRB}{Physical Resource Block}
\acro{RSRP}{Reference Signal Received Power}
\acro{RSS}{Received Signal Strength}
\acro{RET}{Remote Electrical Tilt}
\acro{TTT}{Time to Trigger}
\acro{CRE}{Cell Range Expansion}
\acro{COD}{Cell Outage Detection}
\acro{eNodeB}{Evolved Node B}
\acro{C-SON}{Centralised Self-Organising Network}
\acro{D-SON}{Distributed Self-Organising Network}
\acro{WG2}{Working Group 2}
\acro{WG3}{Working Group 3}
\acro{B5G}{Beyond Fifth Generation}
\acro{OSC}{O-RAN Software Community}
\acro{OAI}{OpenAirInterface}
\acro{OTA}{Over-the-Air}
\acro{OTIC}{Open Testing and Integration Centre}
\acro{OAIC}{Open AI Cellular}
\acro{UPF}{User Plane Function}
\acro{MEC}{Mobile Edge Computing}
\acro{ONAP}{Open Network Automation Platform}
\acro{NF}{Network Function}
\acro{OAM}{Operations, Administration and Maintenance}
\acro{PM}{Performance Management}
\acro{CM}{Configuration Management}
\acro{CTMC}{Continuous-Time Markov Chain}
\acro{ILP}{Integer Linear Programming}
\acro{FWA}{Fixed Wireless Access}
\acro{PELT}{Pruned Exact Linear Time}
\acro{FSM}{Finite-State Machine}
\acro{NR}{New Radio}
\acro{NRMSE}{Normalised Root Mean Square Error}
\acro{TS}{Technical Specification}
\acro{DCD}{Direct Conflict Detection}
\acro{ICD}{Indirect Conflict Detection}
\acro{ImCD}{Implicit Conflict Detection}
\acro{MaxTS}{Maximum Throughput-based Selection}
\acro{MinPS}{Minimum Power-based Selection}
\acro{EES}{Energy Efficiency-based Selection}
\acro{TVS}{Throughput Violation-based Selection}
\acro{PIOR}{Platform for Interoperable O-RAN}
\acro{IDNAFO}{Intent-Driven Network Automation Function Orchestrator}
\acro{CBM}{Cognitive Bargaining Mechanism}
\acro{OOTB}{Out-of-the-Box}
\acro{RNN}{Recurrent Neural Network}
\acro{gNB}{Next Generation Node B}
\acro{HeNB}{Home Evolved Node B}
\acro{eNB}{Evolved Node B}
\acro{NSA}{Non-Standalone}
\acro{SA}{Standalone}
\acro{CU}{Central Unit}
\acro{xNIB}{xApp Network Information Base}
\acro{DAG}{Directed Acyclic Graph}
\acro{ORCA}{Open RAN Conflict Agents}
\acro{ZODIAC}{Zero-shot Offline Diffusion for Inferring Multi-xApps Conflicts in Open Radio Access Networks}
\acro{TACIT}{Trustworthy Arbitration through Calibrated Inference Tracking}
\acro{ORACLE}{Open RAN Arbitration via Consensus Ledger Enforcement}
\acro{CAMINO}{Contextually Aware Mediation of Intent for Network Orchestration}
\acro{MACOL-x}{Multi-Agent Conflict Learning for xApp Coordination}
\acro{ACCoRD}{Actor-Critic Conflict Resolution with Deep Learning}
\acro{CoDR}{Conflict Detection and Resolution}
\acro{MABCR}{Multi-Armed-Bandit Conflict Resolution}
\acro{M2O-IBNS}{Multi-Objective Intent-Based Network Slicing}
\acro{ES-NAM}{Energy-Saving Neural Additive Model}
\acroplural{UE}[UEs]{User Equipment}
\end{acronym}

\begin{document}
%

\title{{\normalfont\footnotesize This version of the paper has been submitted for publication to IEEE and may be removed upon request or copyright issues.}\\[1.2ex]
Conflict Management in Autonomous Radio Access Networks: A Lifecycle Perspective from SON to AI-Native 6G}

%
%

\author{Abdul Wadud,~\IEEEmembership{Graduate Student Member,~IEEE,}
Rashmi Sharan Sinha,~\IEEEmembership{Member,~IEEE,}
Ali Deylamy,
Fatemeh Golpayegani,~\IEEEmembership{Senior Member,~IEEE,}
and~Nima Afraz,~\IEEEmembership{Senior Member,~IEEE}
\thanks{Abdul Wadud is with the School of Computer Science, University College Dublin (UCD), Ireland, and the Bangladesh Institute of Governance and Management (BIGM), Dhaka, Bangladesh.}
\thanks{Rashmi Sharan Sinha, Fatemeh Golpayegani, and Nima Afraz are with the School of Computer Science, University College Dublin, Ireland.}
\thanks{Ali Deylamy is with University College Dublin, Ireland.}
\thanks{Manuscript submitted for review, June 2026.}}

%
%

\markboth{IEEE Communications Surveys \& Tutorials,~Vol.~, No.~, 2026}%
{Wadud \MakeLowercase{\textit{et al.}}: Conflict Management in Autonomous Radio Access Networks}
%



\maketitle

\acresetall
\begin{abstract}
The autonomous management of \acp{RAN} depends on the automatic tuning of the control parameters to optimise different network functions through network control application. Conflict occurs among these control applications when  multiple applications try to modify the same parameter, associated parameter, or radio resource through successive requests. These conflicts cause performance degradation if not handled properly. Conflict detection and management can be performed proactively, before actions are executed, or reactively, after execution. This survey systematically examines conflict management in \acp{SON}, \ac{CAN}, and \ac{O-RAN} using a unified classification scheme and considering the application lifecycle perspective. It discusses direct, indirect, and implicit conflicts across these three architectures. As the level of accessibility and control for control applications in different architectures is different, we consider a seven-stage lifecycle perspective for conflict management across these architectures. These stages are design-time avoidance, conflict model construction, runtime detection and classification, resolution and arbitration, enforcement, verification/attribution/feedback, and governance related to application lifecycle management. This survey make a distinction in case of \ac{O-RAN} as the \ac{RIC} interface for application deployements can be in two different logical layers. The first one is horizontal conflicts, which occur within and between \acp{RIC}, and the second one is vertical conflicts, which span control layers. This survey studies conflict detection, classification, and mitigation approaches ranging from rule-based graph-based, game-theoretic, reinforcement learning, federated learning, to intent-driven approaches. A total of {106} papers are reviewed and classified across different life-cycle and management approaches. This survey finds that simulation-based evaluation methods dominates in \ac{SON} and \ac{CAN} conflict studies, whereas testbeds and data-driven validation are frequently used in \ac{O-RAN}. We observed that there is lacking for labelled conflict dataset and end-to-end evaluation of implicit conflict in the literature. Also, there are significant gaps in cross-layer conflict management and post-resolution verification.
\end{abstract}
\acresetall 


%
\IEEEpeerreviewmaketitle

\section{Introduction}
%
%
%
%
\IEEEPARstart{T}{he} wireless \ac{RAN} has evolved immensely to meet the requirements of bandwidth-hungry and latency-sensitive applications such as \ac{IoT}-based automation, connected and autonomous vehicles, high-definition video streaming, and immersive real-time services \cite{singh2020evolution}. These applications have heterogenous requirements and often set conflicting requirements for \ac{RAN} management. For example, connected vehicles demand ultra-reliable low-latency communication \cite{cano2018evolution}, whereas massive \ac{IoT} deployments prioritise wide coverage with low power consumption \cite{ohn2016looking}. Similarly, mobile broadband applications seek peak throughput in dense urban scenarios, whereas energy-saving applications for \ac{RAN} control prefers lower transmission power. Meeting all these requirements within a single cost-effective network management architecture has been the central issue in every generation of mobile communication systems.

\par The simplest solution to satisfy these requirements is to create a separate network for each of these applications. However, operating many small networks simultaneously is complex and requires high \ac{CapEx} to create them separately for different services. Therefore, industry and academia are working together to establish a single networking system that can meet a wide range of requirements, is easy to operate, and has lower \ac{CapEx}. Researchers from industry and academia found a feasible solution to meet these aforementioned requirements. Splitting the \ac{RAN} base station into several entities by their functionality makes the architecture more centralized, flexible, and smarter \cite{arnold20175g}. The new architecture was cloud-based, easy to operate, and energy-efficient, and was named \ac{C-RAN}, often called a centralised \ac{RAN}. The \ac{C-RAN} architecture splits the traditional \ac{BS} of \ac{RAN} into two different components: the \ac{BBU} pool and the \ac{RRH} \cite{gavrilovska2020cloud}. The \ac{BBU} pool is the collection of \acp{BBU} that connects all the \acp{RRH} via the fronthaul, which consists of optical fiber cables. The \ac{BBU} pool is located at the centralized data center (the cloud), which manages all the \acp{RRH} using \acp{VNF}. \acp{RRH} are located at the cell side that connects \acp{UE} and performs digital-to-analogue conversions or vice versa. Multiple \acp{RRH} can be connected to a single \ac{BBU} via a fiber-optic cable, which provides low latency and higher bandwidth while reducing \ac{OpEx} \cite{tonini2019cost, gavrilovska2017flexible}. However, \ac{C-RAN} has strict fronthaul capacity and latency requirements. This raises security and trust concerns in shared cloud environments and creates tight coupling between hardware and software that limits operational flexibility \cite{tian2018trusted, checko2014cloud}.

\par \ac{vRAN} took the centralisation concept further by replacing dedicated baseband hardware with software-defined \acp{VNF} running on the \ac{COTS} servers. This introduction to software-defined \acp{VNF} separates hardware from software and enables flexible resource scaling and cloud-native management. Although \ac{vRAN} reduces hardware dependence, it introduces new challenges of real-time processing, multi-vendor orchestration, and synchronisation. This motivates the need for further architectural innovation towards fully open, programmable, and disaggregated networks \cite{checko2014cloud, polese2022understanding}.

\par The \ac{O-RAN} is the new architecture that addresses the aforementioned limitations of \ac{vRAN} and introduces new functional splits, disaggregating the \ac{BBU}. The new components are called the \ac{O-CU}, \ac{O-DU}, and \ac{O-RU}. \ac{O-RAN} introduces programmable network management through a set of logical layers called \acp{RIC} \cite{polese2022understanding}. The \ac{Near-RT} \ac{RIC} hosts \acp{xApp} that close radio control loops on timescales of 10\,ms to 1\,s. At the same time, the \ac{Non-RT} \ac{RIC} manages \ac{AI}/\ac{ML} model training, policy, and intent translation through \acp{rApp} on timescales above 1\,s. All these applications can be developed by any vendors, which solves the vendor lock-in issue in the \ac{RAN} ecosystem. These new features reduces dependence on proprietary hardware and give flexibility to the \acp{MNO} to choose from various options.  However, it simultaneously brings a new challenge of creating a coordination among these control applications. Otherwise, they often conflict with each other due to conflicting objectives and shared radio resources that degrades network performance \cite{ric_oran_alliance, wadud2026ai}. 

\par The problem of control decision conflict didn't originate with \ac{O-RAN} though. In 4G \ac{LTE}, the introduction of \ac{SON} first brings this problem into light. When multiple \acp{SF}, each optimising a distinct \ac{KPM} operates together, conflict may occur among them during network operation if a shared \ac{NCP} is modified in subsequent request by two different \acp{SF} \cite{lateef2015lte}. In this paper, we use \ac{KPM} and \ac{KPI} interchangeably as both indicate key performance metrics. \ac{KPM} is mostly used for \ac{SON} and \ac{KPI} is used for \ac{O-RAN}. Also, \ac{NCP} and \ac{ICP} are interchangeably used in the paper, as they mean the same thing. \ac{CAN} extends \ac{SON} by replacing rule-based functions with learning-based \acp{CF} that adapt their policies based on the network state. This escalates conflicts further as independent \ac{AI}-agents pursuing competing objectives can reach unstable equilibria that static priority rules cannot resolve \cite{banerjee2020ran, banerjee2021intent}. \ac{O-RAN} takes this conflict to a higher level as it deploys multiple control applications for the same purpose in different logical hierarchies based on latency requirements, such as \acp{rApp}, \acp{xApp}, and \acp{dApp} \cite{ORANSC2020, ORANSC2025}. According to ORAN ALLIANCE's definition, these conflicts are categorised into direct conflicts (two \acp{xApp} writing the same parameter to different values), indirect conflicts (\acp{xApp} affecting the same \ac{KPI} through different parameters), and implicit conflicts (a \ac{KPI} degraded as an unintended side-effect of another \ac{xApp}'s optimisation) \cite{ric_oran_alliance, wadud2023conflict, adamczyk2023conflict}.

\par If the conflicts are not managed properly, the foundational promise of self-optimising \ac{AI}-\ac{RAN} will not be achieved. Rather than achieving the performance gain, a conflict scenario degrades the network performance because of sustained \ac{KPI} degradation \cite{adamczyk2023conflict, wadud2024qacm}. This problem is expected to grow even more rapidly with mature \ac{O-RAN} deployment for 6G and \ac{AI}-native \ac{RAN} as hundreds of vendor-diverse \acp{xApp} interact over a shared radio environment \footnote{Operator-grade \ac{RIC} clusters are already dimensioned for ``potentially hundreds of \acp{xApp}'' in dense deployments \cite{cormoran25}; Ericsson's public \ac{rApp} Directory markets itself as a one-stop shop for hundreds of \acp{rApp} \cite{ericssonRappDir24}; and industry analysts count more than 50 independent vendors actively building commercial \acp{xApp} and \acp{rApp} \cite{snstelecom24}.}. Using exhaustive manual coordination for conflict in these settings becomes more challenging. The existing works prove how uncoordinated conflicts affects handover performance \cite{salmi2026ai}, reduces throughput \cite{wadud2025xApp}, and how misconfigurations amplify rather than resolve \ac{KPI} violations \cite{yungaicela2024misconfiguration}. These findings establish that conflict management is not a peripheral problem; rather, it is a foundational concern for \ac{AI}-native \ac{RAN} automation.

\par Existing works cover only part of this problem. \ac{SON} surveys discuss self-coordination within an integrated \ac{RAN} automation, \ac{CAN} papers focus on controller-based coordination among \acp{CF}, and \ac{O-RAN} considers conflict as an architectural challenge. Also, the \ac{O-RAN} Alliance specifies conflict management in their working group specifications without surveying their cross-paradigm lineage. Till we finish writing this survey, we found one important \ac{O-RAN}-specific conflict survey that we compare with the positioning of this survey below \cite{almeida2026survey}. We emphasise more on the complete management lifecycle and its evaluation evidence across \ac{SON}, \ac{CAN}, and \ac{O-RAN} by reviewing published papers from 2012 to November 2026.

\subsection{Positioning Against Prior Surveys and Related Syntheses}
\label{subsec:prior_surveys}
The novelty and scope of this survey is illustrated by Table~\ref{tab:prior_survey_comparison}. Previous works focusing on \ac{SON} only consider conflict among \acp{SF} and doesn't cover \ac{CAN} and \ac{CF}-based conflicts \cite{lateef2015lte,bayazeed2021survey}. Also, multi-tier \ac{RIC} applications like \acp{xApp}, \acp{rApp}, and \acp{dApp} are not considered altogether. Banerjee et~al. studies coordination among \acp{CF} in \ac{CAN} extensively. However, most of the works focus on conflict-specific conflict management design rather than a cross-architecture survey~\cite{banerjee2022toward}. Alam et~al. provide a broad \ac{O-RAN} tutorial in which conflicts are one challenge among many~\cite{alam2025comprehensive}. Almeida et~al.provide the first dedicated \ac{O-RAN} conflict-management survey and cover both detection and mitigation but neither trace the \ac{SON}/\ac{CAN} lineage nor organise evidence across the full lifecycle~\cite{almeida2026survey}.

\begin{table*}[!t]
\centering
\small
\caption{Comparison with prior surveys, standards, and paradigm-level syntheses. ``Partial'' indicates that the topic is discussed but is not a dedicated analytical dimension.}
\label{tab:prior_survey_comparison}
\setlength{\tabcolsep}{3.5pt}
\renewcommand{\arraystretch}{1.18}
\begin{adjustbox}{max width=\textwidth}
\begin{tabular}{lcccccc}
\hline
\textbf{Study} & \textbf{Paradigm(s)} & \textbf{Conflict categorisation} & \textbf{Detection} & \textbf{Mitigation} & \textbf{Conflict scope} & \textbf{Evaluation-method analysis} \\
\hline
Lateef~et~al.~\cite{lateef2015lte} & \ac{SON} & \ac{SON} \ac{NCP}/\ac{KPM} classes & Partial & Yes & \ac{SON}-function level & No \\
Bayazeed~et~al.~\cite{bayazeed2021survey} & \ac{SON} & \ac{SON}-specific & Partial & Yes & \ac{SON}-function level & Partial \\
Banerjee~et~al.~\cite{banerjee2022toward} & \ac{CAN} & \ac{CAN}-specific & Partial & Yes & \ac{CF}--Controller & No \\
Alam~et~al.~\cite{alam2025comprehensive} & \ac{O-RAN} & Brief/architectural & Brief & Brief & \ac{RIC} architecture & No \\
\ac{O-RAN} \ac{WG3}~\cite{oran_wg3_confmit_draft} & \ac{O-RAN} & Direct/indirect/implicit & Procedural & Procedural & Mainly intra-\ac{Near-RT} \ac{RIC} & No \\
Almeida~et~al.~\cite{almeida2026survey} & \ac{O-RAN} & \ac{O-RAN}-specific & Yes & Yes & Horizontal/vertical & Partial \\
\textbf{This survey} & \textbf{\ac{SON}, \ac{CAN}, \ac{O-RAN}} & \textbf{Unified structural mapping} & \textbf{Yes} & \textbf{Yes} & \textbf{Intra-\ac{RIC}/inter-\ac{RIC}/vertical} & \textbf{Dedicated analysis} \\
\hline
\end{tabular}
\end{adjustbox}
\end{table*}

Our previous work in \cite{wadud2026ai} introduces a \ac{GenC} for conflict classification to handle conflict when there is data scarcity. It trains and evaluates \ac{AI}-models on both the \ac{GenC}-generated dataset and the ns3-oran dataset. \ac{AI}-models used are \ac{GNN}, \ac{Bi-LSTM} and \ac{SMOTE}-\ac{GNN} models for conflict classification of all three types of conflicts. This survey uses all our previously published works and other related state-of-the-art literature as units of analysis and synthesizes their core concepts in a life-cycle perspective. We did not use any dataset, classifier results or performance claims from any of our previously published papers.

\subsection{Contributions}
To the best of our knowledge, this survey is the first of its kind that covers conflict in wireless radio access networks across \ac{SON}, \ac{CAN}, and \ac{O-RAN} as one evolutionary problem. Also, this is the first survey that combines conflict management literature across these architectures in a life-cycle perspective. The main contributions of this survey are as follows:

\begin{itemize}
    \item \textbf{Unified conflict taxonomy:} We show that conflicts in wireless radio access networks are structurally the same across \ac{SON}, \ac{CAN}, and \ac{O-RAN} and can be categorised as direct, indirect and implicit. The only difference across these architectures are observability and enforceability issues. But the core idea of \ac{SON}'s \ac{NCP}/\ac{KPM} conflicts, \ac{CAN}'s configuration/measurement/characteristic conflicts, and the \ac{O-RAN} Alliance's \ac{xApp} conflict types are the same. We construct a unified four-axis conflict descriptor $C=(T,S,L,P)$ that captures different structural types $T$, architectural score $S$, lifecycle steps $L$, and application contexts $P$ across these various architectures for an individual conflict to establish a common representation of cross-paradigm representations.

    \item \textbf{Critical analysis of detection methods:} We review and categorise different conflict detection methods found in the existing literature across these three paradigms.

    \item \textbf{Critical analysis of mitigation strategies:} We analyse mitigation techniques that are used across these three paradigms ranging from  policy-based coordination and game-theory to \ac{AI}-driven approaches.

    \item \textbf{Conflict scope classification:} We present a conflict scope classification that distinguishes horizontal intra-\ac{RIC} conflicts (between applications within a single \ac{RIC} tier), horizontal inter-\ac{RIC} conflicts (across separate \ac{RIC} instances), and vertical cross-layer conflicts (between \ac{Non-RT} \ac{RIC} \acp{rApp}, \ac{Near-RT} \ac{RIC} \acp{xApp}, and \ac{RAN} \acp{dApp}). This provides a clear conflict scope, particularly for \ac{O-RAN}, that complements the existing direct/indirect/implicit taxonomy.

    \item \textbf{Structured classification of the literature:} We present a structured classification of all the reviewed 106 papers. We categorise each work by domain, conflict types, detection and mitigation scope, \ac{AI}/\ac{ML} methods employed, and validation methodology adopted in each of these reviewed papers.

    \item \textbf{Evaluation methodology analysis:} We analyse the simulation and testbed used for evaluating conflicts across these three paradigms and identify the absence of a cross-paradigm conflict benchmark.

    \item \textbf{Open challenges and research directions:} We identify the challenges and future research directions for conflict management issues focusing on the management lifecycle that spans from scalability in future \ac{AI}-native \acp{RAN}, agentic-\ac{AI}-based governance, benchmarking, post-resolution verification, cross-layer conflict management, and implicit conflict handling, as it is the one studied less in the literature. 
\end{itemize}

\subsection{Organisation}
The remaining part of the paper is organised as follows. Section~\ref{sec:back} discusses the evolution of \ac{RAN} architectures from traditional distributed deployments to cloud-based \ac{RAN} automation.  Section~\ref{sec:oran_son_can} investigates the cross-paradigm conflict similarity and dissimilarity of conflicts across \ac{SON}, \ac{CAN}, and \ac{O-RAN}. Section~\ref{sec:lifecycle} reviews the literature by conflict-management lifecycle stages. Section~\ref{sec:eval_challenges} analyses evaluation methodologies and benchmarking gaps. Section~\ref{sec:lessons} synthesises lessons learned and future research directions. Finally, Section~\ref{sec:conclusion} concludes the paper.

\begin{table*}[!ht]
\centering
\caption{Abbreviations used in this paper.}
\label{tab:abbrev}
\begin{tabular}{|l|l||l|l|}
\hline
\textbf{Abbreviation} & \textbf{Full Form} & \textbf{Abbreviation} & \textbf{Full Form} \\
\hline
3GPP  & 3rd Generation Partnership Project   & KPI   & Key Performance Indicator \\
\hline
5G    & Fifth Generation                      & KPM   & Key Performance Metric \\
\hline
6G    & Sixth Generation                      & LTE   & Long Term Evolution \\
\hline
AI    & Artificial Intelligence               & MARL  & Multi-Agent Reinforcement Learning \\
\hline
API   & Application Programming Interface    & MLB   & Mobility Load Balancing \\
\hline
BBU   & Baseband Unit                        & ML    & Machine Learning \\
\hline
CAN   & Cognitive Autonomous Network         & MNO   & Mobile Network Operator \\
\hline
CADC  & Conflict-Aware Distributed Coordination & MRO & Mobility Robustness Optimisation \\
\hline
CCO   & Coverage and Capacity Optimisation    & NCP   & Network Configuration Parameter \\
\hline
CF    & Cognitive Function                    & Near-RT & Near Real-Time \\
\hline
CIO   & Cell Individual Offset                & NFV   & Network Function Virtualisation \\
\hline
CMF   & Conflict Mitigation Framework         & Non-RT & Non-Real-Time \\
\hline
COTS  & Commercial Off-The-Shelf              & O-RAN & Open Radio Access Network \\
\hline
C-RAN & Cloud Radio Access Network            & PG    & Parameter Group \\
\hline
D-RAN & Distributed Radio Access Network      & PPO   & Proximal Policy Optimisation \\
\hline
DDQN  & Double Deep Q-Network                 & QACM  & QoS-Aware Conflict Mitigation \\
\hline
DRL   & Deep Reinforcement Learning           & QoS   & Quality of Service \\
\hline
DT    & Digital Twin                          & RAN   & Radio Access Network \\
\hline
DU    & Distributed Unit                      & RIC   & RAN Intelligent Controller \\
\hline
EGS   & Eisenberg-Gale Solution               & RLF   & Radio Link Failure \\
\hline
ES    & Energy Saving                         & RRH   & Remote Radio Head \\
\hline
FL    & Federated Learning                    & SDN   & Software Defined Networking \\
\hline
GCN   & Graph Convolutional Network           & SLA   & Service Level Agreement \\
\hline
GNN   & Graph Neural Network                  & SLC2  & Scalable Low-Complexity Coordination \\
\hline
GT    & Game Theory                           & SON   & Self-Organising Network \\
\hline
HOAP  & Handover Aggregate Performance        & SONCO & SON Coordinator \\
\hline
ICP   & Input Control Parameter               & UE    & User Equipment \\
\hline
IBN   & Intent-Based Networking               & SMO   & Service Management and Orchestration \\
\hline
IoT   & Internet of Things                    & vRAN  & Virtualised Radio Access Network \\
\hline
ISP   & Intent Specification Platform         & xApp  & Near-RT RIC Application \\
\hline
LLM   & Large Language Model                  & RAG   & Retrieval-Augmented Generation \\
\hline
XAI   & Explainable Artificial Intelligence   & NSWF  & Nash Social Welfare Function \\
\hline
SHAP  & SHapley Additive exPlanations         & SA-RAG & Segmented Active RAG \\
\hline
T-RAN & Traditional Radio Access Network       & SF     & SON Function \\
\hline
OpEx & Operating Expenditure & rApp & Non-RT RIC Application \\
\hline
VNF & Virtual Network Function & dApp & Distributed RAN Application \\
\hline
\end{tabular}
\end{table*}

\begin{figure*}[!ht]
\centering
\includegraphics[scale=0.48]{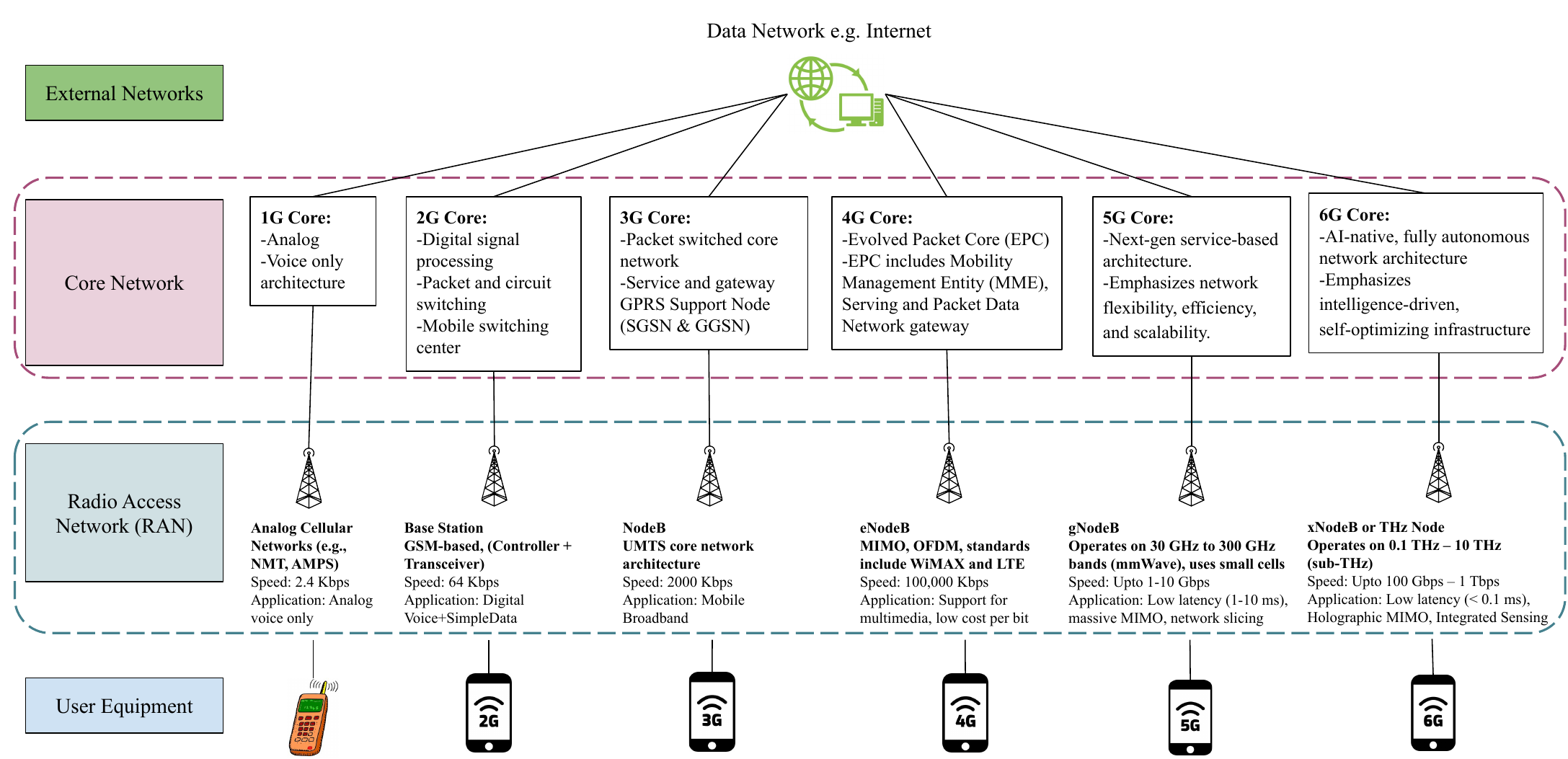}
\caption{Evolution of mobile communication systems and \ac{RAN} architectures from 1G to 6G.}
\vspace{-0.5cm}
\label{fig:mobsys_evolution}
\end{figure*}

\section{Generational Evolution, \ac{RAN} Architectures, and Control Conflicts}
\label{sec:back}

The \ac{RAN} has gone through significant transformation since the introduction of mobile communication systems. In the first two generations of the mobile communication systems, the \ac{BBU} in the base station use to be a blackbox coming from a handful of vendors and all radio functions used to be managed from the \ac{BBU} site. However, it required heterogenous and denser \ac{BBU} placements in later generations of \ac{RAN} like 3G and 4G, therefore, there was a need for architectural centralisation (i.e., \ac{C-RAN}/\ac{H-CRAN}) and automation layers like \ac{SON}. During late 4G and 5G, network functions became virtualised (\ac{vRAN}) with the motivation of reducing hardwares and to save \ac{OpEx} and \ac{CapEx}. Later, the concept of \ac{O-RAN} arrived with \acp{RIC} and \acp{xApp} for \ac{AI}-dirven network automation and management. \ac{O-RAN} also enables multi-vendor programmability, reducing dependence on a handful of vendors. Figure~\ref{fig:mobsys_evolution} illustrates the evolution of mobile communication systems from 1G to 6G along with the evolution of core and \ac{RAN} architectures.

\subsection{Traditional Radio Access Network}
\label{trad_ran}
\begin{figure}[!ht]
 \centering
 \vspace{-0.1in}
	\includegraphics[scale=0.37]{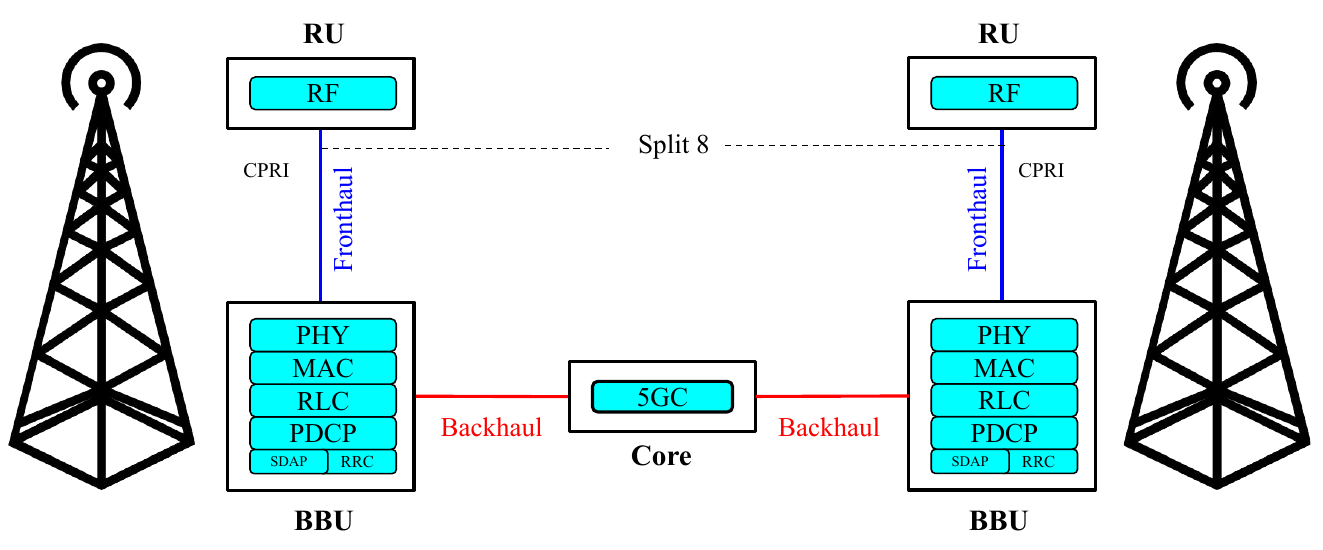}
	\vspace{-0.1in}
	\caption{Architecture of \acs*{D-RAN}.}
 \vspace{-0.2in}
	\label{fig:d-ran}
\end{figure}

In traditional \ac{RAN} or \ac{D-RAN}, the \ac{BBU} and the \ac{RU} are located in the same geographical location called the base station, as illustrated in Fig.~\ref{fig:d-ran}. Only functional split 8 is used to separate them from each other. There were no standardised network automation functions and only an implicit form of conflict due to conflicting optimisation objectives used to incur during the \ac{RAN} operation.  These conflicts were managed through periodic manual retuning by the \ac{MNO}, and there were no automated conflict management mechanisms.

\subsection{Cloud Radio Access Network}
\label{c_ran}
\begin{figure}[!ht]
 \centering
 \vspace{-0.1in}
	\includegraphics[scale=0.37]{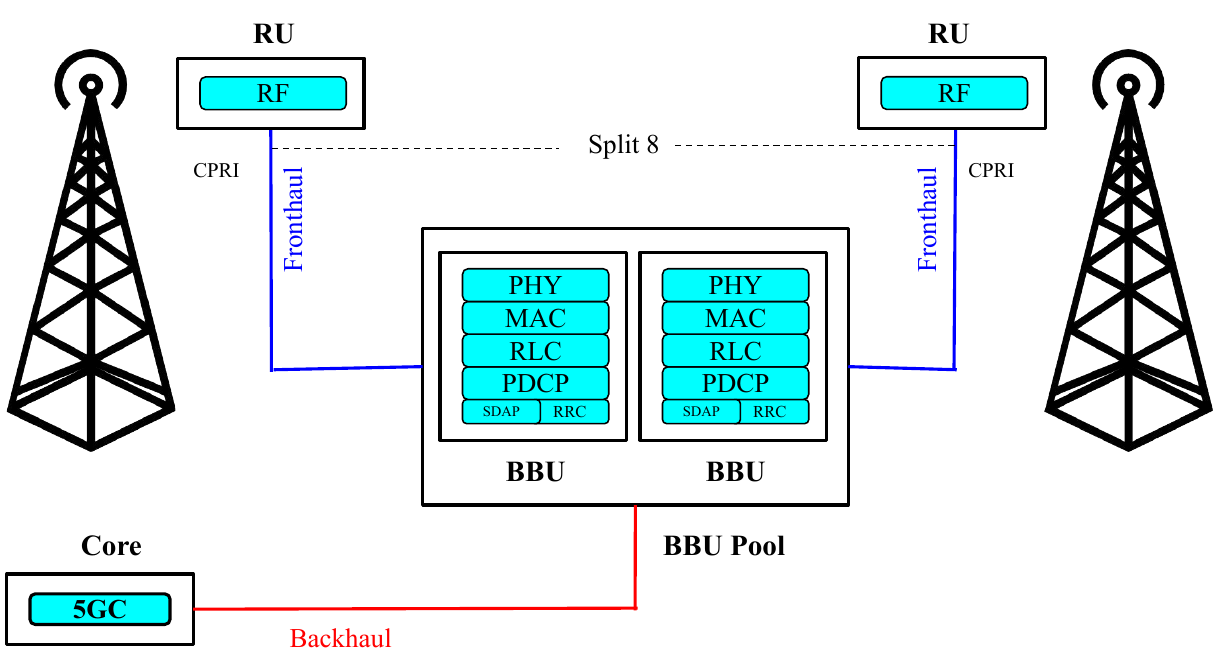}
	\vspace{-0.1in}
	\caption{Architecture of \acs*{C-RAN} with a centralised \acs*{BBU} pool.}
 \vspace{-0.25cm}
	\label{fig:c-ran1}
\end{figure}

Unlike traditional \ac{RAN} architecture, the \ac{C-RAN} places a central \ac{BBU} pool for baseband processing across multiple base stations. Only the \ac{RRH} remains at the cell site. The \ac{BBU} pool connects the \ac{RRH} via a high-speed front-haul fiber link, as shown in Fig.~\ref{fig:c-ran1}. To reduce latency requirements, some of the lower-layer \ac{BBU} processing are kept at the \ac{RRH} in some variants of the \ac{C-RAN} architectures. 

However, \ac{C-RAN} is mostly considered sensitive to strict fronthaul bandwidth and latency requirements. The main benefits it offers are reducing per-site \ac{CapEx} for the \ac{MNO} and improving resource utilization using statistical multiplexing. Neither of the \ac{C-RAN} variants we state here has a centralized conflict management mechanism. An extra automation layer added as logical components on top of these architectures, primarily \ac{SON} and later \ac{CAN}, to automate network management was the first to introduce conflict in \ac{C-RAN}. 

\subsection{Virtualised Radio Access Network}
\label{v_ran}
The \ac{vRAN} architecture extends the \ac{C-RAN} architecture concept. It replaces dedicated baseband hardware with virtualized software functions that run on general-purpose cloud or edge platforms. \ac{vRAN} makes operations more flexible and reduces \ac{CapEx}, but it can also introduce timing jitter and complicate orchestration. \ac{vRAN}’s software-based baseband control inspired the future \ac{O-RAN} and \ac{AI}-\ac{RAN}. 

\subsection{Self-Organising Networks as the First Explicit Conflict Context}
\label{subsec:son_overview}
The discussion above highlights the architectural shifts and evolution of baseband functions across generations and their physical deployment. The concept of conflict in \ac{RAN} first emerged with \ac{SON} \cite{4g_americas_son_2013}, introduced by the \ac{3GPP}. With 4G \ac{LTE}, there was a growing need for a logical automation layer to handle densification and operational complexity. They introduced the concept of \acp{SF} to enable autonomous network self-optimization, self-configuration, and self-healing, such as handover management (\ac{MRO}), load balancing in the cell (\ac{MLB}), \ac{COD}, etc. These \ac{SON} functions could be deployed as logical layers in a distributed form at \acp{eNodeB}, a \ac{NMS}, or a \ac{C-RAN} pool. They can also act as an \ac{OSS} for centrally \ac{SON} (\ac{C-SON}). Therefore, the concepts of \ac{SON} and \acp{SF} are not tied to a single  \ac{RAN} architecture \cite{3gpp_son_2024}.  When multiple \acp{SF} concurrently adjust overlapping \acp{NCP} (e.g., power, tilt, handover offsets) while optimizing different \acp{KPM}, this can lead to conflicts due to competing objectives. Thereafter, the early conflict taxonomy for characterizing parameter conflicts, \ac{KPI} conflicts, and logical dependency conflicts was studied, along with coordination methods. 

\subsection{Open Radio Access Network}
\label{o_ran}
\begin{figure}[!ht]
 \centering
 \vspace{-0.1in}
	\includegraphics[scale=0.37]{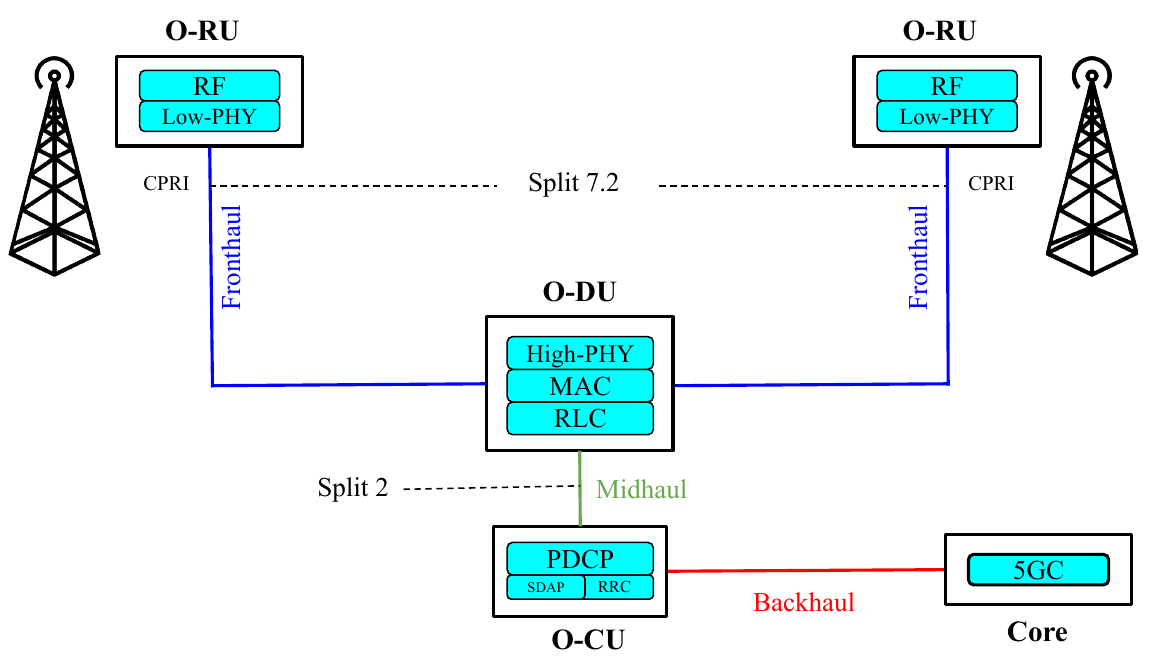}
	\vspace{-0.1in}
	\caption{Architecture of Open \ac{RAN} with disaggregated \ac{RU}/\ac{DU}/\ac{CU} and \ac{RIC}-based control.}
 \vspace{-0.15cm}
	\label{fig:o-ran}
\end{figure}

The concept of \ac{O-RAN} evolved through the virtualized and programmable network management ideas of \ac{vRAN} and the automation ideas of \ac{SON} functions. The \ac{O-RAN} architecture splits the \ac{BBU} into three separate components: the \ac{O-RU}, which handles radio-frequency signal processing at the cell site; the \ac{O-DU}, which handles layer-2 processing and real-time scheduling; and the \ac{O-CU}, which handles higher-layer functions such as mobility management and session handling. These three components are created by replacing the old split 8 with split 7.2 and introducing a new functional split 2 as illustrated in Fig.~\ref{fig:o-ran}. The \ac{O-RAN} enables multiple vendors to develop and deploy \ac{RAN} management components through open interfaces, which is unique compared to previous architectures discussed above \cite{alam2025comprehensive}. \ac{O-RAN} introduces the concept of the \ac{RIC}, a dedicated logical interface that enables network management applications to be deployed for automated \ac{RAN} management. There are two variants of \ac{RIC}, namely the \ac{Near-RT} \ac{RIC} and the \ac{Non-RT} \ac{RIC}. The former manages time-critical tasks with a latency requirement of 10~ms to 1~s, and the latter handles longer-term operations above 1~s, including \ac{AI}/\ac{ML} model training, intent handling, and policy management. The \ac{Near-RT} \ac{RIC} connects to the \ac{RAN} nodes via a dedicated E2 interface; the \ac{Non-RT} \ac{RIC} communicates via the A1 and O1 interfaces. These dedicated interfaces, along with the emerging R1 interface, collectively enable developers to deploy \acp{xApp} and \acp{rApp} independently on the \ac{RAN} hardware.

The main benefit of vendor neutrality is that it gives the \acp{MNO} freedom to source hardware and software from competing suppliers instead of choosing from a handful of them. Also, the design and development of third-party \ac{AI}/\ac{ML} algorithms for the \ac{RAN} control loop can become faster and smoother with more contributors. The only tradeoff is the need for a coordination mechanism to ensure smooth cooperation among components from various sources. Therefore, coordination and conflict management are important issues in \ac{O-RAN} compared to previous generations. This survey investigates this issue. 

\subsection{Origin and Evolution of Conflicts across Generations}
\label{sec:conf_evo}

\begin{table*}[!t]
\centering
\small
\caption{Evolution of \ac{RAN} Architectures, Automation Layers, and Conflict Notions}
\label{tab:era_arch_conflicts}
\begin{tabular}{p{2.6cm} p{3.4cm} p{3.6cm} p{4.4cm}}
\hline
\textbf{Era / Generation} & \textbf{\ac{RAN} Architecture (informal)} & \textbf{Automation Layer} & \textbf{Conflict Notion Status} \\
\hline
1G--early 3G &
\ac{T-RAN} / \ac{D-RAN} monolithic &
None (manual planning) &
Conflicts implicit (misconfiguration and parameter mis-tuning), not formalised as multi-loop control conflicts. \\
\hline
\ac{LTE} (early) &
\ac{D-RAN}, starting \ac{C-RAN} &
\ac{SON} (\ac{3GPP}) &
First formal conflict definitions (e.g., \ac{NCP} and \ac{KPI} conflicts between concurrent \ac{SON} functions); centralised coordination mechanisms proposed. \cite{3gpp_son_2024} \\
\hline
\ac{LTE}-A / 5G \ac{NSA} &
\ac{C-RAN} / \ac{H-CRAN} / \ac{vRAN} &
\ac{SON} + emerging \ac{CAN} &
Learning-based conflicts among \acp{CF} and \ac{SON} Functions; need for dynamic controllers and arbitration beyond static priorities. \cite{yu2018self, banerjee2020ran, banerjee2021intent} \\
\hline
5G \ac{SA} / towards 6G &
\ac{O-RAN} + \ac{vRAN} &
\acp{rApp}/\acp{xApp}, \ac{CAN}-like frameworks &
Direct, indirect, and implicit \ac{xApp} conflicts; multi-vendor, multi-\ac{RIC}, multi-time-scale conflicts in open, programmable \acp{RAN}. \cite{alam2025comprehensive} \\
\hline
\end{tabular}
\end{table*}

Table~\ref{tab:era_arch_conflicts} summarises the evolution of conflicts across architectures and generations of mobile communication systems. It illustrates how conflict evolved from implicit \ac{D-RAN} misconfigurations to complex parameter and \ac{KPI} conflicts in \ac{LTE}, \ac{LTE}-A, 5G, and beyond. This table highlights the growing complexity of conflict and the need for scalable conflict management mechanisms \cite{wadud2023conflict, wadud2024qacm}.

\section{\ac{RAN} Automation Paradigms and Conflict Taxonomy}
\label{sec:oran_son_can}

The discussion in Section~\ref{sec:back} clearly points to three automation paradigms that are involved with the concept of conflict and conflict management in \ac{RAN}. Each of these paradigms and their conflicts are considered separate research areas. In this survey, we ask the question of a unified conflict definition across these three paradigms, as they all operate on the same \ac{RAN} hardware and ecosystem. So why should we have a three-way taxonomy instead of a single unified one? We investigated further to answer this question and found shared features of conflicts in all three of these paradigms. For example, we found that all three paradigms define conflicts as a contest among two entities controlling the same radio resources. These conflicts can be categorized based on a parameter group definition or through affect on observed \acp{KPI}. The main idea remains the same across all three paradigms and only differs in how the conflict can be managed and resolved. Based on that, Section~\ref{sec:son_sf} outlines the four variables that affect these aspects. Later, Section~\ref{subsec:unified_taxonomy} shows how the conflict vocabularies fit into a unified structure. Finally, Section~\ref{sec:lifecycle} reviews the literature from all three paradigms together, focusing on a conflict management lifecycle perspective. 

\subsection{What Varies Across Paradigms, and What Does Not}
\label{sec:son_sf}
\label{sec:can_cf}
\label{sec:xapp_rapp_dapp}

As stated earlier, the concept of conflict and the need for conflict management originated with the introduction of \ac{SON} by \ac{3GPP} for  \ac{LTE} and \ac{LTE}-Advanced. Studies in \cite{yu2018self, 3gpp_son_2024} show clearly how \acp{SF} causes conflicts while adjusting overlapping \acp{NCP} to optimize their own \ac{KPI} objectives. The \ac{SON} conflicts are mainly categorized into \ac{NCP} conflicts and \ac{KPM} conflicts for the centralized \ac{SONCO}~\cite{4g_americas_son_2013}.  Later, \ac{CAN} extends the idea of \ac{SON} replacing rule-based \acp{SF} with learning-based \acp{CF} and inheriting the same conflict issues \cite{banerjee2020ran}.  Moving from rule-based to learning-based control makes the conflict problem more complex in two ways. Firstly, the relationship between a \ac{CF}'s parameters and its \ac{KPI} is learned, so the conflicts cannot be cataloged at design time. Secondly, independently trained \acp{CF} may develop policies that make sense individually but are suboptimal when used together. This is a unique situation that was not seen in \ac{SON}'s fixed rule-based logic~\cite{banerjee2021intent}. Later, \ac{O-RAN} came with new hierarchical \ac{RAN} management ideas with applications deployed in several logical layers based on latency sensitivity \cite{polese2022understanding}. \ac{O-RAN} Alliance Working Group~2 technical reports~\cite{oran_wg2_plc_conflict_tr, ORAN_A1PCM_TR, oran_wg2_r1servconfmit_tr} highlight how conflict can appear in different layers with the possibility of inter-layer conflicts. This makes conflicts more complex and challenging to address in \ac{O-RAN} compared to all previous generations. 

Table~\ref{tab:paradigm_variables} defines the four variables that differentiate conflict among these three paradigms. Conflict in \ac{SON} can be resolved by design because a single vendor implemented all functions. Whereas \ac{CAN} needs a cognitive autonomous controller to resolve and mediate conflict. \ac{O-RAN} neither has a shared design nor a guaranteed coordination framework; therefore, conflict detection and management became a separate research problem only in \ac{O-RAN}. Even though conflict is a decade older than the \ac{O-RAN} architecture, as shown in Sections~\ref{subsec:unified_taxonomy}--\ref{sec:lifecycle}. The evolution of conflict conditions is shown in Fig.~\ref{fig:conflict_taxonomy}(a). 

\begin{table*}[!t]
\centering
\caption{What varies across \ac{SON}, \ac{CAN}, and \ac{O-RAN}, and what stays similar.}
\label{tab:paradigm_variables}
\renewcommand{\arraystretch}{1.25}
\begin{tabular}{p{2.6cm} p{4.6cm} p{4.6cm} p{5.0cm}}
\hline
\textbf{Variable} & \textbf{\ac{SON}} & \textbf{\ac{CAN}} & \textbf{\ac{O-RAN}} \\
\hline
Decision authority & Function-local, or a single \ac{SONCO} & A central cognitive Controller & \ac{RIC} platform, split across \ac{Near-RT} and \ac{Non-RT} tiers \\
\hline
Control logic & Rule-based / heuristic & Learned (\ac{RL}, neural policy) & Learned, heterogeneous across vendors \\
\hline
Vendor boundary & Single vendor implements every function & Single vendor & \textbf{Multi-vendor, black-box} across \acp{xApp}, \acp{rApp}, and \acp{dApp} \\
\hline
Loop structure & One control loop & One Controller--\ac{CF} loop & Two \ac{RIC} tiers plus emerging real-time/lower \ac{dApp} loops \\
\hline
Enforceability of arbitration & High: one implementer can enumerate every interaction & High: one implementer, opaque internal logic & \textbf{Low}: arbitration must be intercepted at the E2 or A1 interface between independently owned components \\
\hline
\end{tabular}
\end{table*}

\subsection{Unified Conflict Taxonomy Across Paradigms}
\label{subsec:unified_taxonomy}

Each of the three paradigms we discussed has its own terms to define conflicts. But the main idea of conflict structure remains the same and originates in a similar manner across these paradigms. Conflict occurs in all three paradigms when multiple control applications or functions compete for the same control parameter, interact through a set dependency, or are linked by an unregistered \ac{KPI} effect. Table~\ref{tab:conflict_taxonomy} maps the relationships among applications and control functions across \ac{SON}, ac{\ac{CAN}}, and \ac{O-RAN} and links them to the core unified direct, indirect, and implicit types. Timing, lifecycle stage, application tier, and architectural scope are used to define the context for a conflict across paradigms. 

\begin{table*}[!t]
\centering
\small
\caption{Mapping paradigm- and tier-specific terminology onto the invariant structural conflict taxonomy.}
\label{tab:conflict_taxonomy}
\setlength{\tabcolsep}{4pt}
\renewcommand{\arraystretch}{1.2}
\begin{tabular}{p{2.0cm} p{4.2cm} p{2.5cm} p{8.0cm}}
\hline
\textbf{Context} & \textbf{Source terminology} & \textbf{Unified placement} & \textbf{Mapping rule} \\
\hline
\ac{SON} & \ac{NCP} conflict & Direct & Two \ac{SON} Functions request incompatible values for the same \ac{NCP} or actuator. \\
\hline
\ac{SON} & \ac{KPM}/objective conflict & Indirect or implicit & Indirect when the parameter--\ac{KPM} dependency is known; implicit when the interaction is visible only through degradation. \\
\hline
\ac{CAN} & Configuration conflict & Direct & \acp{CF} submit incompatible values or optimal ranges for the same configuration parameter. \\
\hline
\ac{CAN} & Characteristic/direct dependency conflict & Indirect & Different configurations are connected by an explicitly modelled functional dependency. \\
\hline
\ac{CAN} & Measurement/logical-dependency conflict & Indirect or implicit & Indirect if the chain is represented in the model; implicit if the coupling is hidden and inferred from its \ac{KPI} effect. \\
\hline
\ac{xApp} & Direct / parameter-group / unregistered \ac{KPI} effect & Direct / indirect / implicit & The \ac{O-RAN} \ac{WG3} terms instantiate the three structural classes directly. \\
\hline
\ac{rApp} & Parameter conflict or shared-capacity contention & Direct & The contested object is a managed parameter, actuator, platform resource, or radio capacity. \\
\hline
\ac{rApp} & Objective interference & Indirect or implicit & Indirect when the policy--parameter--\ac{KPI} coupling is registered; implicit when it is observed only as a side effect. \\
\hline
\ac{rApp} & Temporal conflict & Condition on any type & Timing changes when a direct, indirect, or implicit conflict manifests; it is not a fourth structural class. \\
\hline
\ac{dApp} & Scheduling contention & Direct & Co-located \acp{dApp} request incompatible use of the same scheduling or radio resource. \\
\hline
\ac{dApp} & Co-location interference & Direct, indirect, or implicit & Direct for shared-resource contention, indirect for a known pipeline dependency, and implicit for an unmodelled physical-layer side effect. \\
\hline
\end{tabular}
\end{table*}

\subsection{Conflict Scope: Horizontal, Vertical, Intra- and Inter-\ac{RIC}}
\label{subsec:conflict_scope}
\begin{figure*}[!htbp]
  \centering
  \includegraphics[width=0.85\textwidth]{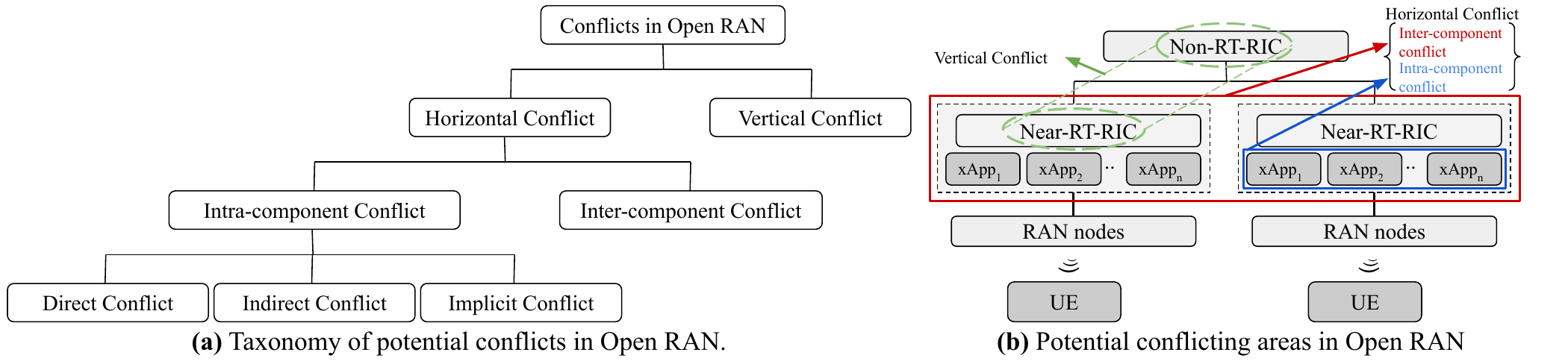}
  \caption{\ac{O-RAN} conflict scope classification: horizontal intra-\ac{RIC}, horizontal inter-\ac{RIC}, and vertical cross-layer conflicts \cite{wadud2023conflict}.}
  \vspace{-0.5cm}
  \label{fig:conflict_scope}
\end{figure*}

Table~\ref{tab:conflict_taxonomy} classifies conflicts by type across different paradigms. Another important dimension is scope, where conflict originates across different paradigms. Fig.~\ref{fig:conflict_taxonomy}(c) defines four separate axes for conflicts that include type, scope, lifecycle stage, and application context. As stated in previous sections, conflicts in \ac{O-RAN} are more complex and unique because of its hierarchical network management architecture, multi-vendor programmability, and multi-\ac{RIC} setups. For instance, a solution for conflicts inside one \ac{Near-RT} \ac{RIC} instance might not work for conflicts that happen across different \ac{RIC} deployments in separate locations. Also, cross-layer conflicts among \acp{xApp} and \acp{rApp} cannot be solved using a single mechanism deployed in just the \ac{Near-RT} \ac{RIC} layer or \ac{Non-RT} \ac{RIC}. Therefore, this section particularly focuses on \ac{O-RAN} intra-layer and cross-layer conflicts.  Figure~\ref{fig:conflict_scope} shows the three different scope-based conflicts in the \ac{O-RAN} architecture. We also map conflicts happening in \ac{SON} and \ac{CAN} to these scope types. 

\textbf{Horizontal Intra-\ac{RIC} Conflicts.} It indicates conflicts happening among two or more control applications deployed within the same \ac{RIC} instance. This is the most investigated conflict type across three paradigms, including \ac{O-RAN}. The typical conflict across these paradigms is Horizontal Intra-\ac{RIC} conflicts. In \ac{O-RAN}, horizontal conflicts can be between \ac{xApp}--\ac{xApp} in one \ac{Near-RT} \ac{RIC} or \ac{rApp}--\ac{rApp} in one \ac{Non-RT} \ac{RIC}.  The current \ac{O-RAN} \ac{WG3} \ac{CMF} and \ac{WG2} A1 policy arbitration procedures mainly address this type of conflict.

\textbf{Horizontal Inter-\ac{RIC} Conflicts.} These conflicts happen among control applications residing in between two different \ac{Near-RT} \acp{RIC} or between two different \ac{Non-RT} \acp{RIC}.  This conflict often occurs in shared-spectrum deployments, neutral-host setups, and split-\ac{RAN} deployments where cells are managed by \ac{Near-RT} \ac{RIC} clusters in different geographical locations. These types of conflict are not found in either \ac{SON} or \ac{CAN} paradigms. Right now, there is no \ac{O-RAN} specification for resolving conflicts between \acp{RIC}, so coordination depends on O1/O2 management interfaces or separate operator agreements~\cite{i14y}. 

\textbf{Vertical Cross-Layer Conflicts.}  These conflicts happen among control applications residing in two different hierarchical layers of the \ac{O-RAN} architectures. These types of conflicts are not found in either \ac{SON} or \ac{CAN} paradigms. A typical example of this conflict in \ac{O-RAN} is \acp{rApp} that work on timescales longer than one second, conflicting with \acp{xApp} that react within 10 ms to 1 s. An \ac{xApp} action can quickly contradict an \ac{rApp} policy before the \ac{rApp} updates its policy~\cite{oran_wg3_confmit_tr}. This conflict can also happen between \acp{xApp} and new \acp{dApp} if an E2 scheduling command clashes with a lower-loop \ac{PRB} assignment. These types of conflicts are still not investigated in detail in the existing literature and \ac{O-RAN} Alliance specifications. 

The above are scope-based conflict classifications. However, based on type, conflicts can be categorized as direct, indirect, and implicit. These are basically considered as horizontal scope of conflict. When two control applications or control functions contest a shared control parameter, it is called a direct control. When one control application's or function’s control decision impacts negatively on a \ac{KPI} owned by another control application or function, it can be categorized as either indirect or implicit. If the control parameter-\ac{KPI} is known prior and the indirect relationship can be explicitly identified and associated, then it is called indirect. If the relationships cannot be rationally justified but the control parameter affects the \ac{KPI} by accident during runtime, it is called an implicit conflict. 

\subsection{Formal Definitions of Conflict Types}
\label{subsec:conflict_defs}
\begin{figure*}[!htbp]
  \centering
  \includegraphics[width=\textwidth, keepaspectratio]{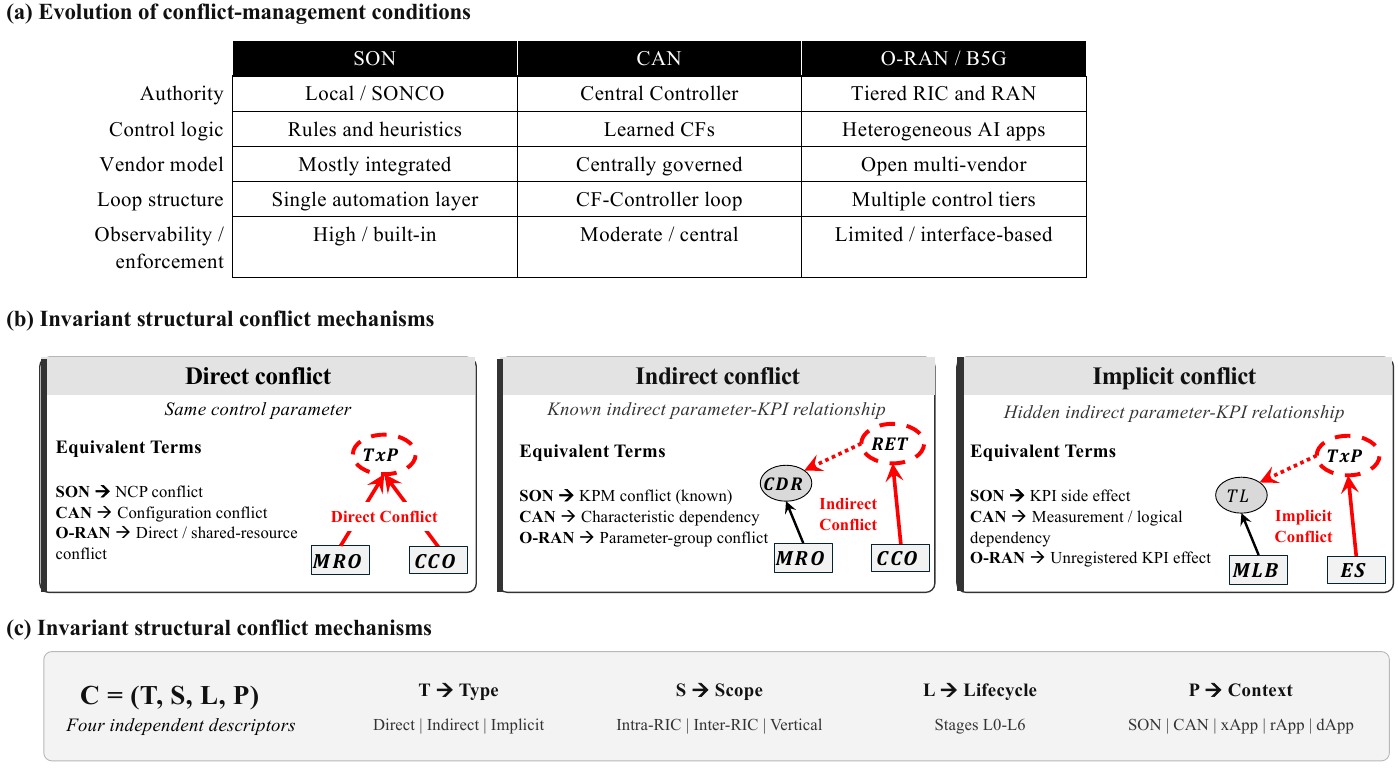}
  \caption{Conflict-management conditions, invariant structural mechanisms, and the four-axis conflict descriptor across \ac{SON}, \ac{CAN}, \ac{O-RAN}, and \ac{B5G} networks.}
  \vspace{-0.5cm}
  \label{fig:conflict_taxonomy}
\end{figure*}

The definitions given below highlight the formal definitions of conflict types used in this survey. The definitions are synthesized from existing literature across the three paradigms and are used as a reference in the lifecycle-stage review in Section~\ref{sec:lifecycle}. Fig.~\ref{fig:conflict_taxonomy} shows the taxonomy of conflicts in \ac{RAN} across paradigms. 

Figure~\ref{fig:conflict_taxonomy}(a) shows the evolution of conflict management conditions across the three paradigms, while the main types of conflict stay the same. Figure~\ref{fig:conflict_taxonomy}(b) illustrates the direct, indirect, and implicit conflicts with real examples of control applications or functions having conflicts over shared parameters or associated \acp{KPI}. Figure~\ref{fig:conflict_taxonomy}(c) shows the unified definition of a conflict that we introduce in this survey. We represent each conflict as $C=(T,S,L,P)$, where $T$ is the structural type, $S$ is the architectural scope, $L$ is the lifecycle stage, and $P$ is the application context. These four descriptors are independent. Using this definition, a direct conflict can be represented whether it happens within a single \ac{RIC}, between \acp{RIC}, or across hierarchical layers at any lifecycle stage from L0 to L6. As stated earlier, this definition covers conflict among \ac{SON} functions, \ac{CAN} cognitive functions, \acp{xApp}, \acp{rApp}, or \acp{dApp}. 

\subsubsection*{\ac{SON} Conflict Types}
Conflict in \ac{SON} appeared a decade before the concept of \ac{O-RAN} even crystallized. Therefore, the \ac{O-RAN} Alliance's use of the terms direct, indirect, and implicit to represent different types of conflict is not used in either \ac{SON} or \ac{CAN}. An \ac{NCP} conflict in \ac{SON} where two \ac{SON} Functions try to control the same parameter is mapped as a direct conflict. A \ac{KPM} (measurement) conflict, where the link is through a shared or affected \ac{KPI} instead of a shared parameter, is mapped as an indirect conflict. Finally, if the link is only with an unshared and incidentally affected \ac{KPI} with no parameter-group relationship, it is mapped as an implicit conflict. We use this mapping for all \ac{SON} and \ac{CAN} classifications in Tables~\ref{tab:survey_class_l0}--\ref{tab:survey_class_adj} and the same is also used in Fig.~\ref{fig:conflict_taxonomy}(b).

An \ac{NCP} conflict happens when two or more \acp{SF} try to set different values for the same \ac{NCP} on the same cell or user equipment at the same time. Since a control parameter can have only one value at a time, one function's control decision is replaced by the other one’s. The function whose update is replaced does not see the expected \ac{KPI} change and tries the update again in the next control cycle. This causes an ongoing recurring conflict situation. For example, suppose \ac{MLB} and \ac{MRO} \acp{SF} run at the same time in the network. A direct conflict happens if \ac{MLB} lowers the \ac{CIO} to move users from a crowded cell, while \ac{MRO} raises the same \ac{CIO} to stop early handovers. As a result, neither function can reach convergence, and conflict will keep recurring in the network~\cite{zia2014policy,chen2012difference}.

A \ac{KPI} measurement conflict is also called a \ac{KPM} or objective conflict. It happens when one \ac{SF} changes a parameter that affects a \ac{KPM} another function is trying to optimize. This relationship is indirect. Suppose function A changes parameter $p_1$, which affects \ac{KPI} $k_2$. If function B notices the drop in $k_2$ and tries to change parameter $p_2$ in response, that affects $k_1$ associated with function A. Then this $p_2$--$k_1$ relationship can be termed indirect, and the induced conflict is called indirect conflict.  A real example of this in the network can be among \ac{MLB} and \ac{MRO} functions over \ac{RLF} rate. Suppose \ac{MLB} speeds up handovers to reduce congestion, but this raises the \ac{RLF}. \ac{MRO} sees the higher \ac{RLF} and makes handover thresholds stricter. As a result, it becomes harder for \ac{MLB} to offload the congested cell and keeps the conflict going over several cycles~\cite{mwanje2012selforganized,moysen2018conflict}.

\subsubsection*{\ac{CAN} Conflict Types}
A configuration conflict happens when two or more \acp{CF} send different values for the same \ac{NCP} during the same control cycle. Two sub-types are identified: an \emph{input parameter conflict} (type A1), where both \acp{CF} use the same \ac{NCP} but send different optimal-range requests to the Controller; and an \emph{output parameter conflict} (type A2), where two \acp{CF} assign conflicting values to the same managed object~\cite{banerjee2021game}. An instance of this is an \ac{MLB} \ac{CF} requesting a \ac{CIO} decrease. If an \ac{MRO} \ac{CF} requests a \ac{CIO} increase for the same cell in the same cycle, then it is an output parameter conflict.

A measurement conflict (type B) happens when one \ac{CF} changes a parameter that affects the \ac{KPI} of another \ac{CF} \cite{banerjee2021game}. For example, a \ac{CCO} \ac{CF} that increases the transmit power of a macro cell to extend coverage simultaneously increases the \ac{RSRP} related to an interference management \ac{CF}.  As a result, the interference management \ac{CF} then misunderstands the environment and reduces power on an adjacent cell, undermining the \ac{CCO} objective. 

A characteristic conflict (type C) happens because of an indirect link between two \acp{CF} through the network transfer function~\cite{banerjee2021game}. In a \emph{direct characteristic conflict} (type C1), one \ac{CF}'s output parameter is directly used as input by another \ac{CF}'s model. In a logical dependency conflict (type C2), the link is indirect. Suppose one \ac{CF}'s output changes a network state variable, which then affects the \ac{KPI} optimized by another \ac{CF}. For example, a coverage \ac{CF} that adjusts antenna tilt changes the cell's physical footprint. As a result, it alters the load distribution observed by an \ac{MLB} \ac{CF}.

\subsubsection*{\ac{xApp} Conflict Types (\ac{Near-RT} \ac{RIC})}
A direct conflict happens when two or more \acp{xApp} send conflicting control decisions over the E2 control messages for the same \ac{RAN} parameter and control object \cite{adamczyk2023detection}. As stated earlier for \ac{SON}, a direct conflict replaces one \ac{xApp}'s instruction with another without telling the affected \ac{xApp}. For example, \ac{MRO} \ac{xApp} increases the control parameter \ac{CIO} to prevent early handovers. At the same time, the \ac{MLB} decreases the \ac{CIO} to balance the load. If this happens in the same cell, the most recent control decision replaces the previous one without informing the affected \ac{xApp}. 

An indirect conflict happens when \acp{xApp} change different parameters within the same \ac{PG}. These parameters work together to control a \ac{RAN} function and certain \acp{KPI}. Changing any of the parameters in this group can cause conflicting effects on other \ac{KPI} associated with parameters that belong to the same group~\cite{adamczyk2023detection,wadud2025xApp}. For example, \acp{xApp} that adjust \ac{RET} and \ac{CIO} both affect the handover boundary. So \ac{CIO} and \ac{RET} can belong to the same parameter group. If both are adjusted at the same time, it can cause instabilities and lower handover success rates~\cite{adamczyk2023detection}.

An implicit conflict happens when a control decision coming from one \ac{xApp} accidentally worsens the \ac{KPI} of another \ac{xApp}. The difference between an indirect conflict and an implicit conflict is that the involved \acp{xApp} do not share  any parameters or a Parameter Group in the case of an implicit conflict~\cite{adamczyk2023detection}. Therefore, this type of conflict can only be observed through \ac{KPI} violation. For example, an \ac{ES} \ac{xApp} turns off a cell carrier to save energy, but it reduces the number of available \acp{PRB} in the cell. As a result, the cell experiences more congestion and lower throughput, which are usually optimized by the \ac{QoS} \ac{xApp} \cite{10287529}. In such a case, the \ac{QoS} \ac{xApp} has no clue why its performance is getting degraded all of a sudden. 

\subsubsection*{\ac{rApp} Conflict Types (\ac{Non-RT} \ac{RIC})}
Similar to \ac{xApp} conflicts, \ac{rApp} conflicts use the same categories: direct, indirect, and implicit. A direct conflict among \acp{rApp} happens when multiple \acp{rApp} send conflicting A1 policies or O1 configuration policies for the same managed resources. For example, if two \acp{rApp} have competing objectives for computing power, transport, data service, or \ac{RAN} capacity of the same cell, it will lead to a direct conflict among them \cite{oran_wg2_plc_conflict_tr,oran_wg2_r1servconfmit_tr}. Objective inference is considered an indirect conflict among \acp{rApp}. For example, if the connection between policy, parameter, and \ac{KPI} is known during an indirect association of one \ac{rApp}'s policies causing \ac{KPI} degradation to another \ac{rApp}. Otherwise, it is called implicit for unknown association among policy, parameter, and \ac{KPI}. A new type called temporal conflict exists among \acp{rApp}. It refers to timing problems where a conflict happens before the network has stabilized after a previous action~\cite{oran_wg2_plc_conflict_tr,ORAN_A1PCM_TR}.

\subsubsection*{\ac{dApp} Conflict Types (\ac{O-CU}/\ac{O-DU}/\ac{O-RU})}
\acp{dApp} can also have the same direct, indirect, and implicit conflicts among them. A direct conflict happens when \acp{dApp} located together request incompatible use of the same \acp{PRB} or time-frequency resources. To be more specific, two \acp{dApp} co-located together and competing for shared buffers, precoding chains, or antenna ports can lead to direct conflict among them. Indirect conflict can happen among two \acp{dApp} if their actions affect each other through known processing links. It is implicit if it only appears through unexpected radio or \ac{KPI} effects. \ac{dApp} conflicts usually happen in the lower-layer control loops of the \ac{O-RAN} architecture, as \acp{dApp} are deployed at the edge and work in real-time. Coordination among \acp{dApp} and conflict mitigation require extra effort because of the sub-millisecond limit. However, there are no standardized coordination interfaces present for \acp{dApp} till we finish writing this survey. \subsubsection*{\ac{rApp} Conflict Types (\ac{Non-RT} \ac{RIC})}
Similar to \ac{xApp} conflicts, \ac{rApp} conflicts use the same categories: direct, indirect, and implicit. A direct conflict among \acp{rApp} happens when multiple \acp{rApp} send conflicting A1 policies or O1 configuration policies for the same managed resources. For example, if two \acp{rApp} have competing objectives for computing power, transport, data service, or \ac{RAN} capacity of the same cell, it will lead to a direct conflict among them \cite{oran_wg2_plc_conflict_tr,oran_wg2_r1servconfmit_tr}. Objective inference is considered an indirect conflict among \acp{rApp}. For example, if the connection between policy, parameter, and \ac{KPI} is known during an indirect association of one \ac{rApp}'s policies causing \ac{KPI} degradation to another \ac{rApp}. Otherwise, it is called implicit for unknown association among policy, parameter, and \ac{KPI}. A new type called temporal conflict exists among \acp{rApp}. It refers to timing problems where a conflict happens before the network has stabilized after a previous action~\cite{oran_wg2_plc_conflict_tr,ORAN_A1PCM_TR}.

\subsubsection*{\ac{dApp} Conflict Types (\ac{O-CU}/\ac{O-DU}/\ac{O-RU})}
\acp{dApp} can also have the same direct, indirect, and implicit conflicts among them. A direct conflict happens when \acp{dApp} located together request incompatible use of the same \acp{PRB} or time-frequency resources. To be more specific, two \acp{dApp} co-located together and competing for shared buffers, precoding chains, or antenna ports can lead to direct conflict among them. Indirect conflict can happen among two \acp{dApp} if their actions affect each other through known processing links. It is implicit if it only appears through unexpected radio or \ac{KPI} effects. \ac{dApp} conflicts usually happen in the lower-layer control loops of the \ac{O-RAN} architecture, as \acp{dApp} are deployed at the edge and work in real-time. Coordination among \acp{dApp} and conflict mitigation require extra effort because of the sub-millisecond limit. However, there are no standardized coordination interfaces present for \acp{dApp} till we finish writing this survey. 

\section{Conflict Management Lifecycle}
\label{sec:lifecycle}

In this section, we review the \ac{SON}, \ac{CAN}, and \ac{O-RAN} paradigms as one unified conflict issue in wireless \ac{RAN} and organise them by the conflict management lifecycle. The lifecycle has seven separate states. These are design-time avoidance (L0), conflict model construction (L1), runtime detection and classification (L2), resolution and arbitration (L3), enforcement and actuation control (L4), verification, attribution and feedback (L5), and governance and application lifecycle management (L6). Each of these lifecycle stages uses the unified conflict descriptor $L$ presented in Fig.~\ref{fig:conflict_taxonomy}(c). Each paper that we reviewed is listed only once in the referred lifecycle tables in Tables~\ref{tab:survey_class_l0}--\ref{tab:survey_class_adj}. However, if a particular paper has contributions in other lifecycle stages, we mention them in the text of those sections. We highlight the similarities of conflicts across the three paradigms and investigate how they differ in conflict observation and coordination implementation. The earlier reviews that cover conflict separately in any of these three paradigms paid little attention to conflict avoidance (L0), enforcement (L4), verification (L5), and governance (L6). In this survey, we discuss these stages in detail along with the detection (L2) and resolution (L3) stages, addressing the gap discussed in Section~\ref{sec:lessons}.

\begin{figure*}[!t]
  \centering
  \includegraphics[width=\textwidth]{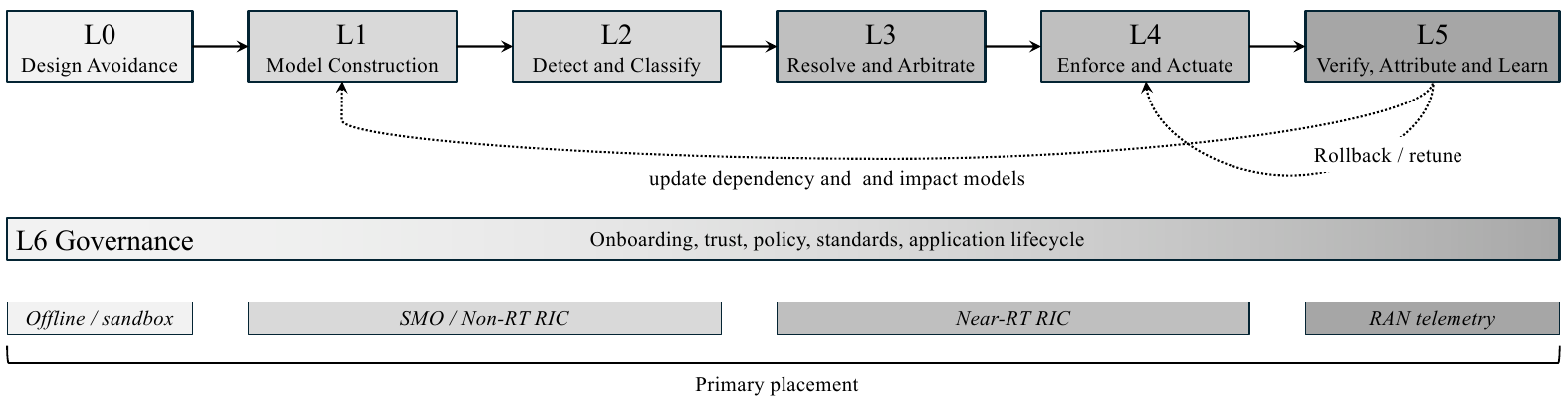}
  \caption{Conflict-management lifecycle (L0--L6), showing architectural placement, feedback paths, and cross-stage governance.}
  \label{fig:lifecycle_arch}
\end{figure*}

The conflict management lifecycle figure shown in Fig.~\ref{fig:lifecycle_arch} highlights where each stage lies in the \ac{RAN} architecture. The preference for each stage to be deployed is carefully considered based on the latency sensitivity and requirements of each stage. For example, design time avoidance (L0) can be done before the applications are deployed to \ac{RAN}, so it is considered an offline/sandbox stage. L1 and L2 may require an \ac{AI} model to be trained for anomaly detection (i.e., \ac{KPI} degradation) and conflict classification; therefore, we consider them to be preferably deployed in the \ac{SMO} or \ac{Non-RT} \ac{RIC}. However, rule-based detection and classification can be placed in the \ac{Near-RT} \ac{RIC} for real-time detection and classification. L3 and L4 stages are for conflict resolution or mitigation, so methods designed for these stages need to make real-time interventions to mitigate or resolve the conflicts. Therefore, we consider them to be deployed in the \ac{Near-RT} \ac{RIC}. L5 stage supports the whole conflict management process by supplying necessary \ac{RAN} telemetry to other stages of the management lifecycle; the dashed line shows its connections with other stages in Fig.~\ref{fig:lifecycle_arch}. Finally, L6 discusses the conflict management governance across different components of the \ac{RAN} architecture. 

Tables~\ref{tab:survey_class_l0}--\ref{tab:survey_class_adj} summarise our analysis across all reviewed papers and categorise them according to the above-mentioned lifecycle stages. The shorter column titles are used only to fit the table within the page margin. The full forms of these column titles are: Det. (Detection), Mit. (Mitigation), Dir. (Direct Conflict), Ind. (Indirect Conflict), Imp. (Implicit Conflict), Cen. (Centralised), Dis. (Distributed), Hier. (Hierarchical), \ac{RL}, \ac{MARL}, \ac{GT}, \ac{GNN}, and Int. (Intent-based). A blank or dash value in any table field of a reviewed paper indicates that the particular column field is not covered or established in that reviewed paper. The following section discusses each of the lifecycle stages in detail. 

\subsection{L0: Design-Time Avoidance}
\label{subsec:l0}
Design-time avoidance refers to techniques for designing the network management application in such a way that they do not create conflicts while deployed during the \ac{RAN} operation. In \ac{SON}, Zia~et~al. in \cite{zia2014policy} designed the \ac{MLB} that partitions \ac{CIO} control in such a way that \ac{MRO} does not counteract \ac{MLB} during traffic congestion. Later, Huang and Chen in~\cite{huang2018conflict} develop a distributed linear programme that constrains \ac{MLB}'s directional \ac{CIO} adjustments to \ac{MRO}-safe bounds. In \ac{CAN}, Mwanje~et~al. in~\cite{mwanje2018synchronized} design a coordinated learning framework among \ac{MRO}--\ac{MLB} \acp{CF} in such a way that it reduces their cross-effects during training. They formulate the reward function with peer effects and schedule training accordingly to ensure learning as spatially separated function instances. Their earlier study named QMRO in \cite{mwanje2014distributed} evaluates a single \ac{MRO} learner with a future possibility to incorporate load balancing while handover decision-making.

The \ac{O-RAN} is the one paradigm where conflict-avoidance by design is studied extensively compared to the other two paradigms. This is because independently trained agents in a multi-agent system may always introduce conflict if they are not jointly trained or have a mechanism for combining their policies into one model. Iturria-Rivera~et~al. in~\cite{iturria2022multi} study a \ac{MATL} framework for joint power-allocation and resource-allocation. They show that these two applications work better together when jointly trained rather than being trained independently. However, they highlight that communication overhead and changing peer behavior are challenges of this joint training procedure. Also, scaling the joint training for a large number of control applications is the biggest challenge of them all. Zhang~et~al. in \cite{zhang2022team} take this idea of joint training to a higher level by introducing a memory buffer to retain a two-round exchange of planned actions during operation. They also consider the power and resource allocation \acp{xApp}. Later, Erdöl~et~al. in~\cite{erdol2024xapp} combine independently trained \ac{xApp} policies into one model through policy distillation. Although this avoids conflicts between two trained \acp{xApp}, it requires retraining if a new \ac{xApp} is to be added into the \ac{Near-RT} \ac{RIC}. He~et~al. in~\cite{he2025heterogeneous} train \acp{xApp} jointly offline before deployment using synthetic data collected from a digital twin using VIAVI. However, this is not tested on a realistic environment or a testbed. Kouchaki and Marojevic in~\cite{kouchaki2022actor} propose a coordinated actor-critic multi-\ac{xApp} resource allocation method. However, the evaluation only uses one \ac{xApp}, and cooperation among multi-\acp{xApp} is never tested.

Another group of studies, apart from the multi-agent learning studies, examines the mechanism of checking conflicts before the applications are deployed. Brach del Prever~et~al. in~\cite{del2024pacifista} proposes the PACIFISTA framework for evaluating conflict severity among \acp{xApp}, considering the measured effects on \acp{KPI} and \ac{MNO}-defined restrictions on choosing which \acp{xApp} may be deployed together. Their later work in~\cite{brachdelprever2026predicting} uses the individual \ac{xApp} profile based on conflict severity score and action frequency to predict the combined effects on control parameters and \acp{KPI}. Another interesting work in cite{fang2026zodiac} by Fang~et~al. proposes a framework named \ac{ZODIAC} to identify network conditions that are likely to cause conflict among two \acp{xApp}.  Another significant work done by Armstrong~et~al. from Ericsson Ireland in \cite{armstrong2024pre} uses historical operator data from approximately 800 cells to predict how a proposed configuration change could affect \acp{KPI}. They use similar cell behavior for this prediction. Tsampazi~et~al. in~\cite{tsampazi2023comparative} compare \ac{xApp} designs and training methods before deployment. They also propose a framework called PandORA in~\cite{tsampazi2024pandora} to automate training-parameter selection and policy transfer for \ac{DRL}-based \acp{xApp}. They show that separately trained policies often cause conflicts in the network when deployed together. However, their work does not evaluate runtime conflict management methods.

Another study worth discussing here is by Bakri~et~al.~\cite{bakri2025mitigating}. Although their result does not highlight conflict-rate or packet-delay-violation metrics, the work is focused on reducing communication overhead between cooperating agents in a multi-agent \ac{RL} slicing setting using \ac{GCN}. Since the result does not highlight conflict mitigation among independently developed \acp{xApp} despite its title, we consider this work as coordinated learning for avoiding conflict by design rather than a mitigation category.

\begin{table*}[!b]
\centering
\footnotesize
\setlength{\tabcolsep}{3pt}
\renewcommand{\arraystretch}{1.05}
\caption{L0: Design-Time Avoidance.}\label{tab:survey_class_l0}
\begin{adjustbox}{max width=\textwidth}
\begin{tabular}{|p{1.07cm}|>{\centering\arraybackslash}p{0.62cm}|>{\centering\arraybackslash}p{0.95cm}|>{\centering\arraybackslash}p{0.6cm}|>{\centering\arraybackslash}p{0.6cm}|>{\centering\arraybackslash}p{0.6cm}|>{\centering\arraybackslash}p{0.6cm}|>{\centering\arraybackslash}p{0.6cm}|>{\centering\arraybackslash}p{0.6cm}|>{\centering\arraybackslash}p{0.6cm}|>{\centering\arraybackslash}p{0.6cm}|>{\centering\arraybackslash}p{0.6cm}|>{\centering\arraybackslash}p{0.86cm}|>{\centering\arraybackslash}p{0.58cm}|>{\centering\arraybackslash}p{0.58cm}|>{\centering\arraybackslash}p{0.58cm}|>{\centering\arraybackslash}p{1.35cm}|p{1.45cm}|}
\hline
\textbf{Paper} & \textbf{Year} & \textbf{Domain} & \textbf{Det.} & \textbf{Mit.} & \textbf{Dir.} & \textbf{Ind.} & \textbf{Imp.} & \textbf{Cen.} & \textbf{Dis.} & \textbf{Hier.} & \textbf{\acs*{RL}} & \textbf{\acs*{MARL}} & \textbf{\acs*{GT}} & \textbf{\acs*{GNN}} & \textbf{Int.} & \textbf{Evidence} & \textbf{Validation} \\
\hline
\cite{zia2014policy} & 2014 & \acs*{SON} & -- & \checkmark & \checkmark &  &  &  & \checkmark &  &  &  &  &  &  & Moderate & Simulation \\
\hline
\cite{huang2018conflict} & 2018 & \acs*{SON} & -- & \checkmark & \checkmark &  &  &  & \checkmark &  &  &  &  &  &  & Moderate & Simulation \\
\hline
\cite{mwanje2018synchronized} & 2018 & \acs*{CAN} &  & \checkmark &  & \checkmark &  &  & \checkmark &  & \checkmark & \checkmark &  &  &  & Moderate & Simulation \\
\hline
\cite{iturria2022multi} & 2022 & \acs*{O-RAN} &  & \checkmark &  & \checkmark &  &  & \checkmark &  & \checkmark & \checkmark &  &  &  & Moderate & Simulation \\
\hline
\cite{kouchaki2022actor} & 2022 & \acs*{O-RAN} &  &  &  &  &  & -- &  &  & \checkmark &  &  &  &  & Weak & Simulation \\
\hline
\cite{zhang2022team} & 2022 & \acs*{O-RAN} &  & \checkmark &  & \checkmark &  &  & \checkmark &  & \checkmark & \checkmark &  &  &  & Moderate & Simulation \\
\hline
\cite{tsampazi2023comparative} & 2023 & \acs*{O-RAN} &  &  &  &  &  &  &  &  & \checkmark &  &  &  &  & Moderate & Testbed \\
\hline
\cite{armstrong2024pre} & 2024 & \acs*{O-RAN} & \checkmark &  &  & \checkmark &  & \checkmark &  &  &  &  &  &  &  & Strong & Data \\
\hline
\cite{tsampazi2024pandora} & 2024 & \acs*{O-RAN} &  &  &  &  &  &  &  &  & \checkmark &  &  &  &  & Strong & Testbed \\
\hline
\cite{del2024pacifista} & 2025 & \acs*{O-RAN} & \checkmark & \checkmark & \checkmark & \checkmark & \checkmark & \checkmark &  &  &  &  &  & -- &  & Strong & Testbed \\
\hline
\cite{he2025heterogeneous} & 2025 & \acs*{O-RAN} &  & \checkmark &  &  & \checkmark &  & \checkmark &  & \checkmark & \checkmark &  &  &  & Weak & Simulation \\
\hline
\cite{brachdelprever2026predicting} & 2026 & \acs*{O-RAN} & \checkmark &  & \checkmark &  &  & \checkmark &  &  &  &  &  &  &  & Strong & Testbed \\
\hline
\cite{erdol2024xapp} & 2026 & \acs*{O-RAN} &  & \checkmark & \checkmark & \checkmark &  & \checkmark &  &  & \checkmark &  &  &  &  & Moderate & Simulation \\
\hline
\cite{fang2026zodiac} & 2026 & \acs*{O-RAN} & -- &  & \checkmark & \checkmark & \checkmark &  &  &  &  &  &  &  &  & Strong & Simulation \\
\hline
\cite{kimlee2026admission} & 2026 & \acs*{O-RAN} & -- & -- & -- & -- & -- & \checkmark & -- & -- & -- & -- & -- & -- & -- & Not rated & Not verified \\
\hline
\end{tabular}
\end{adjustbox}
\end{table*}

\subsection{L1: Conflict Model Construction}
\label{subsec:l1}
Conflict detection and arbitration methodologies require knowledge of how applications, control parameters and \acp{KPI} influence each other. That knowledge is incorporated into a model during this lifecycle stage. The model can be derived from expert knowledge, from an existing conflict classification or from relationships learned from data and represented as a graph. Lateef et al's \ac{SON} conflict classification framework forms the basis of this work. It employs a coordination matrix to describe the relationships between control parameters and \ac{SON} functions. This provides later \ac{SON} coordination work with a point of departure for identifying which functions might be in conflict with one another~\cite{lateef2015lte}. Mwanje et al. make the case for such a model. They propose a self-tuning \ac{MRO} and point to a potential conflict between the handover adjustments that this would make and the load balancing. They point out this interaction and simply state that it is a potential problem, they don't actually go on to test it. Their discussion does underline the need to have some co-ordination of the handover parameters when you have a number of \ac{SON} functions all running at the same time~\cite{mwanje2012selforganized}. Moysen and Giupponi develop a model which can predict the impact of configuration changes and can support the operator's decision. The distributed \ac{SON} functions send their required configurations to a central co-ordinator via the Itf-N interface. A bagged-\ac{SVM} model (a model which is itself a collection of several \acp{SVM}) is used to predict the impact of each configuration on the \acp{KPI}. An \ac{NSGA-II} search is then used to search across the configurations against a number of objectives and to identify the options to present to the operator. The system thus provides recommendations with the operator making the final decision~\cite{moysen2018conflict}.

Preciado Rojas and Mitschele-Thiel look at the problem of coordinating a large number of functions and network conditions. Rather than maintaining a full list of potential conflicts, they employ an \ac{ML} regression model to predict the impact of one function's behaviour on the \acp{KPI} of the other functions. They also make use of a supervised Self Organising Map to cluster similar network states into what they call excursion zones. The clusters contain network states, not the \ac{SON} functions. This clustering reduces the amount of coordination required as the scheduling decisions can be made within each cluster~\cite{preciado2020scalable}. Lee et al's \ac{CADC} covers both model construction and arbitration. It first builds a graph with a node for each base station and an edge for each possible conflict relationship. The weights in the graph are determined via \ac{PCA}. The graph is then split up using spectral clustering into groups of base stations. Each group is then given its own \ac{DDQN} coordinator. This effectively breaks the network wide coordination problem down into a number of smaller local problems~\cite{lee2025conflict}. In \ac{O-RAN}, Adamczyk et al.'s \ac{PG} approach is for understanding known relationships among conflicting control parameters \cite{adamczyk2023detection}. Parameters are grouped together if they have an effect on the same radio \acp{KPI}. For example, both the \ac{RET} and the \ac{CIO} control the handover boundary and affect handover-related \acp{KPI}, although they are entirely different parameters and are controlled by different \acp{xApp}. This known effect on each other is mapped in the form of a parameter group so that detecting indirect conflict becomes easier. 

Zolghadr et al. in \cite{zolghadr2025learning} learn part of this relationship structure from the data. Their approach is to reconstruct the graph of connections between parameters and \acp{KPI}. More precisely, they learn the parameter to \ac{KPI} portion of a conflict adjacency matrix (which records which entities are connected). The learning is done over a temporal chain graph, $\mathcal{G}_T$, where each node is a time step. The neighbors of node $v_t$ are the previous and next time steps $\{v_{t-1},v_{t+1}\}$. The method does not learn which applications control which parameters or which applications monitor which \acp{KPI}. Those connections are added in afterwards using data from the \ac{Near-RT} \ac{RIC} Subscription Manager. The full model is therefore a combination of learned relationships and known subscription information. The evaluation reported is based on synthetic data generated from Banerjee et al.'s \ac{CAN} conflict model \cite{banerjee2020ran} and does not use an ns-\ac{O-RAN} simulation. Al Shami et al. use a graph to classify \ac{xApp} interactions at each scheduling interval as either conflicting or non-conflicting. The graph consists of various types of nodes (\acp{xApp}, parameters and \acp{KPI}). It is then broken down into smaller graphs with the state of each node being a binary value. These graphs are fed into a two layer \ac{GCN}. Global mean pooling is used to aggregate the node information into a graph level representation and focal loss is used to train the classifier as the number of conflict examples varies between the different classes. They also track the incoming graph edges in order to identify potential root causes. They report F1 scores on five synthetic datasets with the proportion of conflicts ranging from 10\% to 40\%. They do not compare their method to another detection method~\cite{shami2025ran}. Santos et al. break the process down into three stages: learning the interactions, reconstructing the graph and identifying the conflicts. A two-tower encoder is used to learn the parameter-\ac{KPI} relationships using a ground truth matrix which defines the expected interactions. A sparse graph is then constructed which contains the selected interactions. The existing rules based on the graph topology are then used to identify the conflicts. The method does create and use a graph, the two-tower encoder itself is not a \ac{GNN} \cite{santos2026conflict}. Sharma~et~al. in~\cite{sharma2025towards} are attempting to model the causal relationships between parameters and \acp{KPI}. They use XGBoost and \ac{SHAP} explanations to identify potential relationships and then use those to build a \ac{DAG}. The authors use DoWhy and EconML to estimate the effects of the parameter-\ac{KPI} relationships  under different conditions as well as the average effect. However, the paper does not report detection accuracy on conflicting \acp{xApp} in a realistic environment. The \ac{SHAP}-based connection graph is useful, but the synthetic data experiment does not establish the assumptions used in this study. 

\begin{table*}[!tbp]
\centering
\footnotesize
\setlength{\tabcolsep}{3pt}
\renewcommand{\arraystretch}{1.05}
\caption{L1: Conflict Model Construction.}\label{tab:survey_class_l1}
\begin{adjustbox}{max width=\textwidth}
\begin{tabular}{|p{1.07cm}|>{\centering\arraybackslash}p{0.62cm}|>{\centering\arraybackslash}p{0.95cm}|>{\centering\arraybackslash}p{0.6cm}|>{\centering\arraybackslash}p{0.6cm}|>{\centering\arraybackslash}p{0.6cm}|>{\centering\arraybackslash}p{0.6cm}|>{\centering\arraybackslash}p{0.6cm}|>{\centering\arraybackslash}p{0.6cm}|>{\centering\arraybackslash}p{0.6cm}|>{\centering\arraybackslash}p{0.6cm}|>{\centering\arraybackslash}p{0.6cm}|>{\centering\arraybackslash}p{0.86cm}|>{\centering\arraybackslash}p{0.58cm}|>{\centering\arraybackslash}p{0.58cm}|>{\centering\arraybackslash}p{0.58cm}|>{\centering\arraybackslash}p{1.35cm}|p{1.45cm}|}
\hline
\textbf{Paper} & \textbf{Year} & \textbf{Domain} & \textbf{Det.} & \textbf{Mit.} & \textbf{Dir.} & \textbf{Ind.} & \textbf{Imp.} & \textbf{Cen.} & \textbf{Dis.} & \textbf{Hier.} & \textbf{\acs*{RL}} & \textbf{\acs*{MARL}} & \textbf{\acs*{GT}} & \textbf{\acs*{GNN}} & \textbf{Int.} & \textbf{Evidence} & \textbf{Validation} \\
\hline
\cite{lateef2015lte} & 2015 & \acs*{SON} & \checkmark & \checkmark & \checkmark & \checkmark &  & \checkmark & \checkmark &  &  &  &  &  &  & Moderate & Simulation \\
\hline
\cite{moysen2018conflict} & 2018 & \acs*{SON} & -- & \checkmark & \checkmark &  &  & \checkmark & \checkmark & \checkmark &  &  &  &  &  & Moderate & Simulation \\
\hline
\cite{preciado2020scalable} & 2020 & \acs*{SON} & -- & \checkmark &  &  &  & \checkmark &  &  &  &  &  &  &  & Moderate & Simulation \\
\hline
\cite{lee2025conflict} & 2025 & \acs*{SON} & -- & \checkmark &  & \checkmark &  &  & \checkmark &  & \checkmark &  &  & -- &  & Strong & Simulation \\
\hline
\cite{adamczyk2023detection} & 2023 & \acs*{O-RAN} & \checkmark & \checkmark &  & \checkmark &  & \checkmark &  &  &  &  &  &  &  & Moderate & Simulation \\
\hline
\cite{sharma2025towards} & 2025 & \acs*{O-RAN} & -- &  & -- & -- & -- & \checkmark &  &  &  &  &  &  &  & Weak & Simulation \\
\hline
\cite{zolghadr2025learning} & 2025 & \acs*{O-RAN} & \checkmark &  & \checkmark & \checkmark & \checkmark &  &  &  &  &  &  & \checkmark &  & Weak & Simulation \\
\hline
\cite{santos2026conflict} & 2026 & \acs*{O-RAN} & \checkmark &  & \checkmark & \checkmark & \checkmark & \checkmark &  &  &  &  &  & -- &  & Weak & Simulation \\
\hline
\cite{shami2025ran} & 2026 & \acs*{O-RAN} & \checkmark &  & \checkmark & \checkmark & \checkmark & \checkmark &  &  &  &  &  & \checkmark &  & Weak & Simulation \\
\hline
\cite{zolghadr2026compatibility} & 2026 & \acs*{O-RAN} & -- & \checkmark & \checkmark & -- & -- & -- & -- & -- & -- & -- & -- & -- & -- & Moderate & Simulation \\
\hline
\end{tabular}
\end{adjustbox}
\end{table*}

Another question in the area of conflict modelling is whether two agents controlling the same parameter must necessarily be in conflict. Zolghadr et al. \cite{zolghadr2026compatibility} look at this issue in the context of agents which act when a \ac{KPI} moves outside of its acceptable range. The agents are modelled as sampled deadband controllers (they sample the system at regular intervals and will only take action if the \ac{KPI} has moved outside of a deadband around the target). They use the \ac{KPI} target and the allowable deadband for each agent to create tests to determine whether a pair of agents are compatible. Agents can share control of a parameter if their acceptable operating regions overlap. Arbitration is required if the agents' objectives are incompatible. Their ns3-oran simulations confirm the approach under the stated assumptions of deadband and the ability to reach the acceptable operating region. The paper does not establish that the method will be robust for any learned policy or in the face of changing network dynamics. The paper in question is a preprint.

\subsection{L2: Runtime Detection and Classification}
\label{subsec:l2}

L1 builds a model of the relationships between the applications, the parameters and the \acp{KPI}. L2 uses that model during operation to determine whether a proposed or already applied action creates a conflict. Detection can occur prior to the action being applied (pre-action detection) or following an observed \ac{KPI} change (post-action detection).

Adamczyk et al. define pre-action procedures for the detection of both direct and indirect conflicts (\ac{DCD} and \ac{ICD} respectively) and also propose a separate detector \ac{ImCD} for the detection of implicit conflicts. \ac{ImCD} operates post-action (i.e. after the action has had an effect on the network) and so cannot prevent the original conflicting action \cite{adamczyk2023conflict}.

Wadud~et~al. start again with an observed effect: a \ac{QoS} threshold violation leads to negotiation over a new control value. Their evaluation is based on four synthetic \acp{xApp} with Gaussian utility functions. They don't deploy \ac{MRO} or energy saving \acp{xApp} in a radio network~\cite{wadud2023conflict}.

De Oliveira et al's AGIR system looks for conflicts between operator intents. It takes the operator's intent in the form of natural language and converts that into network policies. Word2Vec is used to represent the text numerically and \ac{LSTM}/\ac{GRU} classifiers are used to detect any contradictions prior to the intent being accepted. The evaluation is based on 10,000 synthetic intent pairs. It doesn't evaluate the actual \ac{xApp} in operation or the application of a mitigation decision~\cite{de2023agile}.

Giannopoulos~et~al.'s COMIX uses a network digital twin to validate the proposed \ac{xApp} actions via a range of rollout scoring techniques (\ac{MaxTS}, \ac{MinPS}, \ac{EES} and \ac{TVS}) rather than learning a policy to select the action. It validates the action prior to its application and can also issue an alarm once an action has produced an implicit effect. The case considered is a direct power control conflict. The paper does go on to describe how it would deal with indirect conflicts, though it doesn't validate that in any separate case. To deal with implicit effects you'd need to issue an alarm once the effect has occurred and then have a manual update to the relevant cluster~\cite{giannopoulos2025comix}.

Nasim et al. combine Riemannian geometry with an \ac{RNN} based detector. They represent the \acp{xApp}, the parameters, the \acp{KPI} and the relationships between them as a conflict graph which evolves over time. The graph Laplacian (a matrix which describes the connections in the graph) is represented as a symmetric positive definite matrix. The Log-Euclidean Metric is then used to compare these matrix representations. The data generated contains direct, indirect and implicit conflicts, which the authors combine into a single conflict class for training. The detector is thus trained to make a binary distinction between conflict and no conflict. It achieves the oracle reference F1 score by epoch 20 of training \cite{nasim2025conflict}.

Kwon and Zhang's \ac{ORCA} tackles a different problem: a detector can become less reliable when the operating environment differs from that in which it was developed. \ac{ORCA} employs a two stage feature selection process coupled with \ac{SHAP} analysis to determine which features are important in a reference environment. An Environment Encoder is then used to detect changes in the environment. When such a change is detected, an \ac{LLM} based $\beta$-Reasoning Cascade is used to update the feature importance estimates without the need to collect any new data. Segmented Active \ac{RAG} is used to provide retrieved information in order to reduce the amount of unsupported \ac{LLM} output. The authors evaluate \ac{ORCA} across 32 environment categories in a MATLAB 5G simulator and report a 60.7\% reduction in cosine error and a 65.1\% reduction in \ac{NRMSE} against both the extrapolation and pure \ac{LLM} baselines. The \ac{ORCA} framework proposed by these authors evaluates individual parameter requests and makes accept-or-reject decisions accordingly. However, it does not analyze interactions between pairs of conflicting control applications~\cite{kwon2026orca}. The results reported above pertain to feature importance as highlighted in the paper.

Some studies examine conflicts at the service intent level before the service is instantiated by the application. Bikos~et~al.'s 6G-ACCORD compares each incoming service intent against a database of previously admitted intents prior to the creation of the new service. The paper describes the approach as \ac{LLM}-based for detection, while actual conflict checking is performed using pairwise regular expression comparisons and filtering rules. The \ac{LLM} is utilized to explain the results. However, the paper does not evaluate an implementation without the \ac{LLM}, making it unclear what proportion of the reported 93\% accuracy is attributable to the rules versus the \ac{LLM}~\cite{bikos2025accord}. A one-page companion paper from the same authors discusses potential methods for resolving the identified conflicts. The authors mention reinforcement learning-based methods and bargaining-based solutions, but provide no mathematical formulation, algorithm, or evaluation. While the papers collectively offer some evidence for an intent screening approach, the proposed conflict resolution methods remain unformalized and unevaluated~\cite{bikos2024accord}.

Gelban et al. are looking at a completely different problem (one to do with the order of control messages). They've added monitoring to the FlexRIC \ac{Near-RT} \ac{RIC} to detect messages that are arriving or being processed out of order. This is a control plane fault rather than one of the \ac{WG3} conflict types, which are about competing \ac{RAN} control decisions. None of the monitoring \acp{xApp} actually issue a \ac{RAN} control action, so there's no scenario in which they are in any way creating competing requests for a shared parameter. The principal contribution is therefore the monitoring infrastructure and the resulting dataset of 22.6 million labeled events. The detection results need some care in the way they are interpreted, as in the vast majority of the cases where the positive class is the majority, the F1 scores are no more than 0.001 different from a classifier which simply always predicts positive. These results in no way at all demonstrate the detection of competing \ac{xApp} control decisions~\cite{gelban2026detecting}.

\begin{table*}[!b]
\centering
\footnotesize
\setlength{\tabcolsep}{3pt}
\renewcommand{\arraystretch}{1.05}
\caption{L2: Runtime Detection and Classification.}\label{tab:survey_class_l2}
\begin{adjustbox}{max width=\textwidth}
\begin{tabular}{|p{1.07cm}|>{\centering\arraybackslash}p{0.62cm}|>{\centering\arraybackslash}p{0.95cm}|>{\centering\arraybackslash}p{0.6cm}|>{\centering\arraybackslash}p{0.6cm}|>{\centering\arraybackslash}p{0.6cm}|>{\centering\arraybackslash}p{0.6cm}|>{\centering\arraybackslash}p{0.6cm}|>{\centering\arraybackslash}p{0.6cm}|>{\centering\arraybackslash}p{0.6cm}|>{\centering\arraybackslash}p{0.6cm}|>{\centering\arraybackslash}p{0.6cm}|>{\centering\arraybackslash}p{0.86cm}|>{\centering\arraybackslash}p{0.58cm}|>{\centering\arraybackslash}p{0.58cm}|>{\centering\arraybackslash}p{0.58cm}|>{\centering\arraybackslash}p{1.35cm}|p{1.45cm}|}
\hline
\textbf{Paper} & \textbf{Year} & \textbf{Domain} & \textbf{Det.} & \textbf{Mit.} & \textbf{Dir.} & \textbf{Ind.} & \textbf{Imp.} & \textbf{Cen.} & \textbf{Dis.} & \textbf{Hier.} & \textbf{\acs*{RL}} & \textbf{\acs*{MARL}} & \textbf{\acs*{GT}} & \textbf{\acs*{GNN}} & \textbf{Int.} & \textbf{Evidence} & \textbf{Validation} \\
\hline
\cite{banerjee2021contradiction} & 2022 & \acs*{CAN} & \checkmark & \checkmark & \checkmark &  &  & \checkmark &  &  &  &  &  &  & \checkmark & Moderate & Prototype \\
\hline
\cite{adamczyk2023conflict} & 2023 & \acs*{O-RAN} & \checkmark & \checkmark & \checkmark & \checkmark & -- & \checkmark &  &  &  &  &  &  &  & Moderate & Simulation \\
\hline
\cite{de2023agile} & 2023 & \acs*{O-RAN} & \checkmark & -- &  &  &  & \checkmark &  &  &  &  &  &  & \checkmark & Moderate & Data \\
\hline
\cite{wadud2023conflict} & 2023 & \acs*{O-RAN} & \checkmark & \checkmark & \checkmark & \checkmark & \checkmark & \checkmark &  &  &  &  & \checkmark &  &  & Weak & Simulation \\
\hline
\cite{bikos2025accord} & 2025 & \acs*{O-RAN} & \checkmark &  &  &  &  &  &  &  &  &  &  &  &  & Weak & Data \\
\hline
\cite{giannopoulos2025comix} & 2025 & \acs*{O-RAN} & \checkmark & \checkmark & \checkmark & -- & -- & \checkmark &  &  &  &  &  &  &  & Moderate & Simulation \\
\hline
\cite{nasim2025conflict} & 2025 & \acs*{O-RAN} & \checkmark &  & \checkmark & \checkmark & \checkmark &  &  &  &  &  &  &  &  & Weak & Simulation \\
\hline
\cite{kwon2026orca} & 2026 & \acs*{O-RAN} & \checkmark & \checkmark & -- & -- & -- & \checkmark &  &  &  &  &  &  &  & Weak & Simulation \\
\hline
\cite{riggio2026codr} & 2026 & \acs*{O-RAN} & \checkmark & \checkmark & -- & -- & -- & \checkmark & -- & -- & -- & -- & -- & -- & \checkmark & Moderate & Simulation \\
\hline
\cite{sinha2026experimental} & 2026 & \acs*{O-RAN} & \checkmark & -- & -- & -- & -- & \checkmark & -- & -- & -- & -- & -- & -- & -- & Moderate & Testbed \\
\hline
\cite{wadud2026ai} & 2026 & \acs*{O-RAN} & \checkmark &  & \checkmark & \checkmark & \checkmark &  &  &  &  &  &  & \checkmark &  & Weak & Simulation \\
\hline
\end{tabular}
\end{adjustbox}
\end{table*}

Banerjee~et~al. demonstrate a contradiction management prototype in their work in \cite{banerjee2021contradiction}. The authors use Nash social welfare to find the balanced configuration of the conflicted parameter. They use a Python-based evaluation of 30 manually created intents and do not perform network-level \ac{KPI} experiments. A recent testbed-based conflict classification for direct and indirect conflict is studied by Sinha~et~al. in \cite{sinha2026experimental}. The authors use rule-based detection of cross-granularity infeasibility between slice \ac{PRB} budgets and per-\ac{UE} scheduling floors. The experiment is conducted using both the simulator and an OCUDU/Open5GS/\ac{USRP} testbed with five commercial \acp{UE}. 

\subsection{L3: Resolution and Arbitration}
\label{subsec:l3}
\begin{figure*}[!htbp]
  \centering
  \includegraphics[width=0.85\textwidth]{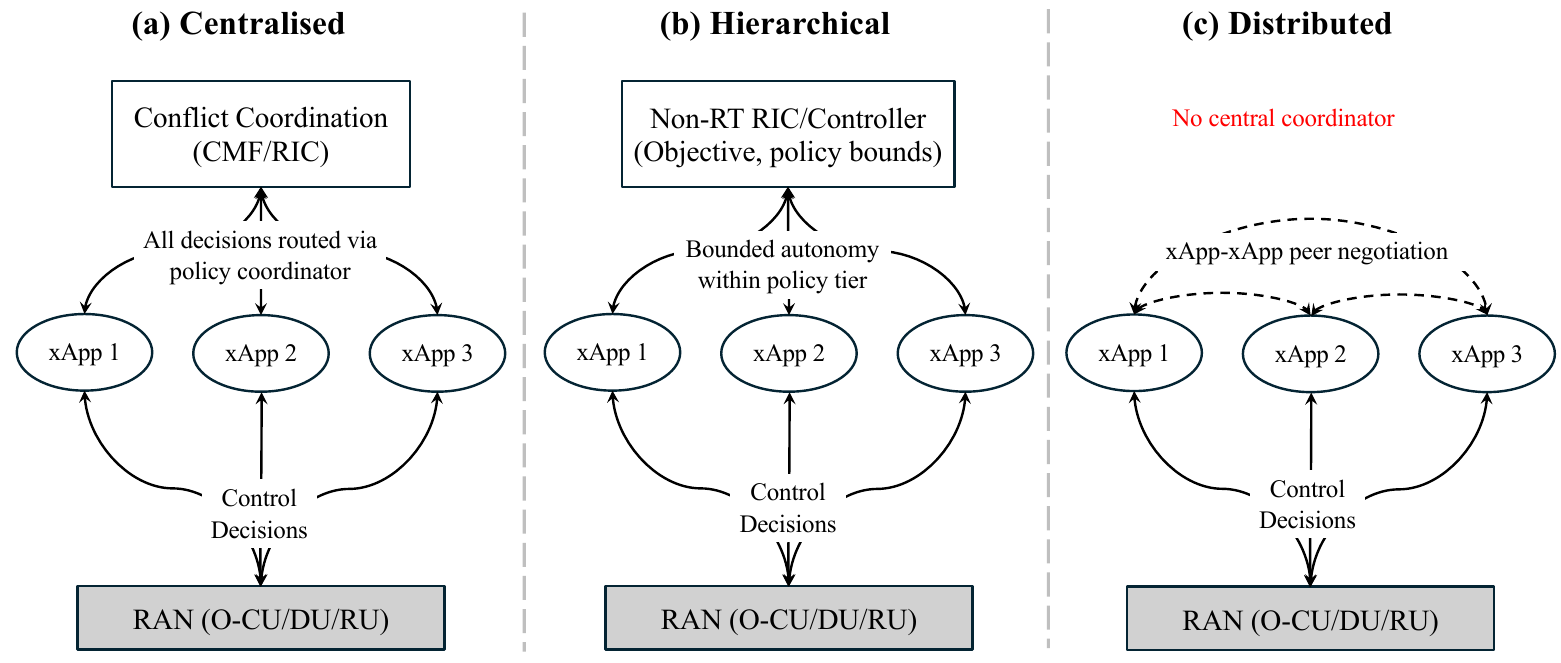}
  \caption{Centralised, hierarchical, and distributed conflict resolution architectures across \ac{SON}, \ac{CAN}, and \ac{O-RAN}.}
  \vspace{-0.5cm}
  \label{fig:coord_comparison}
\end{figure*}

\begin{figure*}[!htbp]
  \centering
  \includegraphics[width=\textwidth]{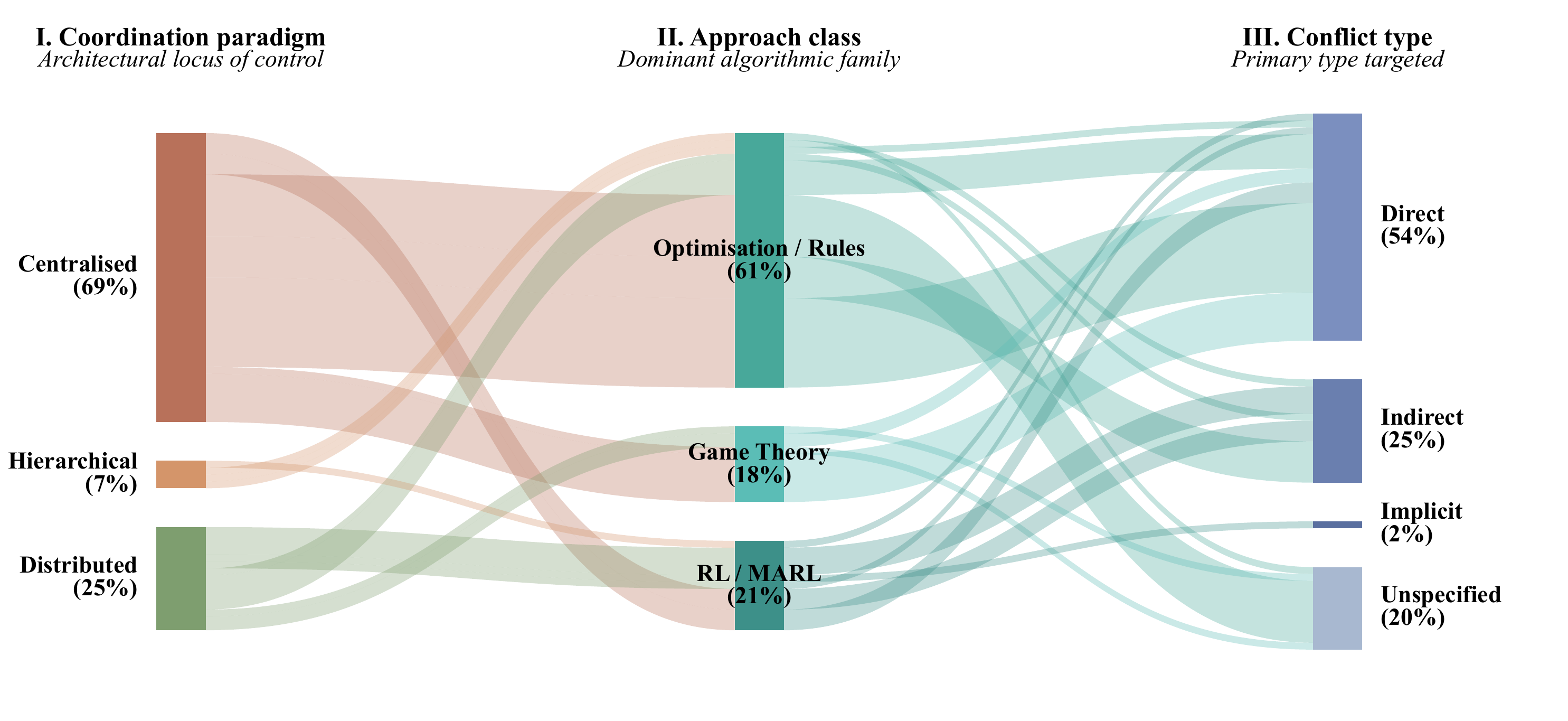}
  \vspace{-0.9cm}
  \caption{Alluvial mapping of evaluated mitigation studies by coordination architecture, method, and conflict type.}
  \label{fig:sankey}
  \vspace{-0.5cm}
\end{figure*}

This is the stage with the most work done in the existing literature. Studies mostly cover conflict management using priority rules, game-theoretic bargaining, learning and planning approaches and negotiated solutions. We categorise each of the mitigation studies into three different coordination architectures, as shown in Fig.~\ref{fig:coord_comparison}. These are: the centralised coordinator, which selects actions based on the operator's policy; the hierarchical approach in which a higher level sets the goals and the lower level applications operate within those constraints; and the distributed approach in which the applications themselves negotiate directly (this obviously requires the applications to be able to communicate with each other via a common interface across vendors).  We visually represent all the work in Fig.~\ref{fig:sankey} across these three different architectures.

\subsubsection{Formal Game-Theoretic Model}
A single conflicting control decision can be represented as a game $\mathcal{G}=\langle\mathcal{N},\{\mathcal{S}_i\}_{i\in\mathcal{N}},\{u_i\}_{i\in\mathcal{N}}\rangle$~\cite{fudenberg1991game,altman2006survey,lasaulce2011game,han2011game}. The players $i\in\mathcal{N}$ are the active \ac{SON} functions, \acp{CF}, \acp{xApp} and \acp{rApp}. Each player selects a strategy $s_i\in\mathcal{S}_i$ (this could be an allowed parameter value, an action, a resource request or the reported optimal configuration range). The payoff to player $i$, $u_i(s_i,s_{-i};x)$, is a measure of how well the resulting \ac{KPI} performs in meeting its objective. The game considers the strategies of the other players $s_{-i}$ and the network state $x$. In a non-cooperative game, a Nash equilibrium $s^*$ satisfies $u_i(s_i^*,s_{-i}^*;x)\geq u_i(s_i,s_{-i}^*;x)$ for every player and every allowed alternative $s_i$. This means that no player benefits by changing its choice alone. Cooperative or centrally coordinated methods instead select a feasible combination of actions through Nash or weighted Nash bargaining, Kalai--Smorodinsky bargaining, Nash social welfare maximisation, or an Eisenberg--Gale programme. The Shapley value serves a different purpose: it measures each player's contribution or influence. Similarly, analysing the stability of interacting control loops is a control-theory problem. We distinguish these roles when reviewing the methods, rather than treating every utility-based approach as the same form of game theory.

Bargaining also raises a trust issue in that applications may misreport their own utilities or their acceptable configuration ranges in order to get a better outcome. Mechanism design is all about how you can prevent this. The \ac{VCG} mechanism takes the allocation which maximises the reported sum of utilities and then uses payments which are based on the impact of each participant on the others. Under the usual assumptions (quasilinear utilities etc), truthful reporting is the optimal strategy~\cite{vickrey1961counterspeculation,clarke1971multipart,groves1973incentives}. None of the \ac{RAN} conflict management studies we surveyed actually implement \ac{VCG} or some other allocation mechanism which makes truthful reporting the optimal strategy. This is significant as a lot of the \ac{CAN} and \ac{O-RAN} bargaining work is predicated on the assumption that parties will truthfully report their utilities (although it's been shown that a manipulative \ac{CF} can gain from making false reports~\cite{banerjee2023trust}). Applying \ac{VCG} to \ac{RAN} control is in itself a very difficult problem as actions are interdependent and often take on continuous values, utilities are a function of \acp{KPI} which are themselves uncertain and monetary payments are generally not an option. Any arbitration based on self reported utilities therefore requires some form of protection, such as incentives for truthful reporting, audit scores or non-monetary penalties.

\subsubsection{Reinforcement and Multi-Agent Learning Model}
Conflicting control decisions that repeat in the network affect the future network condition. Such conflicting control decisions can be modelled as a Markov game $\mathcal{M}=\langle\mathcal{N},\mathcal{X},\{\mathcal{A}_i\},P,\{r_i\},\gamma\rangle$. In this expression, $\mathcal{N}$ is the set of applications, and $x_t\in\mathcal{X}$ is the network state. Also, $a_{i,t}\in\mathcal{A}_i$ represents application $i$'s action, $P(x_{t+1}\mid x_t,\mathbf{a}_t)$ refers to the transition probability which describes how the joint actions affect the state of the network, $r_i(x_t,\mathbf{a}_t)$ indicates application $i$'s \ac{KPI} based reward and finally $\gamma$ represents the discount factor which controls the weight of future rewards~\cite{sutton2018reinforcement,han2011game}. Independent learning control applications or functions get confused as the network condition changes as other applications update their own policies during training. \ac{CTDE} is one way of dealing with this. The critic learns from the joint state and action of the applications, and each application maintains its own execution policy~\cite{lowe2017multiagent}. There are other solutions to this problem, like Team learning, shared critic methods, and policy distillation. However, we need to note that coordinated training does not guarantee safe operation if the network conditions change or if a new application is added. 

Three issues remain. Firstly, exploration can violate \acp{KPI} or safety constraints in a live \ac{RAN}, which in turn will necessitate offline or digital twin training, restricted action spaces or some form of safety mechanism which will simply prevent the agent from making unsafe decisions. Secondly, the convergence guarantees that exist for tabular single-agent \ac{RL} don't necessarily carry over to deep multi-agent \ac{RL} where all the policies are interdependent and all changing together. All of the studies we've looked at report stable training in their experiments but give absolutely no general convergence guarantee~\cite{hernandezleal2019survey}. Thirdly, too little exploration will leave the co-ordinator with a very bad policy, too much exploration will just result in a constant stream of conflicting actions. Performance in the experiment is therefore very much separate from proven convergence or stability.

Early \ac{SON} studies use a number of coordination methods. Karla predicts, ranks and applies local configurations in order to reduce the oscillations that are likely to be repeated~\cite{karla2012resolving}. Chen et al. use a combination of handover problems, signal gain and expected resource savings to prioritise the users and jointly tune the \ac{CIO} and \ac{TTT}~\cite{chen2012difference}. Zhou et al. tune the \ac{CIO} in both directions between cells whilst keeping the changes within \ac{MRO} safe limits~\cite{zhou2013joint}. Iacoboaiea et al.'s \ac{SONCO} learns which requests to accept based on their importance and coordinates \ac{CRE}, \ac{MRO} and interference control without the need to evaluate a runtime conflict classifier~\cite{iacoboaiea2016son}. Tall et al. use a coordination matrix and a matrix inequality condition to stabilise the interacting control loops. Their approach is a combination of a management plane server and distributed updates and provides a control theoretic stability result~\cite{tall2014distributed}. Moysen and Giupponi split a global \ac{MDP} up into smaller models for individual functions and use a coordination game to deal with shared power conflicts~\cite{moysen2015coordination}. The conference paper is the same \ac{CCO}/\ac{ICIC} experiment so provides no independent validation~\cite{moysen2015self}. Their earlier self healing paper is about recovery from faults rather than conflicts between functions~\cite{moysen2014reinforcement}. Mu\~{n}oz et al. jointly tune the load balancing and handover parameters using fuzzy Q-learning~\cite{munoz2015load}. MOHOLA maintains a set of trade-offs between avoiding failures and load distribution. A central controller is used to schedule the neighbouring agents to avoid simultaneous \ac{CIO} changes; \ac{MLB} and \ac{MRO} are not treated as separate functions~\cite{kim2024load}. \ac{SLC2} uses \ac{DDQN} with prioritised experience replay to accept or reject requests, tested in ns-3/ns3-gym with 5 \ac{BS}/40 \ac{UE} and 9 \ac{BS}/64 \ac{UE}~\cite{lee2024scalable}.

In \ac{CAN} each \ac{CF} reports its preferred configuration ranges and utility function. The Controller then chooses a value which maximises Nash social welfare~\cite{banerjee2020ran}. Banerjee et al. introduce configuration weights via an Eisenberg-Gale solution. They report up to a 7.7\% improvement over the unweighted case in a synthetic two \ac{CF} example and a 2\% gain in their Hamburg network simulation with learned \ac{MLB} and \ac{CCO}~\cite{banerjee2021optimal}. A subsequent paper uses Shapley values to determine the weights and reports gains of up to 9.18\% over the \ac{NSWF} only Controller~\cite{banerjee2022toward}. Baktir et al predict how candidate actions affect the \acp{KPI} and pick the action with the lowest intent violation penalty. Their Dockerised 5G emulator with a Kubernetes hosted core is one of the very few \ac{CAN} testbed evaluations that has been carried out to date~\cite{baktir2022intent}. Kattepur et al's CONVERGENCE uses \ac{AI} planning and a \ac{PDDL} domain model to coordinate the \ac{RAN} partition goals, with an operator data example used to demonstrate the approach involving premium and normal \ac{eMBB} slices~\cite{kattepur2024convergence}. Cinemre et al compare the four bargaining methods (\ac{NBS}, \ac{WNBS}, \ac{KSBS} and \ac{SEBS}) in the case of two intents requesting different antenna tilts. \ac{KSBS} is found to be the fairest in that example according to Jain's fairness index~\cite{cinemre2023direct}. In their follow up paper they compare the directions of the two intents' loss gradients. A non-positive cosine similarity indicates conflicting directions and an \ac{SPSA} based Multiple Gradient Descent algorithm is used to select a compromise update~\cite{cinemre2024gradient}.

In \ac{O-RAN}, Wadud~et~al. use \ac{QACM} to search for a shared parameter value which will reduce the number of weighted \ac{KPI}-threshold violations and at the same time increase the number of \acp{xApp} that are satisfied. They use per-\ac{xApp} \ac{KPI} predictors and assume that conflicts have already been detected~\cite{wadud2024qacm}. A companion paper looks at a direct transmit power conflict between \ac{MRO} and energy saving \acp{xApp} in the MATLAB 5G Toolbox, using an \ac{O-RAN} 7.2 split with 4 \acp{gNB} and 100 \acp{UE}~\cite{wadud2025xApp}. Adamczyk and Kliks \ac{ACCoRD} uses a \ac{PPO}-Clip actor-critic which will either leave the decision as is, reject it with a cooldown or make incremental adjustments to it. It achieves the lowest Relative Penalty Score at both medium and high loads, though it only evaluates indirect \ac{MRO}-\ac{MLB} conflicts in simulation~\cite{adamczyk2026accord}. Zafar~et~al. use Double \ac{DQN} to balance the energy efficiency and load balancing objectives via \ac{RU} activation decisions. Their evaluation is in Python simulation and the proposed E2 control path is not implemented~\cite{zafar2025conflict}. Armstrong~et~al.'s \ac{TACIT} enables \acp{rApp} to evaluate each other's proposals based on predicted \acp{KPI} and utility votes. Acceptance of a proposal is determined via reputation-weighted Nash social welfare, with no central arbiter or explicit conflict classifier required. Reputation is updated on the basis of observed outcomes. A 19 cell, 7 \ac{rApp} ns-3 experiment is used to demonstrate arbitration and the mitigation of influence from a failing \ac{rApp}~\cite{armstrong2026tacit}. Al-Hameed~et~al.'s \ac{ES}-NAM makes use of interpretable traffic, coverage and history contributions to inform energy saving decisions. A peer objection is only overridden if both the traffic and coverage contributions are above the threshold values derived from the training data. On synthetic scenarios resolution accuracy increases from 58.3\% to 72.9\% with 99.5\% override precision and only 35.2\% recall. The baseline is dependent on the distribution of the sampled scenario and arbitration remains within the energy saving \ac{xApp} rather than in an integrated \ac{WG3} \ac{CMF}~\cite{alhameed2026explainable}. Almohammedi~et~al. resolve a direct power conflict by testing 21 weighted combinations of the two requested values. A digital twin ranks the candidates and a moving average of the prediction errors is used to track the reliability. If the error exceeds a threshold the method simply falls back to the best action that was previously evaluated in the simulated network~\cite{almohammedi2026twin}. Kose's \ac{MACOL-x} coordinates the traffic steering, power control and handover \acp{xApp} via separate memories which take into account the actions of the peers. An \ac{xApp} will simply skip its preferred action if the conflict level exceeds a running threshold~\cite{kose2026macolx}. Kurtulan~et~al.'s \ac{ML} Orchestrator intercepts the request prior to execution, uses an XGBoost model to predict the outcome of each candidate action subset over a 5 second period and selects the subset which has the highest utility. Requests that are rejected are simply blocked for 5 seconds. The two papers are from the same study family and both show an indirect TxPower--A3 interaction~\cite{kurtulan2026orchestrator,kurtulan2026proactive}. Sultana~et~al. take a much simpler approach with a Central Controller Agent which simply orders the requests by arrival (without any regard to application importance) on an srsRAN/\ac{USRP} X310 testbed~\cite{sultana2024software}. Two related intent management papers provide wider comparisons. Karakaya et al.'s \ac{MABCR} uses a hierarchical multi-armed bandit: the child agents suggest configurations and the parent chooses from them using the Upper Confidence Bound \cite{karakaya2025online}. Ojaghi et al.'s \ac{M2O-IBNS} uses priority based re-optimisation and Benders decomposition with a 1\% stopping gap to balance slice throughput, energy cost and creation cost. A high priority \ac{SLA} can be protected at the expense of the throughput of other slices. Their experiments vary the \ac{SLA} thresholds externally and they do not present an online policy which would automatically relax them \cite{ojaghi2026intent}. Neither of these papers provides an \ac{O-RAN} \ac{RIC} implementation.

Two other adjacent papers look at problems that fall outside the \ac{O-RAN} \ac{xApp} conflict taxonomy. The first one is Li et al., who use an \ac{MDP} with Rainbow \ac{DQN} to allocate shared edge computing resources amongst user intents, with a proposed placement in the \ac{ONAP} Service Orchestrator. Their action space precludes infeasible allocations, so there is no need for any separate conflict detection. The problem does not involve conflicting \ac{RAN} control applications~\cite{li2024conflict}. Bikos and Verikoukis in their one-page 6G-ACCORD proposal (discussed in Section~\ref{subsec:l2}) identify Weighted Nash, Kalai-Smorodinsky, and Shannon entropy bargaining as potential resolution methods. However, the paper gives no bargaining formulation, no disagreement point, and no evaluation~\cite{bikos2024accord}.

Table~\ref{tab:survey_class_l3} also includes the more general intent coordination methods. For example, Banerjee~et~al. in~\cite{banerjee2021game} employ learned synthetic utilities and a central Nash social welfare Controller in \ac{CAN}. Later, Dehghan Biyar et al. in~\cite{biyar2025autonomous} predict the impact of actions on various \acp{KPI} within a containerized intent management emulator. But they do not show any \ac{O-RAN} deployment at all. Similarly, Cinemre et al. in~\cite{cinemre2025xapp} use an \ac{A2C} scheduler to select between power control and resource allocation policies in an only indirect conflict scenario. Karakaya et al. in~\cite{karakaya2025online} propose a hierarchical bandit, and Ojaghi et al. in~\cite{ojaghi2026intent} propose priority-based slice re-optimization are \ac{CAN}/\ac{IBN} contributions, with no \ac{O-RAN} \ac{RIC} implementations demonstrated by either of these works.

\begin{table*}[!tbp]
\centering
\footnotesize
\setlength{\tabcolsep}{3pt}
\renewcommand{\arraystretch}{1.05}
\caption{L3: Resolution and Arbitration.}\label{tab:survey_class_l3}
\begin{adjustbox}{max width=\textwidth}
\begin{tabular}{|p{1.07cm}|>{\centering\arraybackslash}p{0.62cm}|>{\centering\arraybackslash}p{0.95cm}|>{\centering\arraybackslash}p{0.6cm}|>{\centering\arraybackslash}p{0.6cm}|>{\centering\arraybackslash}p{0.6cm}|>{\centering\arraybackslash}p{0.6cm}|>{\centering\arraybackslash}p{0.6cm}|>{\centering\arraybackslash}p{0.6cm}|>{\centering\arraybackslash}p{0.6cm}|>{\centering\arraybackslash}p{0.6cm}|>{\centering\arraybackslash}p{0.6cm}|>{\centering\arraybackslash}p{0.86cm}|>{\centering\arraybackslash}p{0.58cm}|>{\centering\arraybackslash}p{0.58cm}|>{\centering\arraybackslash}p{0.58cm}|>{\centering\arraybackslash}p{1.35cm}|p{1.45cm}|}
\hline
\textbf{Paper} & \textbf{Year} & \textbf{Domain} & \textbf{Det.} & \textbf{Mit.} & \textbf{Dir.} & \textbf{Ind.} & \textbf{Imp.} & \textbf{Cen.} & \textbf{Dis.} & \textbf{Hier.} & \textbf{\acs*{RL}} & \textbf{\acs*{MARL}} & \textbf{\acs*{GT}} & \textbf{\acs*{GNN}} & \textbf{Int.} & \textbf{Evidence} & \textbf{Validation} \\
\hline
\cite{chen2012difference} & 2012 & \acs*{SON} &  & \checkmark & \checkmark &  &  &  & \checkmark &  &  &  &  &  &  & Moderate & Simulation \\
\hline
\cite{karla2012resolving} & 2012 & \acs*{SON} & \checkmark & \checkmark & \checkmark & \checkmark &  &  & \checkmark &  &  &  &  &  &  & Weak & Simulation \\
\hline
\cite{zhou2013joint} & 2013 & \acs*{SON} &  & \checkmark & \checkmark &  &  &  & \checkmark &  &  &  &  &  &  & Moderate & Simulation \\
\hline
\cite{tall2014distributed} & 2014 & \acs*{SON} &  & \checkmark &  & \checkmark &  &  & \checkmark & \checkmark &  &  & -- &  &  & Moderate & Simulation \\
\hline
\cite{moysen2015coordination} & 2015 & \acs*{SON} & \checkmark & \checkmark & \checkmark &  &  &  & \checkmark &  & \checkmark & \checkmark & \checkmark &  &  & Moderate & Simulation \\
\hline
\cite{moysen2015self} & 2015 & \acs*{SON} & -- & \checkmark & \checkmark &  &  &  & \checkmark &  & \checkmark & \checkmark & \checkmark &  &  & Moderate & Simulation \\
\hline
\cite{munoz2015load} & 2015 & \acs*{SON} &  & \checkmark & \checkmark &  &  &  & \checkmark &  & \checkmark &  &  &  &  & Moderate & Simulation \\
\hline
\cite{iacoboaiea2016son} & 2016 & \acs*{SON} & -- & \checkmark & \checkmark & \checkmark &  & \checkmark &  &  & \checkmark &  &  &  &  & Weak & Simulation \\
\hline
\cite{kim2024load} & 2024 & \acs*{SON} &  & \checkmark & \checkmark &  & -- &  & \checkmark & \checkmark & \checkmark & \checkmark &  &  &  & Moderate & Simulation \\
\hline
\cite{lee2024scalable} & 2024 & \acs*{SON} &  & \checkmark &  & \checkmark &  & \checkmark &  &  & \checkmark &  &  &  &  & Strong & Simulation \\
\hline
\cite{banerjee2020ran} & 2020 & \acs*{CAN} & -- & \checkmark & \checkmark &  &  & \checkmark &  &  &  &  & \checkmark &  &  & Weak & Simulation \\
\hline
\cite{banerjee2021game} & 2020 & \acs*{CAN} & \checkmark & \checkmark & \checkmark & \checkmark & \checkmark & \checkmark &  &  &  &  & \checkmark &  &  & Weak & Simulation \\
\hline
\cite{banerjee2021optimal} & 2021 & \acs*{CAN} & \checkmark & \checkmark & \checkmark &  &  & \checkmark &  &  &  &  & \checkmark &  &  & Moderate & Simulation \\
\hline
\cite{banerjee2022toward} & 2021 & \acs*{CAN} & \checkmark & \checkmark & \checkmark &  &  & \checkmark &  &  &  &  & \checkmark &  &  & Strong & Simulation \\
\hline
\cite{baktir2022intent} & 2022 & \acs*{CAN} & \checkmark & \checkmark &  & \checkmark &  & \checkmark &  & -- &  &  &  &  & \checkmark & Weak & Testbed \\
\hline
\cite{cinemre2023direct} & 2023 & \acs*{CAN} &  & \checkmark & \checkmark &  &  & \checkmark &  &  &  &  & \checkmark &  & \checkmark & Strong & Simulation \\
\hline
\cite{cinemre2024gradient} & 2024 & \acs*{CAN} & \checkmark & \checkmark & \checkmark &  &  & \checkmark &  &  &  &  &  &  & \checkmark & Strong & Simulation \\
\hline
\cite{kattepur2024convergence} & 2024 & \acs*{CAN} & -- & -- &  &  &  & \checkmark &  &  &  &  &  &  & \checkmark & Weak & Data \\
\hline
\cite{biyar2025autonomous} & 2025 & \acs*{CAN} & \checkmark & \checkmark &  &  &  & \checkmark &  &  &  &  &  &  & \checkmark & Moderate & Testbed \\
\hline
\cite{karakaya2025online} & 2025 & \acs*{CAN} & -- & \checkmark &  & -- &  &  &  & \checkmark &  &  &  &  & \checkmark & Moderate & Simulation \\
\hline
\cite{ojaghi2026intent} & 2026 & \acs*{CAN} & -- & \checkmark & -- & -- &  & \checkmark &  &  &  &  &  &  & \checkmark & Strong & Simulation \\
\hline
\cite{sultana2024software} & 2024 & \acs*{O-RAN} & \checkmark & \checkmark & \checkmark &  &  & \checkmark &  &  &  &  &  &  &  & Moderate & Testbed \\
\hline
\cite{wadud2024qacm} & 2024 & \acs*{O-RAN} &  & \checkmark & \checkmark & \checkmark & \checkmark & \checkmark &  &  &  &  & -- &  &  & Moderate & Simulation \\
\hline
\cite{cinemre2025xapp} & 2025 & \acs*{O-RAN} &  & \checkmark &  & \checkmark &  & \checkmark &  &  & \checkmark &  &  &  &  & Moderate & Simulation \\
\hline
\cite{wadud2025xApp} & 2025 & \acs*{O-RAN} & \checkmark & \checkmark & \checkmark & -- & -- & \checkmark &  &  &  &  & -- & -- &  & Moderate & Simulation \\
\hline
\cite{zafar2025conflict} & 2025 & \acs*{O-RAN} &  & \checkmark & \checkmark &  &  & \checkmark &  &  & \checkmark &  &  &  &  & Weak & Simulation \\
\hline
\cite{adamczyk2026accord} & 2026 & \acs*{O-RAN} &  & \checkmark &  & \checkmark &  & \checkmark &  &  & \checkmark &  &  &  &  & Strong & Simulation \\
\hline
\cite{alhameed2026explainable} & 2026 & \acs*{O-RAN} &  & \checkmark & \checkmark &  &  &  &  &  &  &  &  &  &  & Moderate & Simulation \\
\hline
\cite{almohammedi2026twin} & 2026 & \acs*{O-RAN} &  & \checkmark & \checkmark &  &  & \checkmark &  &  &  &  &  &  &  & Strong & Simulation \\
\hline
\cite{armstrong2026tacit} & 2026 & \acs*{O-RAN} &  & \checkmark & -- & -- & -- &  & \checkmark &  &  &  & \checkmark &  &  & Moderate & Simulation \\
\hline
\cite{kim2026mobilityenergy} & 2026 & \acs*{O-RAN} & -- & \checkmark & -- & \checkmark & -- & \checkmark & -- & -- & -- & -- & -- & -- & -- & Moderate & Simulation \\
\hline
\cite{kose2026macolx} & 2026 & \acs*{O-RAN} &  & \checkmark &  & \checkmark &  &  & \checkmark &  &  &  &  &  &  & Moderate & Simulation \\
\hline
\cite{kurtulan2026orchestrator} & 2026 & \acs*{O-RAN} &  & \checkmark &  & \checkmark &  & \checkmark &  &  &  &  &  &  &  & Moderate & Simulation \\
\hline
\cite{kurtulan2026proactive} & 2026 & \acs*{O-RAN} &  & \checkmark &  & \checkmark &  & \checkmark &  &  &  &  &  &  &  & Moderate & Simulation \\
\hline
\cite{natanzi2026aura} & 2026 & \acs*{O-RAN} & -- & \checkmark & -- & -- & -- & \checkmark & -- & -- & -- & -- & -- & -- & -- & Moderate & Testbed \\
\hline
\cite{wadud2026confidence} & 2026 & \acs*{O-RAN} & -- & \checkmark & \checkmark & -- & -- & \checkmark & -- & -- & -- & -- & -- & -- & -- & Moderate & Simulation \\
\hline
\cite{wadud2026costtriage} & 2026 & \acs*{O-RAN} & -- & \checkmark & -- & -- & -- & \checkmark & -- & -- & -- & -- & -- & -- & -- & Moderate & Simulation \\
\hline
\cite{xia2026xtruce} & 2026 & \acs*{O-RAN} & -- & \checkmark & -- & -- & -- & \checkmark & -- & -- & -- & -- & -- & -- & -- & Moderate & Testbed \\
\hline
\end{tabular}
\end{adjustbox}
\end{table*}

Recent mitigation mechanisms cover different conflict scopes apart from the \acp{xApp} in the \ac{Near-RT} \ac{RIC}. The author in~\cite{riggio2026codr} proposes \ac{CoDR} that mitigates conflict among \acp{rApp} in the \ac{Non-RT} \ac{RIC}.  Another work named AURA~\cite{natanzi2026aura} covers \ac{rApp} coordination by constraining the  autonomous \ac{rApp} actions through feasibility invariants, dwell times, and deadbands. They use \ac{OAI} evaluation to show convergence of the proposed solution on a bounded shared-resource scenario. We can see from the literature that there are not much study done that covers \ac{rApp} conflicts till now. Therefore, let us focus more on \ac{xApp} conflicts. A recent study named xTRUCE in~\cite{xia2026xtruce} uses hard physical/operator constraints, priority-ordered relaxation of \ac{xApp} targets, conflict certificates, and a delay-safe action path for \ac{xApp} conflict mitigation. Their evaluation is conducted on both simulation and \ac{OAI}/FlexRIC \ac{OTA}. Kim and Lee in~\cite{kimlee2026admission} investigate joint handover and energy saving using a priority-based preemption for conditional handover while minimizing energy consumption. A later work by the same author in~\cite{kim2026mobilityenergy} models handover to a sleeping target through stochastic residence and wake-up times and a weighted arbitration rule. Neither of these works particularly focuses on any of the direct, indirect, or implicit conflicts. A recent interesting idea of triaging a conflicting control decision has been presented by Wadud~et~al. in~\cite{wadud2026costtriage,wadud2026confidence}. They presented two different triage techniques, such as cost-aware and confidence-aware. The first one in~\cite{wadud2026costtriage} analyses mitigation intervention costs and benefits on the detected conflict before taking the mitigation actions. The second one in~\cite{wadud2026confidence} takes intervention decisions based on prediction confidence on the affected \ac{KPI}. They use a closed-loop ns3-oran/5G-LENA and \ac{CCO}/\ac{ES} \acp{xApp} for the experiment.

\subsection{L4: Enforcement and Actuation Control}
\label{subsec:l4}
A conflict management decision is only useful if it can be applied to mitigate a conflict. In this stage, we define which applications can change the control parameters, which control decision requests are accepted, when further mitigation action is required to be deferred, and when changes need to be undone. Sultana~et~al. in \cite{sultana2025experimental} propose passing all the E2 CONTROL messages through the conflict mitigation function so that control decisions coming from non-prioritized \acp{xApp} can be blocked before they reach the E2 node. They evaluate their concept on the \ac{OSC}/srsRAN \ac{OOTB} testbed and consider a direct conflict over slice \ac{PRB} allocation. Their reported result shows improvement in throughput after mitigation, although they did not report latency or processing overhead of their proposed method. Armstrong et al's proposed method \ac{CAMINO} in~\cite{armstrong2025camino} screens control decisions a layer up in the \ac{SMO}. It works by grouping together cells that are similarly configured and then predicting the impact of a proposed change on the \acp{KPI}. A control decision is accepted or rejected based on the combined observation of weather, traffic, events, and operator intent. The authors evaluated using historical data collected from 800 \ac{LTE} cells. However, the authors did not use a realistic environment to test their proposed method in a live network. Also, they did not cover indirect and implicit conflict types. Jawad~et~al. in~\cite{jawad2026runtime} looks at \ac{AI}-based control decisions prior to them impacting the \ac{RAN}. Their proposed Safety Copilot works as an intermediary layer to check current limits, the predicted next-step conditions, and the constraints over time windows to take mitigation action. Unsafe control decisions can be clipped, partially blocked, reversed, or replaced with a fallback. The evaluation is based on a single controller with no second \ac{xApp}. Rahman~et~al.~\cite{rahman2026origami} propose ORIGAMI that coordinates requests via a single service bus. \acp{xApp} publish their desired actions as services, and the \ac{PIOR} mediator checks the requests against an intent registry. Only the highest priority valid decision is then passed on to E2 signaling. The authors evaluated their proposed method using a 5G \ac{NR} \ac{O-RAN} testbed covering only direct conflict. 

If mitigation decisions are enforced in an unsafe order across various interfaces, this can degrade the network performance even further. Hou and Yang in~\cite{hou2026veraran} take this into consideration and check the complete action plan prior to execution. The authors use a \ac{DAG} with bounded delays to represent each mitigation action. If a mitigation plan is found to be unsafe, it checks other alternative orderings to find the optimal plan. In their evaluation, the authors consider 144 stratified plans. Their selective ordering establishes the same set of ordering constraints as full sequential execution and leaves 87--89\% of the action pairs with no ordering constraint at all. The authors claim that the completion time bounds are about ten times better, while the median latency between messages coming from the control application and actual acceptance of it remains 0.455 ms. This work shows the significance of evaluating a mitigation decision before execution so that it remains safe across interfaces and timescales and always guarantees an improvement in network performance. However, this is not a typical \ac{WG3} conflict between competing application decisions; rather, it should be considered an execution ordering problem.

\begin{table*}[!b]
\centering
\footnotesize
\setlength{\tabcolsep}{3pt}
\renewcommand{\arraystretch}{1.05}
\caption{L4: Enforcement and Actuation Control.}\label{tab:survey_class_l4}
\begin{adjustbox}{max width=\textwidth}
\begin{tabular}{|p{1.07cm}|>{\centering\arraybackslash}p{0.62cm}|>{\centering\arraybackslash}p{0.95cm}|>{\centering\arraybackslash}p{0.6cm}|>{\centering\arraybackslash}p{0.6cm}|>{\centering\arraybackslash}p{0.6cm}|>{\centering\arraybackslash}p{0.6cm}|>{\centering\arraybackslash}p{0.6cm}|>{\centering\arraybackslash}p{0.6cm}|>{\centering\arraybackslash}p{0.6cm}|>{\centering\arraybackslash}p{0.6cm}|>{\centering\arraybackslash}p{0.6cm}|>{\centering\arraybackslash}p{0.86cm}|>{\centering\arraybackslash}p{0.58cm}|>{\centering\arraybackslash}p{0.58cm}|>{\centering\arraybackslash}p{0.58cm}|>{\centering\arraybackslash}p{1.35cm}|p{1.45cm}|}
\hline
\textbf{Paper} & \textbf{Year} & \textbf{Domain} & \textbf{Det.} & \textbf{Mit.} & \textbf{Dir.} & \textbf{Ind.} & \textbf{Imp.} & \textbf{Cen.} & \textbf{Dis.} & \textbf{Hier.} & \textbf{\acs*{RL}} & \textbf{\acs*{MARL}} & \textbf{\acs*{GT}} & \textbf{\acs*{GNN}} & \textbf{Int.} & \textbf{Evidence} & \textbf{Validation} \\
\hline
\cite{armstrong2025camino} & 2025 & \acs*{O-RAN} & \checkmark & -- &  & -- & -- & \checkmark &  &  &  &  &  &  & \checkmark & Weak & Data \\
\hline
\cite{sultana2025experimental} & 2025 & \acs*{O-RAN} & \checkmark & \checkmark & \checkmark &  &  & \checkmark &  &  &  &  &  &  &  & Moderate & Testbed \\
\hline
\cite{jawad2026runtime} & 2026 & \acs*{O-RAN} &  & -- &  &  &  &  &  &  &  &  &  &  &  & Moderate & Simulation \\
\hline
\cite{rahman2026origami} & 2026 & \acs*{O-RAN} & \checkmark & \checkmark & \checkmark &  &  & \checkmark &  &  &  &  &  &  &  & Moderate & Testbed \\
\hline
\end{tabular}
\end{adjustbox}
\end{table*}

\subsection{L5: Verification, Attribution, and Feedback}
\label{subsec:l5}
This section discusses the impact of a mitigation action, the attributes of the \ac{KPI} change, and the mitigation action's effectiveness. These three are all related but separate steps. The authors in \cite{varshney2025xai4c} present XAI4C that uses KernelSHAP distances and a random forest classifier to identify post-action \ac{KPI} conflicts. The authors proposed both mitigation and feedback, but did not evaluate both. Later, the same authors in \cite{varshney2025explainable} use \ac{SHAP} features with logistic regression and evaluate the changes to the corrective parameters in simulation. Hou~et~al. in \cite{hou2025if} present a counterfactual conformal estimator. Their method uses the logged emulator and the analytical data to estimate the outcomes under alternative application choices. It can be used for impact estimation but does not demonstrate an implicit-conflict detector or concurrent-app recovery. Yungaicela~et~al.  in~\cite{yungaicela2024misconfiguration} propose linking \ac{KPI} anomalies to past \ac{xApp} actions. This is placed in the \ac{xNIB}. However, there is no implementation or evaluation. Riggio in~\cite{riggio2026observability} identifies potential causes of \ac{KPI} changes when multiple \ac{rApp} actions take place. Their method combines \ac{PELT} change detection, conditional Granger causality, and a difference-in-differences analysis to rank candidate causes. Where the evidence is insufficient to separate the causes, it reports shared responsibility. Hung et al. in~\cite{hung2025anomaly} check system integrity in two stages. First, they validate state transitions via a finite state machine and, second, check E2 packets against protocol requirements. Then they monitor message rates and filter. It's designed to deal with malicious \acp{xApp} and malicious E2 nodes rather than conflicts between legitimate control applications. 

\begin{table*}[!tbp]
\centering
\footnotesize
\setlength{\tabcolsep}{3pt}
\renewcommand{\arraystretch}{1.05}
\caption{L5: Verification, Attribution, and Feedback.}\label{tab:survey_class_l5}
\begin{adjustbox}{max width=\textwidth}
\begin{tabular}{|p{1.07cm}|>{\centering\arraybackslash}p{0.62cm}|>{\centering\arraybackslash}p{0.95cm}|>{\centering\arraybackslash}p{0.6cm}|>{\centering\arraybackslash}p{0.6cm}|>{\centering\arraybackslash}p{0.6cm}|>{\centering\arraybackslash}p{0.6cm}|>{\centering\arraybackslash}p{0.6cm}|>{\centering\arraybackslash}p{0.6cm}|>{\centering\arraybackslash}p{0.6cm}|>{\centering\arraybackslash}p{0.6cm}|>{\centering\arraybackslash}p{0.6cm}|>{\centering\arraybackslash}p{0.86cm}|>{\centering\arraybackslash}p{0.58cm}|>{\centering\arraybackslash}p{0.58cm}|>{\centering\arraybackslash}p{0.58cm}|>{\centering\arraybackslash}p{1.35cm}|p{1.45cm}|}
\hline
\textbf{Paper} & \textbf{Year} & \textbf{Domain} & \textbf{Det.} & \textbf{Mit.} & \textbf{Dir.} & \textbf{Ind.} & \textbf{Imp.} & \textbf{Cen.} & \textbf{Dis.} & \textbf{Hier.} & \textbf{\acs*{RL}} & \textbf{\acs*{MARL}} & \textbf{\acs*{GT}} & \textbf{\acs*{GNN}} & \textbf{Int.} & \textbf{Evidence} & \textbf{Validation} \\
\hline
\cite{yungaicela2024misconfiguration} & 2024 & \acs*{O-RAN} & -- &  & -- & -- & -- &  &  &  &  &  &  &  &  & None & Conceptual \\
\hline
\cite{hou2025if} & 2025 & \acs*{O-RAN} & -- &  &  &  &  & \checkmark &  &  &  &  &  &  &  & Strong & Simulation \\
\hline
\cite{varshney2025explainable} & 2025 & \acs*{O-RAN} & \checkmark & \checkmark &  & \checkmark & \checkmark & \checkmark &  &  &  &  &  &  &  & Moderate & Simulation \\
\hline
\cite{varshney2025xai4c} & 2025 & \acs*{O-RAN} & \checkmark &  &  & \checkmark & \checkmark & \checkmark &  &  &  &  &  &  &  & Moderate & Simulation \\
\hline
\cite{riggio2026observability} & 2026 & \acs*{O-RAN} & -- &  &  &  &  & \checkmark &  &  &  &  &  &  &  & Strong & Simulation \\
\hline
\end{tabular}
\end{adjustbox}
\end{table*}

\subsection{L6: Governance and Application Lifecycle Management}
\label{subsec:l6}
This is the final stage of the conflict lifecycle. This stage covers application onboarding, trust, and the policies and standards required for ongoing conflict management. Banerjee~et~al. in \cite{banerjee2023trust} show that a \ac{CF} reporting false utility values can lead to a 10\% reduction in overall performance in a 5-cell 5G simulation. Their proposed CoDeRa method introduces a controlled amount of randomness into the Controller's decisions in order to reduce the potential for manipulation. However, their method does not work to fix measurements which are corrupted by real data manipulation by a \ac{CF}. Their intent-driven orchestration framework in~\cite{banerjee2021intent} uses an Intent Specification Platform and an Intent-Driven Network Automation Function Orchestrator to translate the operator's intent into Controller and \ac{CF} instructions. Corici et al. in \cite{corici2024towards} identify conflict arbitration between \ac{Non-RT} and \ac{Near-RT} control as a 6G design requirement. However, the authors offer no arbitration objective, no weighting method and no evaluation at all. Salmi et al. in \cite{salmi2026ai} propose \ac{Non-RT} intent reasoning with \ac{Near-RT} arbitration. Their evaluation is a token-budget and penalty-counter game between two \acp{xApp} and their proposed \ac{DRL} and \ac{LLM} components, and network \ac{KPI} recovery is not evaluated. Li~et~al. in~\cite{li2026multi} use three agents to produce conflict-aware \ac{rApp} policies. Firstly, a Perception Agent identifies actuator contention, parameter coupling, objective interference and vendor interoperability issues. Second, a Reasoning Agent constructs an \ac{xApp} pipeline for each intent. And finally, a Refinement Agent uses its stored experience to validate and improve the pipeline. Their experiment on 14 textual \ac{xApp} descriptions and 7 intents reports fewer orchestration iterations and improved selection accuracy. However, the authors do not report on token cost, latency, \ac{RAN} \acp{KPI} or closed-loop network performance.

Recent work by Armstrong et al. in~\cite{armstrong2026oracle} proposes \ac{ORACLE}, which makes \ac{rApp} arbitration auditable. At first, it records configuration proposals and their predicted \ac{KPI} impact on a permissioned ledger. Later, it uses smart contracts to apply a published decision rule and link the decision to the observed outcome. However, it does not perform conflict detection, and it does not guarantee the accuracy of the predictions. Their evaluation on a four-validator Hyperledger Besu prototype processed 932 decisions from 257 ns-3-derived conflict scenarios, with a mean latency of 15.4s. This is suitable for slower governance tasks rather than \ac{Near-RT} arbitration. Kouchaki et al. in \cite{kouchaki2025federated} propose neuroevolution and federated aggregation to coordinate \ac{xApp} models across the \acp{RIC}. Their \ac{OAIC}/srsRAN software testbed incorporates a central neuroevolution \ac{xApp} with a number of decentralised \ac{DRL} \acp{xApp}. However, the authors make no attempt to demonstrate conflict detection and resolution between them. Adamczyk in \cite{adamczyk2023challenges} examines all three \ac{WG3} conflict types and the conceptual difficulties of managing them, but implements and evaluates no mechanism. Chatzistefanidis et al. in~\cite{chatzistefanidis2024maestro} propose MAESTRO, which uses \ac{LLM} agents and an operator-managed arbiter to negotiate service intents between business operators before translating them into \ac{RAN} actions. It focuses on competing tenant objectives rather than \ac{WG3} \ac{xApp} conflict detection.  There is also work on two agent personality experiments showing that a fairly vulnerable agent can be pushed into a state of extreme throughput demand. Machado et al. in \cite{machado2026orion} introduce ORION, which demonstrates the timing constraints of \ac{LLM}-based orchestration. The authors show that intent-to-policy translation takes around 9 seconds, compared with 2.4ms for A1 mediation and 0.7ms for E2 control processing. This reinforces the need to separate out the relatively slow \ac{LLM} reasoning from the time-critical control path. Testing six commercial \acp{LLM} shows that the primary failure mode is unreliable tool use rather than a failure to understand the intent. However, the authors did not implement conflict resolution between competing resource goals across independently admitted intents. Finally, i14y Lab in~\cite{i14y} proposes three conflict impact metrics (Severity, Reach, and Vulnerability Creation) to assist operators in prioritising simultaneous conflicts. Their white paper describes the checks that would be carried out prior to deployment, prior to action and post-action. However, they did not implement the proposed lifecycle. 

\begin{table*}[!b]
\centering
\footnotesize
\setlength{\tabcolsep}{3pt}
\renewcommand{\arraystretch}{1.05}
\caption{L6: Governance and Application Lifecycle Management.}\label{tab:survey_class_l6}
\begin{adjustbox}{max width=\textwidth}
\begin{tabular}{|p{1.07cm}|>{\centering\arraybackslash}p{0.62cm}|>{\centering\arraybackslash}p{0.95cm}|>{\centering\arraybackslash}p{0.6cm}|>{\centering\arraybackslash}p{0.6cm}|>{\centering\arraybackslash}p{0.6cm}|>{\centering\arraybackslash}p{0.6cm}|>{\centering\arraybackslash}p{0.6cm}|>{\centering\arraybackslash}p{0.6cm}|>{\centering\arraybackslash}p{0.6cm}|>{\centering\arraybackslash}p{0.6cm}|>{\centering\arraybackslash}p{0.6cm}|>{\centering\arraybackslash}p{0.86cm}|>{\centering\arraybackslash}p{0.58cm}|>{\centering\arraybackslash}p{0.58cm}|>{\centering\arraybackslash}p{0.58cm}|>{\centering\arraybackslash}p{1.35cm}|p{1.45cm}|}
\hline
\textbf{Paper} & \textbf{Year} & \textbf{Domain} & \textbf{Det.} & \textbf{Mit.} & \textbf{Dir.} & \textbf{Ind.} & \textbf{Imp.} & \textbf{Cen.} & \textbf{Dis.} & \textbf{Hier.} & \textbf{\acs*{RL}} & \textbf{\acs*{MARL}} & \textbf{\acs*{GT}} & \textbf{\acs*{GNN}} & \textbf{Int.} & \textbf{Evidence} & \textbf{Validation} \\
\hline
\cite{banerjee2021intent} & 2021 & \acs*{CAN} &  & -- &  &  &  & \checkmark &  &  &  &  &  &  & \checkmark & Moderate & Prototype \\
\hline
\cite{banerjee2023trust} & 2022 & \acs*{CAN} & \checkmark & \checkmark & \checkmark &  &  & \checkmark &  &  &  &  & \checkmark &  &  & Moderate & Simulation \\
\hline
\cite{adamczyk2023challenges} & 2023 & \acs*{O-RAN} & -- & -- & -- & -- & -- &  &  &  &  &  &  &  &  & None & Conceptual \\
\hline
\cite{chatzistefanidis2024maestro} & 2024 & \acs*{O-RAN} &  &  &  &  &  & \checkmark &  &  &  &  &  &  & \checkmark & Moderate & Testbed \\
\hline
\cite{corici2024towards} & 2024 & \acs*{O-RAN} &  & -- &  &  &  & \checkmark &  &  &  &  &  &  &  & None & Conceptual \\
\hline
\cite{kouchaki2025federated} & 2025 & \acs*{O-RAN} &  &  &  &  &  &  & \checkmark &  & \checkmark & \checkmark &  &  &  & Moderate & Testbed \\
\hline
\cite{armstrong2026oracle} & 2026 & \acs*{O-RAN} &  & \checkmark &  &  &  & \checkmark &  &  &  &  &  &  &  & Moderate & Testbed \\
\hline
\cite{li2026multi} & 2026 & \acs*{O-RAN} & -- & -- & -- & -- & -- & \checkmark &  &  &  &  &  &  & \checkmark & Weak & Prototype \\
\hline
\cite{salmi2026ai} & 2026 & \acs*{O-RAN} & -- & \checkmark & -- & -- & -- & \checkmark &  &  & -- &  & \checkmark &  &  & Weak & Testbed \\
\hline
\end{tabular}
\end{adjustbox}
\end{table*}

\subsection{\ac{O-RAN} Alliance Standardisation}
\label{subsec:oran_std}
\ac{O-RAN} standards address conflicts in several areas. \ac{WG2} reports look at A1 policy and R1 service conflicts~\cite{ORAN_A1PCM_TR,oran_wg2_r1servconfmit_tr}, and \ac{WG3} looks at \ac{Near-RT} \ac{RIC}/E2 conflict mitigation~\cite{oran_wg3_confmit_draft}. \ac{ETSI} \ac{TS}~104~301 V7.0.0 Section~9.3.3.6 in~\cite{etsi2026nearrtric}  defines how the \ac{RIC} platform can request detection, avoidance or resolution assistance from conflicting or helper \acp{xApp} that provide this capability. It defines the roles and request/response steps, leaving the selection of which assistance to provide to the implementation. The Oct 2026 \ac{O-RAN} tech release summary in~\cite{oran2026techrelease} contains configuration conflict management for the \ac{RAN} \ac{NF} \ac{OAM} \ac{SMO} Service and A1 policy conflict management for the A1-related \ac{SMO} Service. O1 provides management for configuration operations~\cite{oran2026o1interface}; the v17.00 referenced is an earlier version than the v20.00 listed in the release summary. These documents define the responsibilities and interfaces but don't specify a single conflict classifier or arbitration objective for all of the interfaces.

\subsection{Adjacent and Enabling Studies}
\label{subsec:adjacent}

Table~\ref{tab:survey_class_adj} contains relevant applications, co-designs, signalling safety measures, and resource allocation work that do not exhibit the control decision conflict mechanisms reviewed in L0-L6. They are included for context but are not part of the detection and mitigation counts we examine in Section~\ref{sec:lessons}. Barker et al. in~\cite{barker2025real} REAL deploys a single \ac{RL} \ac{xApp} for slice-level \ac{PRB} allocation on an \ac{OSC} \ac{RIC}/srsRAN testbed. It shows the trade-offs between the slices under one policy, without the need for a second \ac{xApp} or any inter-\ac{xApp} conflict resolution. Hoffmann et al. in~\cite{hoffmann2023open} use the \ac{OSC} \ac{Near-RT} \ac{RIC} and the \ac{ONF} SD-\ac{RAN} $\mu$ONOS-\ac{RIC} platforms. However, the absence of any conflict mitigation units means that systematic multi-\ac{xApp} conflict experiments are not possible. Qazzaz~et~al. in~\cite{qazzaz2024machine} use a Random Forest to select from four pre-defined resource block allocation policies; conflict management is addressed only in the related work. Wang~et~al. in~\cite {wang2024energy} introduce a \ac{DQN} \ac{xApp} that saves energy by turning off radio chains, with penalties for poor \ac{RSS} and capacity violations. However, the authors did not use any co-deployed \ac{xApp} or reward term to account for \ac{xApp} interactions.

\begin{table*}[!tbp]
\centering
\footnotesize
\setlength{\tabcolsep}{3pt}
\renewcommand{\arraystretch}{1.05}
\caption{Adjacent and Enabling Studies.}\label{tab:survey_class_adj}
\begin{adjustbox}{max width=\textwidth}
\begin{tabular}{|p{1.07cm}|>{\centering\arraybackslash}p{0.62cm}|>{\centering\arraybackslash}p{0.95cm}|>{\centering\arraybackslash}p{0.6cm}|>{\centering\arraybackslash}p{0.6cm}|>{\centering\arraybackslash}p{0.6cm}|>{\centering\arraybackslash}p{0.6cm}|>{\centering\arraybackslash}p{0.6cm}|>{\centering\arraybackslash}p{0.6cm}|>{\centering\arraybackslash}p{0.6cm}|>{\centering\arraybackslash}p{0.6cm}|>{\centering\arraybackslash}p{0.6cm}|>{\centering\arraybackslash}p{0.86cm}|>{\centering\arraybackslash}p{0.58cm}|>{\centering\arraybackslash}p{0.58cm}|>{\centering\arraybackslash}p{0.58cm}|>{\centering\arraybackslash}p{1.35cm}|p{1.45cm}|}
\hline
\textbf{Paper} & \textbf{Year} & \textbf{Domain} & \textbf{Det.} & \textbf{Mit.} & \textbf{Dir.} & \textbf{Ind.} & \textbf{Imp.} & \textbf{Cen.} & \textbf{Dis.} & \textbf{Hier.} & \textbf{\acs*{RL}} & \textbf{\acs*{MARL}} & \textbf{\acs*{GT}} & \textbf{\acs*{GNN}} & \textbf{Int.} & \textbf{Evidence} & \textbf{Validation} \\
\hline
\cite{mwanje2012selforganized} & 2012 & \acs*{SON} &  &  &  &  &  &  & \checkmark &  &  &  &  &  &  & Weak & Simulation \\
\hline
\cite{moysen2014reinforcement} & 2014 & \acs*{SON} &  &  &  &  &  &  & \checkmark &  & \checkmark &  &  &  &  & Moderate & Simulation \\
\hline
\cite{mwanje2014distributed} & 2014 & \acs*{SON} &  & -- &  &  &  &  & \checkmark &  & \checkmark &  &  &  &  & Weak & Simulation \\
\hline
\cite{hoffmann2023open} & 2023 & \acs*{O-RAN} &  &  &  &  &  &  &  &  &  &  &  &  &  & None & Testbed \\
\hline
\cite{bikos2024accord} & 2024 & \acs*{O-RAN} &  &  &  &  &  &  &  &  &  &  &  &  &  & None & Conceptual \\
\hline
\cite{qazzaz2024machine} & 2024 & \acs*{O-RAN} &  &  &  &  &  & \checkmark &  &  &  &  &  &  &  & None & Simulation \\
\hline
\cite{wang2024energy} & 2024 & \acs*{O-RAN} &  &  &  &  &  & \checkmark &  &  & \checkmark &  &  &  &  & None & Simulation \\
\hline
\cite{bakri2025mitigating} & 2025 & \acs*{O-RAN} &  & -- &  &  &  &  & \checkmark &  & \checkmark & \checkmark &  & \checkmark &  & Weak & Data \\
\hline
\cite{barker2025real} & 2025 & \acs*{O-RAN} &  &  &  &  &  & \checkmark &  &  & \checkmark &  &  &  &  & None & Testbed \\
\hline
\cite{hung2025anomaly} & 2025 & \acs*{O-RAN} & -- & -- &  &  &  & \checkmark &  &  &  &  &  &  &  & Strong & Testbed \\
\hline
\cite{gelban2026detecting} & 2026 & \acs*{O-RAN} & -- &  &  &  &  &  &  &  &  &  &  &  &  & Weak & Testbed \\
\hline
\cite{hou2026veraran} & 2026 & \acs*{O-RAN} &  &  &  &  &  &  &  &  &  &  &  &  &  & Strong & Simulation \\
\hline
\cite{machado2026orion} & 2026 & \acs*{O-RAN} &  &  &  &  &  &  &  &  &  &  &  &  &  & Strong & Testbed \\
\hline
\cite{li2024conflict} & 2024 & \acs*{IBN} &  &  &  &  &  & \checkmark &  &  & \checkmark &  &  &  &  & Weak & Simulation \\
\hline
\end{tabular}
\end{adjustbox}
\end{table*}

\subsection{Cross-Lifecycle Synthesis of the Classification}
\label{subsec:crosslifecycle}
We find that most work to date has focused on detection (L2) and resolution (L3). There is less work on enforcement (L4), verification and attribution (L5) or ongoing governance (L6). To ensure that a conflict management solution is effective, we should not rely solely on detection accuracy or mitigation reward in a paper. We should also verify that the selected decision is interceptable, enforceable, and verified against \ac{RAN} measurements. Mitigation architecture also affects mitigation execution. In a centralised or hierarchical architecture, a \ac{CMF}, a service bus, or an interface interceptor would simply block the conflicting control decision if it was rejected before E2 or A1 execution. In distributed \ac{SON} or \ac{CAN}, conflict management relies on shared parameter definitions, shared utility models, and trust assumptions. This becomes harder to maintain in distributed architectures with multiple control applications from independent vendors. 

We observe that Graph and \ac{GNN} methods are primarily focused on (L1) modelling and (L2) detection. However, we found that none of the graph approaches evaluated actually go the full route from learning the relationships through to resolution, enforcement and verification. Also, complex learning mechanisms do not ensure better performance, as some of the graph models use synthetic graphs or self-generated labels. In contrast, simpler models, such as priority and interception methods, have been tested on actual physical \ac{RIC} platforms in testbeds, providing more reliability and guaranteed performance in real \ac{RAN} deployments. Implicit conflict is the least studied of the three conflict types. Implicit conflict occurs when there is an unregistered relationship and only becomes apparent when an action negatively impacts another \ac{KPI} or operating condition. These conflicts can be better studied by establishing root cause analysis: first isolating the effect of the action from the effects of traffic, channel, and workload changes, then checking what happens on recovery. Some work in the literature addresses aspects of this process. However, none of the work in the corpus we have looked at demonstrates the full detection, resolution, enforcement, and verification loop for implicit conflicts on a testbed.

\section{Simulation, Testbed, and Dataset Generation Challenges}
\label{sec:eval_challenges}

\begin{figure*}[!htbp]
  \centering
  \includegraphics[width=0.55\textwidth]{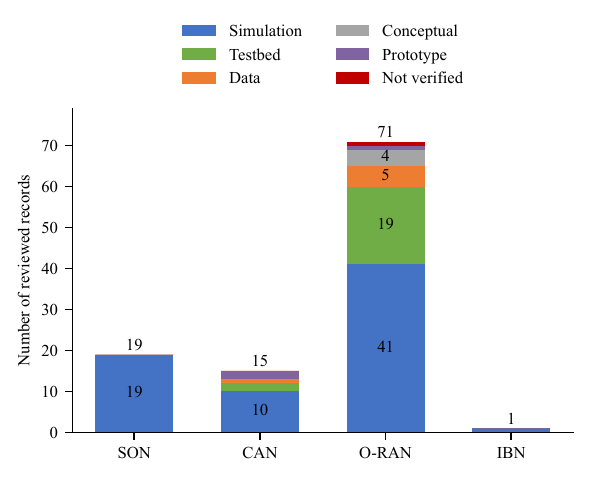}
  \vspace{-0.2cm}
  \caption{Evaluation settings for the reviewed papers across paradigms.}
  \label{fig:validation_bar}
\end{figure*}

\begin{figure*}[!htbp]
  \centering
  \includegraphics[width=\textwidth]{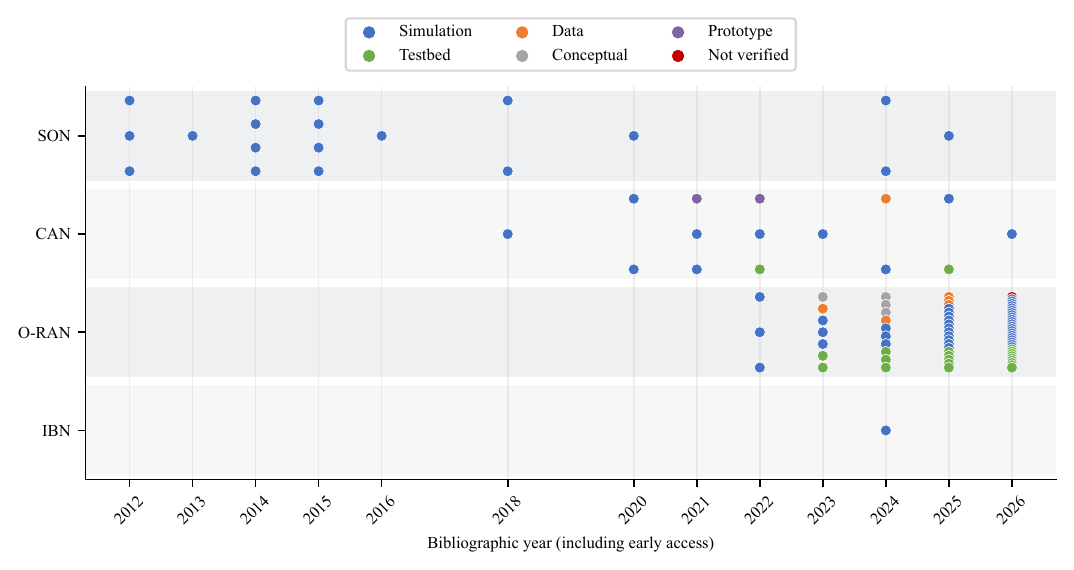}
  \vspace{-0.8cm}
  \caption{Timeline of the reviewed papers across paradigms.}
  \label{fig:timeline}
  \vspace{-0.5cm}
\end{figure*}

Figures~\ref{fig:validation_bar} and~\ref{fig:timeline} illustrate the classification of reviewed papers across three major paradigms covered in this survey, along with one \ac{IBN}-focused paper and a timeline of published papers across these paradigms, respectively. Of the 15 \ac{CAN} papers reviewed in this survey, 10 are simulations, 2 are emulator/testbed evaluations, 2 are functional prototypes, and 1 is an operator-data study. Of the 71 \ac{O-RAN} papers, 41 are simulations, 19 are controller/testbed evaluations, 5 are offline data studies, 4 are conceptual contributions, 1 is a textual prototype, and 1 is a record with an unverified evaluation methodology. There is also 1 \ac{IBN}/\ac{MEC} comparator that is based on simulation. These are all bibliographic records, not separate experiments or full conflict management implementations. Tables~\ref{tab:eval_methodology} and~\ref{tab:eval_recent} map the reviewed papers based on evaluation method, platform, experimental scale, coordination architecture, and evidence limits.

\subsection{\ac{SON}: Simulation-Only Evaluation and Reproducibility Constraints}
\label{subsec:son_eval}
All of the reviewed \ac{SON} studies use a variety of simulation platforms and network sizes. Moysen et al. in ~\cite{moysen2015coordination,moysen2015self,moysen2018conflict} in their coordination papers use ns-3 LENA for the \ac{CCO}/\ac{ICIC} works. Their most recent prediction-based study looks at six cells and 108 \acp{UE}. Iacoboaiea~et~al. in~\cite{lee2024scalable,lee2025conflict} use a proprietary simulator with 21 macro and 6 pico cells~\cite{iacoboaiea2016son}. \ac{SLC2} and \ac{CADC} use ns-3/ns3-gym. Other works in~\cite{munoz2015load,chen2012difference} go as far as 57 cells for the fuzzy Q-learning and 61 cells for the joint per-\ac{UE} handover configuration study. Several papers listed here do not include random seeds; therefore, reproducibility of their results remains a concern. Most of the works test their solution on a few interacting function types under a very limited set of scenarios rather than just cell count. We should note that larger simulations do not guarantee a production-ready application when the applications and their control relationships are all fixed~\cite{munoz2015load,chen2012difference,iacoboaiea2016son,lee2025conflict}.

\subsection{\ac{CAN}: Controller-Level Simulation and the Intent Evaluation Gap}
\label{subsec:can_eval}
\ac{CAN} studies use functional prototypes, synthetic utility models, and network simulations. Early Controller and game-theoretic studies use synthetic utilities~\cite{banerjee2020ran,banerjee2021game}, and the weighted Controller studies also use radio network simulations~\cite{banerjee2021optimal,banerjee2022toward}. A working decision prototype on its own does not show any improvement in radio \acp{KPI} or any improvement in terms of when it would be viable to deploy. Baktir~et~al. in~\cite{baktir2022intent} use an end-to-end emulator with one \ac{gNB}, one \ac{UPF}, six \acp{UE}, and a Kubernetes-hosted core. Dehghan Biyar~et~al. in~\cite{biyar2025autonomous} evaluate a container-based emulator with fifteen \acp{UE} and five intents. Kattepur~et~al. in~\cite{kattepur2024convergence} demonstrate planning using an operator measurement snapshot and predefined examples, without live execution or detector evaluation. These studies highlight the difficulty of translating intents into measurable \acp{KPI} and show why planning towards a goal should be distinguished from demonstrated conflict resolution.

\subsection{\ac{O-RAN}: Testbed Platforms, Dataset Generation, and Remaining Gaps}
\label{subsec:oran_eval}
\ac{O-RAN} studies use controller prototypes, radio simulations, \ac{OTA} testbeds, and offline data (each providing a different kind of evidence). \ac{CoDR} in~\cite{riggio2026codr} takes an \ac{OSC} implementation and combines it with a simulated multi-cell \ac{RAN} to demonstrate its pipeline, with physical \ac{RAN} and carrier-scale testing still to be done. The following discusses the most-used experimental platform in \ac{O-RAN} studies.

\subsubsection{Simulation-Based \ac{O-RAN} Studies}
We observe that most algorithmic work starts with simulation. Wadud et al.~\cite{wadud2023conflict,wadud2024qacm} use synthetic Gaussian utilities. \ac{QACM} also looks at an \ac{ES}/\ac{CCO} case in the MATLAB 5G Toolbox. Neither paper by these authors actually looks at ns3-oran. Adamczyk et al. in~\cite{adamczyk2023conflict,adamczyk2023detection} evaluate their Conflict Mitigation Framework and their indirect conflict detector in simulation. Both approaches are based on predefined parameter groups, which may not be available for independently developed, closed-source \acp{xApp}. Their recent \ac{ACCoRD} study in~\cite{adamczyk2026accord} replaces the rule-based resolution with a \ac{PPO}-Clip actor-critic. It looks at \ac{MRO}-\ac{MLB} coordination across 8 base stations and 3 \ac{UE} load levels, using a normalised Relative Penalty Score. It again only looks at indirect conflicts and doesn't involve a physical \ac{RIC} deployment. Varshney et al. in~\cite{varshney2025xai4c} also evaluate their \ac{SHAP}-based detector in simulation. The \ac{DRL} slicer was trained on Colosseum data, but again the detector evaluation is not a new hardware experiment.

\subsubsection{Testbed Evaluations}
Testbed studies include the Colosseum/OpenRAN Gym experiments and \ac{OTA} deployments. PACIFISTA in~\cite{del2024pacifista} profiles 8 \acp{xApp} using a single \ac{BS}, 6 \acp{UE} and 3 slices. The \ac{OTIC} study in~\cite{sultana2025experimental} examines direct conflict between two slice-\ac{PRB} \acp{xApp} over 20 runs, reporting throughput variation but no latency figures and no assessment of indirect/implicit conflicts. PandORA in~\cite{tsampazi2024pandora} looks at application training and deployment rather than runtime conflict resolution. Hoffmann~et~al. in~\cite{hoffmann2023open} note that the absence of mitigation units meant that multi-\ac{xApp} conflict testing was not possible. The most recent preprints include xTRUCE's \ac{OAI}/FlexRIC arbitration, AURA's \ac{OAI} stability tests and Sinha~et~al.'s OCUDU/Open5GS/\ac{USRP} slicing and scheduling experiment with five commercial \acp{UE}~\cite{xia2026xtruce,natanzi2026aura,sinha2026experimental}. All of these experiments are still very much restricted to a single workload and a single deployment size.

\subsubsection{Data-Driven Studies and the Labelling Challenge}
Data-driven studies use \acp{KPI} taken from experimental platforms or operator networks. Armstrong et al. in~\cite {armstrong2024pre} use historical \ac{RAN} data to predict the impact of proposed configuration changes. Later, \ac{CAMINO} from the same authors in~\cite{armstrong2025camino} takes this a step further by using data from 800 \ac{LTE} cells along with weather, traffic and event data. It does context-aware screening but provides no detector baseline, accuracy figure, or statistical significance test. Reliable conflict labels are still very hard to get, as the traffic and radio conditions can produce \ac{KPI} changes that are very similar to those that would be caused by an application. Controlled conflict injection can be used to help identify the root cause, but that in turn may limit the scale and realism of the experiment. The corpus under review provides no shared benchmark for any of the three \ac{WG3} conflict types and no benchmark for recovery following intervention. Gelban~et~al. in~\cite{gelban2026detecting} labels a dataset that is specifically concerned with control message ordering faults rather than competing valid \ac{RAN} actions. The public Colosseum and OpenRAN Gym datasets also do not yet have a common conflict benchmark with reproducible scenarios, agreed injection methodology and shared evaluation metrics.

\begin{table*}[!ht]
\centering
\caption{Representative studies on evaluation methodology across \ac{SON}, \ac{CAN}, and \ac{O-RAN} conflict management.}
\label{tab:eval_methodology}
\setlength{\tabcolsep}{3.0pt}
\renewcommand{\arraystretch}{1.10}
\footnotesize
\resizebox{\textwidth}{!}{%
\begin{tabular}{|p{2.4cm}|p{1.05cm}|p{1.3cm}|p{2.5cm}|p{1.6cm}|p{2.2cm}|p{3.8cm}|p{3.8cm}|}
\hline
\textbf{Paper} & \textbf{Domain} & \textbf{Method} & \textbf{Platform / Simulator} & \textbf{Scale} & \textbf{Coordination} & \textbf{Key Evaluation Challenge} & \textbf{Implementation Challenges} \\
\hline
\cite{moysen2015coordination,moysen2018conflict} & \ac{SON} & Simulation & ns-3 LENA & \ac{CCO}/\ac{ICIC}: 6 macro + 20 \acp{HeNB}; \ac{MLB}/\ac{MRO}: 6 cells, 108 \acp{UE} & Per-\ac{eNB} \ac{D-SON} / central advisory coordinator & Different experiments; no common benchmark or uncertainty estimates & subMDP composition and surrogate quality depend on the model \\
\hline
\cite{iacoboaiea2016son} & \ac{SON} & Simulation & Proprietary HetNet & 21 macro cells, 6 pico cells & Centralised (\ac{SONCO}) & Simplified mobility; not reproducible without the private simulator & State-aggregation accuracy bounded by manually-tuned fairness weights; no closed-form weight selection \\
\hline
\cite{lateef2015lte} & \ac{SON} & Simulation & Unnamed radio simulator; \ac{ILP} in CPLEX & 7 cells, 70--210 users & X2 coordination plus O\&M & One \ac{MRO}/energy case; no repetitions or numerical performance table & Combinatorial chain-reaction reasoning and model-specific constraints \\
\hline
\cite{lee2025conflict} & \ac{SON} & Simulation & ns-3 + ns3-gym & Small-scale & Distributed (cluster \ac{DRL}) & No open code or data artefacts; limited scenario diversity & Cluster construction sensitive to conflict graph quality; inter-cluster dependencies require dedicated minimisation \\
\hline
\cite{banerjee2020ran,banerjee2021game} & \ac{CAN} & Simulation & Synthetic utility models / Python & \ac{CBM}: up to 6 \acp{CF} & Central Controller & No instantiated radio network in the synthetic study & Utility scaling, disclosure and configuration sampling assumptions \\
\hline
\cite{baktir2022intent} & \ac{CAN} & Testbed & Dockerised emulator; Kubernetes core & 1 \ac{gNB}, 1 \ac{UPF}, 6 \acp{UE} & Central intent manager & One reported trace; no statistical baseline & Grounding intents and predicting cross-\ac{KPI} effects \\
\hline
\cite{kattepur2024convergence} & \ac{CAN} & Data & Operator measurement snapshot; \ac{PDDL}/LPG-TD & Premium / normal \ac{eMBB} and \ac{FWA} exemplar & Central reasoning engine & No live actuation or evaluated conflict baseline & Planning-domain and forecast assumptions \\
\hline
\cite{wadud2023conflict,wadud2024qacm} & \ac{O-RAN} & Simulation & Synthetic Gaussian utilities; one MATLAB 5G Toolbox \ac{QACM} case & 4/5 synthetic \acp{xApp}; radio case separate & Central \ac{Near-RT} \ac{CMS} model & No deployed \ac{RIC} or common radio scenario across the two papers & \ac{KPI} predictors, threshold scaling and search cost \\
\hline
\cite{adamczyk2023conflict,adamczyk2023detection} & \ac{O-RAN} & Simulation & System-level radio simulation & 19 \ac{BS}/380 \ac{UE}; 8 \ac{BS}/310 \ac{UE} respectively & Central \ac{CMF} & Implicit detector untested; no timing or uncertainty measurements & Control-message interception and manually registered parameter groups \\
\hline
\cite{adamczyk2026accord} & \ac{O-RAN} & Simulation & Unnamed system simulator & 8 \ac{BS}, 57--100 \acp{UE} & Central \ac{CR} Agent & Indirect conflicts only; no physical \ac{RIC} validation & \ac{PPO} reward tuning and at most three decisions per inference \\
\hline
\cite{del2024pacifista} & \ac{O-RAN} & Testbed & Colosseum + OpenRAN Gym & 1 \ac{BS}, 6 \acp{UE}, 3 slices & Centralised (\ac{Non-RT} \ac{RIC} / \ac{SMO}) & Single-cell evaluation; profiling requires sandbox and O1/R1 access & Sample-efficiency of statistical profiling; pre-deployment profiling overhead \\
\hline
\cite{sultana2025experimental} & \ac{O-RAN} & Testbed & \ac{OSC} \ac{Near-RT} \ac{RIC}; srsRAN / Open5GS / \ac{USRP} & 1 cell, 2 \acp{xApp}, 3 handsets & Central \ac{CMF} & Direct \ac{PRB} conflict only; no latency/overhead measurement & Limited workload and specialised \ac{OTIC} access \\
\hline
\cite{tsampazi2024pandora} & \ac{O-RAN} & Testbed & Colosseum & 1 \ac{BS}, 6 \acp{UE} & Distributed (\ac{DRL} agents, \ac{Near-RT} \ac{RIC}) & \ac{DRL}-only; no rule-based or hybrid mitigation comparison & Multi-agent training stability; global vs.\ per-slice trade-off; agent black-box behaviour \\
\hline
\cite{hoffmann2023open} & \ac{O-RAN} & Testbed & \ac{OSC} \ac{Near-RT} \ac{RIC} + \ac{ONF} SD-\ac{RAN} $\mu$ONOS-\ac{RIC} & Small-scale & Not evaluated (no mitigation unit) & No conflict-mitigation unit; multi-\ac{xApp} conflict testing blocked entirely & Architectural ambiguity in \ac{O-RAN} specifications; cross-interface gating unsolved \\
\hline
\cite{armstrong2024pre} & \ac{O-RAN} & Data-driven & Operator network (real \ac{PM}/\ac{CM}) & ${\sim}800$ cells & Centralised (\ac{Non-RT} \ac{RIC}) & Cell-similarity metric definition delicate; risk of spurious correlations & Cell-similarity metric design requires care; data-quality dependent \\
\hline
\cite{armstrong2025camino} & \ac{O-RAN} & Data-driven & 800-cell \ac{LTE} operator network + real weather/traffic \acp{API} & 800 cells, real \ac{PM}/\ac{CM} + external data & Centralised (\ac{Non-RT} \ac{RIC} / \ac{SMO}) & Rule logic validated only against historical data, not on a live control loop; real-world \ac{API} reliability unverified; prediction accuracy not compared against other schemes under identical conditions & Static domain-specific rules require manual maintenance; external data latency could affect conflict resolution responsiveness; similarity clustering may not generalise to unseen configurations \\
\hline
\cite{varshney2025xai4c} & \ac{O-RAN} & Simulation & Simulation with a Colosseum-trained \ac{DRL} slicer & Small-scale & Centralised (\ac{Near-RT} + \ac{Non-RT} \ac{RIC}) & Detection only; no mitigation pipeline; dataset not publicly released & \ac{XAI} compute overhead; detector is reactive, not proactive \\
\hline
\end{tabular}%
}
\end{table*}

\begin{table*}[!ht]
\centering
\caption{Recent studies: evaluation settings and evidence limits.}
\label{tab:eval_recent}
\setlength{\tabcolsep}{3.0pt}
\renewcommand{\arraystretch}{1.10}
\footnotesize
\resizebox{\textwidth}{!}{%
\begin{tabular}{|p{2.4cm}|p{1.05cm}|p{1.3cm}|p{2.5cm}|p{1.6cm}|p{2.2cm}|p{3.8cm}|p{3.8cm}|}
\hline
\textbf{Paper} & \textbf{Domain} & \textbf{Method} & \textbf{Platform / Simulator} & \textbf{Scale} & \textbf{Coordination} & \textbf{Key Evaluation Challenge} & \textbf{Implementation Challenges} \\
\hline
\cite{riggio2026codr} & \ac{O-RAN} & Simulation & \ac{OSC} with simulated multi-cell \ac{RAN} & 5 concurrent \acp{rApp} & Central intent pipeline & Observable intents and own conflict taxonomy & Configurable resolution; physical \ac{RAN} untested \\
\hline
\cite{xia2026xtruce} & \ac{O-RAN} & Testbed & Simulation + \ac{OAI}/FlexRIC \ac{OTA} & Multi-\ac{xApp} prototype & Central constrained arbiter & Preprint; encoded-model safety & Feasible fallback, state accuracy and solver-delay handling \\
\hline
\cite{natanzi2026aura} & \ac{O-RAN} & Testbed & OpenAirInterface & Shared-resource \ac{rApp} scenario & Central stability arbiter & Preprint; model-conditional convergence & Feasibility invariants, dwell times and deadbands \\
\hline
\cite{sinha2026experimental} & \ac{O-RAN} & Testbed & OCUDU / Open5GS / \ac{USRP} X310 & 5 commercial \acp{UE} & Command-bus checks & Preprint; slicing/scheduling workload & Core-side throughput proxy and explicit constraint predicates \\
\hline
\cite{kim2026mobilityenergy} & \ac{O-RAN} & Simulation & \ac{CTMC} + raw-event simulation & Sleeping-target handover case & Weighted arbiter & Residence/wake distribution assumptions & Useful-service versus \ac{O-RU}-power trade-off \\
\hline
\cite{wadud2026costtriage} & \ac{O-RAN} & Simulation & Simulator-driven episode evaluation & 4 \acp{xApp}, 30 Monte Carlo runs & Central triage & Preprint; modelled intervention costs & Economic calibration and transfer \\
\hline
\cite{wadud2026confidence} & \ac{O-RAN} & Simulation & ns3-oran / 5G-LENA & \ac{CCO}/\ac{ES} pair & Central confidence triage & Preprint; one direct-conflict scenario & Interval calibration and distribution shift \\
\hline
\end{tabular}%
}
\end{table*}
\section{Lessons Learned and Future Work}
\label{sec:lessons}
Reviewing conflict management research across \ac{SON}, \ac{CAN}, and \ac{O-RAN} paradigms over the past one and a half decades reveals the similarity and dissimilarity of conflict across these paradigms and highlights the existing limitations and future research trends. The following discusses the lesson learned and future research directions. 

\subsection{From Static Arbitration to Adaptive Coordination}
The reviewed methods in this survey consider interactions between control functions or applications in very different ways across paradigms. The static approach simply restricts the parameter space to eliminate any known conflicts~\cite{zia2014policy,huang2018conflict}. On the other hand, Fuzzy Q-learning and Pareto learning methods seek to incorporate the competing objectives into a single policy~\cite{munoz2015load,kim2024load}.  In contrast, \ac{CAN} bargaining takes a different approach: each function is required to report its utility and its preferred range of configuration parameters~\cite{banerjee2021game,banerjee2021optimal}. The evolution from static control to adaptive cooperation is not straightforward. In fact, cooperative learning actually predates several weighted controller studies~\cite{mwanje2018synchronized,banerjee2022toward}. Learning-based methods depend on the rewards, features, and constraints that are chosen. Based on results in the literature, we cannot conclude that learning in general is better than static coordination, or vice versa. 

\subsection{The Detection Gap and the Graph-Based Turn}
Earlier coordination schemes across paradigms tend to assume that the parameter relationships are already known. However, many runtime detection methods identify parameter relationships during operation. For example, Moysen and Giupponi in~\cite{moysen2015coordination} have a resource conflict detector in \ac{D-SON}. In \ac{O-RAN}, graph-based detection methods are used to understand these relationships. However, they still require prior information to build the relationship graphs. Zolghadr et al. in~\cite{zolghadr2025learning} learn the parameter-\ac{KPI} relationships and then add the application relationships from known subscriptions. Later, Santos et al. in~\cite{santos2026conflict} train their model on a ground truth relationship matrix. GRAPHICA in~\cite{shami2025ran} uses synthetic binary state graphs and applies conflict labels inconsistently. All these stated works are promising. However, none demonstrate fully automatic relationship discovery or reliable scaling across vendors when the vendor's internal application behaviour is unknown.

\subsection{Implicit Conflicts as the Unresolved Frontier}
As stated earlier, implicit conflicts are by far the least well-validated conflict type. Their parameter relationships are not pre-established, so they only become apparent after some action leads to an unexpected \ac{KPI} or operating-state degradation. The same kind of degradation may happen due to changes in traffic, radio conditions, mobility or other applications. Therefore, the root cause must be determined by considering what would have happened in the absence of the suspected control decision. A good solution must identify the faulty application or interaction, apply a correction, and verify that the network recovers. Some existing work touches on various aspects of this. Adamczyk and Kliks in~\cite{adamczyk2023conflict} detect the implicit conflicts once the impact has been realised, so they do not actually prevent the initial harm. Zolghadr et al. in~\cite{zolghadr2025learning} work out the parameter-\ac{KPI} relationships from time-series data, but they are limited to a single level of dependency and add the \ac{xApp} relationships after the fact. Counterfactual estimation, \ac{XAI}-based attribution and observability methods also attempt to address this issue partially, and all of them are discussed in L5~\cite{hou2025if,varshney2025xai4c,riggio2026observability}. However, none of the reviewed studies demonstrates the full process on a physical \ac{RIC} (i.e., detect an unregistered side effect, identify the root cause, apply a fix, and verify recovery), which is a major research gap identified in this survey. 

\subsection{Validation Maturity and the Simulation Gap}
Section~\ref{sec:eval_challenges} shows that simulation and functional prototypes are still the norm for handling conflict in \ac{RAN}. There is evidence from hardware-in-the-loop experiments and operator data studies for a number of the lifecycle stages. Armstrong et al. in~\cite{armstrong2025camino} use historical data from 800 cells, which is evidence of large-scale data analysis, but again there is no evidence of live enforcement across multiple \acp{RIC}. xTRUCE in~\cite{xia2026xtruce} and AURA in~\cite{natanzi2026aura} provide some arbitration and stability results from a limited number of \ac{OAI} experiments, respectively. A shared benchmark should be developed that can distinguish between conflicts arising from valid control decisions and signalling faults to avoid false positive conflict flagging. Also, there should be an assessment after each conflict mitigation intervention to check whether the network recovers following the intervention or not.

\subsection{Open Challenges and Future Directions}

Several problems remain open and need further research. These are listed as follows: 

\subsubsection{Standardised Conflict Benchmarking}

The papers reviewed in this survey provide no common benchmark that covers modelling, detection, resolution, enforcement and verified recovery. The use of  various simulators, scenarios and metrics makes it very difficult to compare results. A common benchmark would need to cover two \ac{xApp} conflicts like \ac{MLB}--\ac{MRO} conflicts, multi-\ac{xApp} conflicts with known root causes and larger heterogeneous networks using both synthetic and real traffic traces. Also, use of a common and open-source experimental platform should be encouraged for future studies. 

\subsubsection{Multi-Vendor Black-Box Detection}

Generally, the \acp{MNO} will not have access to the control applications’ code and internal models. Therefore, any methods that are predicated on unavailable parameter relationships are very hard to apply. In such a case, \ac{KPI} based detection is an alternative, but many of the \ac{GNN} and \ac{RNN} papers use scenarios where the conflict graph is known. Developing a reliable detection method handling incomplete parameter--\ac{KPI}--\ac{xApp} relationships and changing network conditions is still an open problem.

\subsubsection{Implicit-Conflict Attribution and Resolution}

Systems must be able to separate out \ac{KPI} degradation due to applications from that due to traffic or radio conditions. They must also be able to report how certain they are about the cause. One possible approach is to combine causal graphs, predictions of what would happen in the absence of the suspected action, action logs and calibrated uncertainty bounds. The tests should introduce previously unobserved relationships and measure attribution error, number of unnecessary interventions, recovery time and whether the \ac{KPI} in question recovers.

\subsubsection{Enforcement, Rollback, and Post-Resolution Verification}

Enforcement and verification (L4-L5) are less mature than detection and resolution (L2-L3). There's still a lot of work to be done on intercepting the request, applying the group action in a consistent manner, blocking the unsafe parts of the action, restoring the saved config and defining cooldown periods. The experiments need to confirm that the changes were actually applied and that any conflicts are either gone or reappear after resolution.

\subsubsection{Cross-\ac{RIC} Conflict Coordination}

The majority of the frameworks that have been reviewed deal with conflict resolution within a single \ac{RIC}. There can be conflicts where applications on different \acp{RIC} have an impact on a shared resource or on a neighbouring cell. Game theoretic approaches and hierarchical \ac{RL} could potentially be used to provide some form of coordination, though they would need to deal with the fact that the operators will have different action times, different \acp{KPI} and different privacy requirements.

\subsubsection{Vertical Cross-Tier and Lower-Loop Coordination}

A control decision could involve a \ac{Non-RT} \ac{rApp}, a \ac{Near-RT} \ac{xApp} and a lower loop \ac{dApp} or \ac{RAN} function. They will have different information and different timing and may not even support a common rollback. There will be a need for clear control ownership, records of where the policy and action originated, deadline aware decisions and clearly defined enforcement authority. The lower loop methods will need to be tested against measured implementation deadlines rather than assuming that every action will need sub-millisecond execution.

\subsubsection{Intent-Aligned Conflict Resolution}

Operator intents get turned into policies and application actions. Much of the work in this area looks at either the contradictions between intents or the conflicts between application actions, with very little attempt to link the two. Runtime conflicts should be traced back to the relevant intents so that the appropriate level of correction can be applied. Conflicting actions don't necessarily mean there are contradictory intents and consistent actions don't necessarily mean that all the intents have been satisfied.

\subsubsection{Adversarial and Trust Dimensions}

Banerjee~et~al. show that a malicious \ac{CF} can manipulate \ac{CAN} arbitration via false utility reports~\cite{banerjee2023trust}. Similarly misleading \ac{xApp} reports could lead to false conflict alarms or the blocking of legitimate actions in \ac{O-RAN}. Possible mitigations would be to check the provenance and integrity of the reports, detect anomalous measurements and to use optimisation methods that are robust to bounded error or manipulation.

\subsubsection{Agentic and Generative-\ac{AI} Governance}

\acp{LLM} can assist with intent translation, application selection and explanation. The reported ORION timings in~\cite{machado2026orion}, however, would suggest that the reasoning should be in the \ac{Non-RT} domain, separate from the time-critical control path. Future systems should check the output format and tool calls, maintain an auditable record of decisions and have safe fallbacks in the case of inconsistent policies. Evaluation should be performed on intent accuracy, repeatability, constraint compliance, latency and any conflicts introduced by the application combinations that are generated.

\subsubsection{Scalability in \ac{AI}-Native 6G \acp{RAN}}

Checking every application pair gives you $O(N^2)$ possible relationships, though the actual cost of processing will be dependent on the method. \ac{CADC} in~\cite{lee2025conflict} reduces the coordination down to base station clusters, and \ac{TACIT} in~\cite{armstrong2026tacit} reports linear complexity per proposal under its own set of assumptions. Scale has to be reported properly. The scale of the test is different in different respects: number of applications, number of base stations, number of graph nodes and number of deployed agents are all different things. For instance, synthetic tests with up to 50 \acp{xApp} don't necessarily establish equivalent multi-vendor \ac{RIC} performance~\cite{wadud2026ai}. Keeping the relationship model and control ownership up to date as applications join is still an open issue.

\section{Conclusion}
\label{sec:conclusion}
This survey reviews conflict management studies covering \ac{SON}, \ac{CAN}, and \ac{O-RAN} paradigms, considering a lifecycle perspective. It covers key conflict management stages like conflict avoidance, modelling, detection, resolution, enforcement, verification, and governance. The investigation shows that learning-based control and independently developed applications make coordination among them much harder. Also, it becomes even more complex when the parameter--\ac{KPI}--\ac{xApp} relationships are unknown or change over time. Key conflict mitigation approaches are rules-based, game-theoretic bargaining, learned graphs, attribution, and uncertainty-aware decisions that offer complementary solutions to this problem~\cite{huang2018conflict,banerjee2021game,zolghadr2025learning,riggio2026observability,wadud2026confidence}. However, comparing and identifying the best mechanism among them is difficult due to different evaluation settings used in different studies and limited comparisons in the existing literature. The main gap identified in this survey is the lack of complete operational validation. Complete operational validation indicates a complete end-to-end feedback loop to check whether detecting a conflict and selecting a mitigation response recovers the network performance effectively or not. Section~\ref{sec:eval_challenges} highlights all the evaluation limitations and challenges found in the existing literature. Future studies in this domain should focus on resolving implicit conflicts, multi-vendor and cross-tier coordination, reliable enforcement and recovery, and scalable governance as discussed in Section~\ref{sec:lessons}. The future of fully functional, dependable autonomous \acp{RAN} will depend on how effective the application coordination framework or mechanisms will be. Therefore, conflict in \ac{RAN} is one of the key areas where future work should focus. 

\section*{Acknowledgment}
This work was supported by the European Union's Horizon Europe research and innovation programme through the Marie Skłodowska-Curie Staff Exchanges grant under Grant Agreement No. 101086343 (RE-ROUTE).



\bibliographystyle{ieeetr}
\bibliography{references}

\end{document}